\documentclass[a4paper,11pt]{article}
\pdfoutput=1 

\usepackage{jheppub} 
                     
\usepackage[utf8]{inputenc}
\usepackage{graphicx} 
\usepackage{grffile}
\usepackage{subcaption}
\usepackage{dcolumn}  
\usepackage{makecell}
\usepackage{colordvi}
\usepackage{color}
\usepackage{epstopdf}
\usepackage{amssymb}
\usepackage{amsmath}
\usepackage{bm}
\usepackage{slashed}
\usepackage{enumerate}
\usepackage{url}
\usepackage{morefloats}
\usepackage{lineno}

\usepackage{natbib}

\usepackage[colorlinks=true,
urlcolor=blue,
linkcolor=blue,
citecolor=blue,
linktocpage=true,
pdfproducer=medialab,
pdfa=true,
anchorcolor=blue]{hyperref}

 \usepackage{footnotebackref}

\usepackage{tabularx,array}

\newcolumntype{C}{>{\centering\arraybackslash}X}
\usepackage{multirow}

\usepackage{lineno}

\usepackage{float}
\usepackage{xcolor}

\usepackage[title]{appendix}

\renewcommand{\thesection}{\arabic{section}}

\usepackage{changepage}
\usepackage{booktabs}

\input belle2sym.tex

\preprint{}

\begin{document}


\title{Analysis note: measurement of energy-energy correlator in \boldmath $e^{+}e^{-}$ collisions at \boldmath $91$~GeV with archived ALEPH data}

\author[a]{Hannah Bossi,}
\author[a]{Yu-Chen Chen,}
\author[b]{Yi Chen,}
\author[b]{Jingyu Zhang,}
\author[a]{Gian Michele Innocenti,}
\author[d]{Anthony Badea,}
\author[e]{Austin Baty,}
\author[c]{Marcello Maggi,}
\author[a]{Chris McGinn,}
\author[a]{Yen-Jie Lee}

\affiliation[a]{Massachusetts Institute of Technology, Cambridge, Massachusetts, USA}
\affiliation[b]{Vanderbilt University, Nashville, Tennessee, USA}
\affiliation[c]{INFN Sezione di Bari, Bari, Italy}
\affiliation[d]{Enrico Fermi Institute, University of Chicago, Chicago IL}
\affiliation[e]{University of Illinois Chicago, Chicago, Illinois, USA}%

\emailAdd{hannah.bossi@cern.ch}
\emailAdd{janice\_c@mit.edu}
\emailAdd{luna.chen@vanderbilt.edu}
\emailAdd{jingyu.zhang@cern.ch}
\emailAdd{ginnocen@mit.edu}
\emailAdd{anthony.badea@cern.ch}
\emailAdd{abaty@uic.edu}
\emailAdd{cfmcginn@mit.edu}
\emailAdd{yenjie@mit.edu}

\date{\today}

\abstract{
Electron-positron ($e^+e^-$) collisions provide a clean environment for precision tests of Quantum Chromodynamics (QCD) due to the absence of hadronic initial-state effects. We present a novel analysis of archived ALEPH data from the Large Electron-Positron  Collider at the $Z$ pole, leveraging energy-energy correlators (EECs) to study hadronic energy flow with unprecedented precision. The two-point EEC is measured as a function of the angular separation between particles spanning from the collinear to the back-to-back limits in a remarkably differential test of perturbative and non-perturbative QCD. The results are consistent with previous LEP measurements and provide significantly improved precision and finer angular binning resolution, especially at small angles and in the back-to-back limit. Comparisons with \textsc{pythia} 6 simulations show overall agreement, with deviations in key kinematic regions offering insights into hadronization. The measurement performed here connects to new opportunities for precision QCD studies in archived and future collider data. 
}


\maketitle
\flushbottom

\section{Introduction}
\label{sec:Introduction}
The $e^{+}e^{-}$ collision environment provides the cleanest setting for studying Quantum Chromodynamics (QCD), as the colliding objects are fundamental particles. Unlike hadronic collisions, $e^{+}e^{-}$ collisions avoid complications such as beam remnants, gluonic initial-state radiation, and parton distribution functions. In recent years, re-analysis of the archived data from the Apparatus for LEp PHysics (ALEPH) experiment at the Large Electron Positron Collider (LEP) has offered a pathway to investigate the current open questions of QCD in a well-understood system. For example, two particle correlations reanalyzed with LEP 1~\cite{Badea:2019vey} and LEP 2~\cite{Chen:2023njr} data reveal a potential excess in the highest multiplicity interval of LEP 2 data not seen in an archived \textsc{pythia} 6 Monte Carlo (MC) sample from ALEPH, providing new insights into the presence of signatures typically attributed to hydrodynamical flow in small systems. The archived data also enables the application of modern experimental techniques developed in recent years. An example of this uses jet reconstruction algorithms developed in the early 2000s~\cite{Cacciari:2008gp} to looks at jets and their substructure in ALEPH data~\cite{Chen:2021uws}, serving as a precision test of analytical calculations and phenomenological models in the collinear limit of QCD.

\begin{figure}[ht!]
    \centering
    \includegraphics[width=0.7\linewidth]{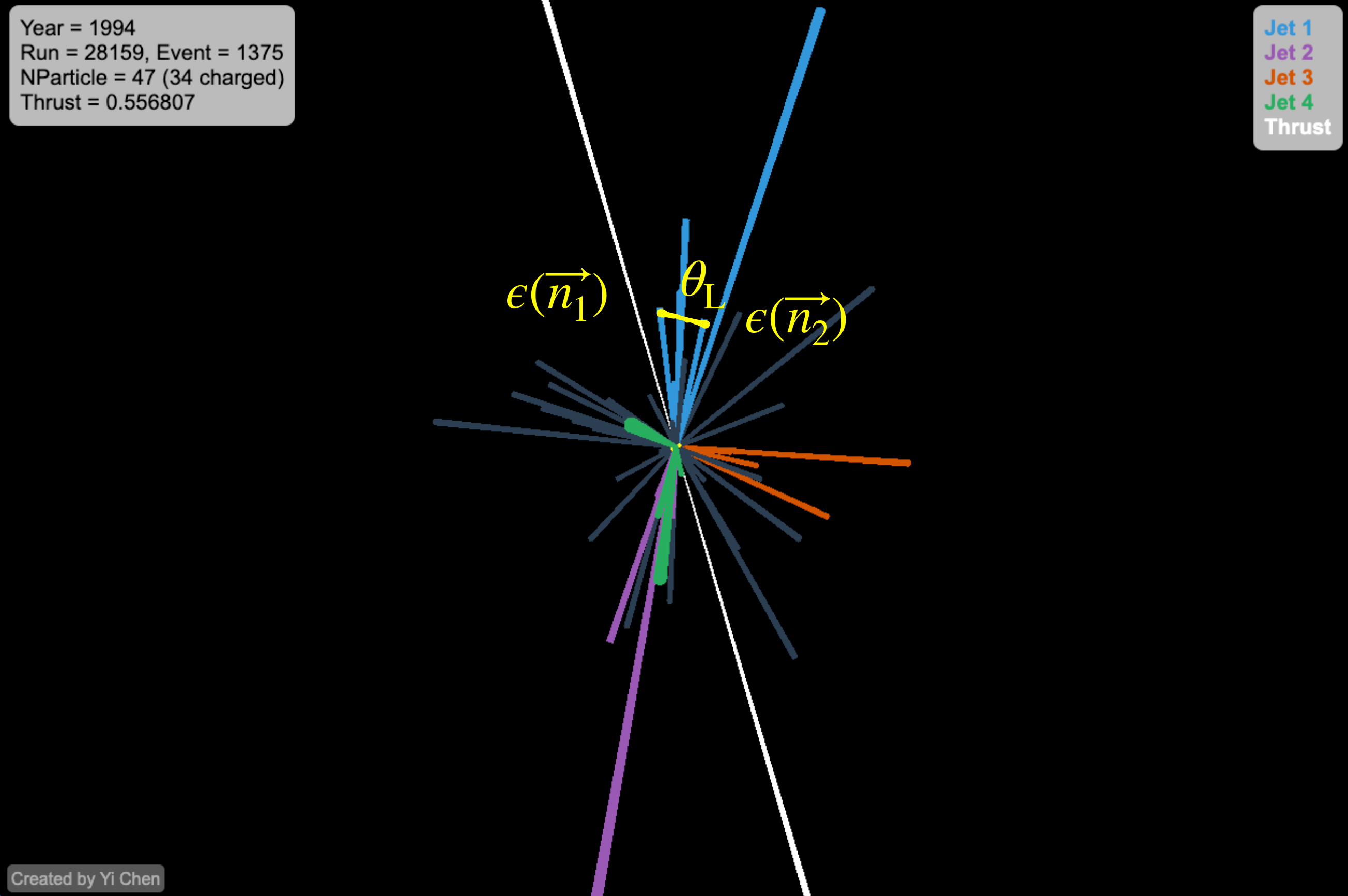}
    \caption{Schematic representation of a single ALEPH event taken in 1994 with a cartoon representation of a two-point energy correlator. The different jets in the event, ordered by energy, are marked in the different colors and the thrust axis is shown in white. Here the correlation functions ($\epsilon$) are written as a function of the unit vectors $\vec{n_{1}}$ and $\vec{n_{\rm 2}}$, which are the vectors on the celestial sphere pointing in the direction of the momentum vectors of the first and second particle being correlated, respectively. The opening angle between $\vec{n_{1}}$ and $\vec{n_{\rm 2}}$, $\theta_{\rm L}$, between the pairs is also shown.}
    \label{fig:introCartoon}
\end{figure}

One technique for the study of energy flow in $e^{+}e^{-}$ collisions theoretically proposed~\cite{PhysRevLett.41.1585,Basham:1978zq, Basham:1979gh,Basham:1977iq} and experimentally measured~\cite{OPAL:1990reb, OPAL:1993pnw,DELPHI:1990sof,L3:1991qlf,L3:1992btq,SLD:1994idb} are N-point correlation functions. This class of observables, which will be properly defined in Section~\ref{sec:EEC}, and herein referred to as ENCs, is useful for studying the energy flow resulting from a hard scattering and is sensitive to the strong coupling constant of QCD, $\alpha_{\rm s}$. A schematic depiction of a two point energy correlator is shown for an ALEPH event display from 1994 in Figure \ref{fig:introCartoon}~\footnote{This event display was created using a publicly-available online software, \url{https://yichen.me/project/EventDisplay}. }. In the original proposals, the energy correlation functions were measured using all particles in the event, resulting in limited precision at very small or very large angles was due to the calorimeter resolution for neutral particles.

Recently, ENCs have been successfully re-imagined in a proton-proton collision environment, where they have been used to measure correlations between the constituent particles of reconstructed jets~\cite{CMS:2024mlf,ALICE:2024dfl,STAR:2025jut,Komiske:2022enw}. Predominantly, these applications measure the so-called  \textit{projected} correlators that integrate out all shape information, keeping the ``longest" side fixed. Note that in the case of the two point correlator, the ``longest" side is trivially the distance between the two particles. The three point correlator is the first case where the determination of the longest side becomes relevant. The projected correlators are useful to isolate the universal scaling behaviors that are associated with QCD, offering a clean separation between the parton shower, hadronization, and the transition between these two regimes. The projected correlators are also an Infrared and Collinear (IRC) safe~\cite{Tkachov:1999py} observable that can be calculated in the perturbative limit from QCD first principles~\cite{Kardos:2018kqj}. ENC results in proton-proton collision environments~\cite{CMS:2024mlf,ALICE:2024dfl,STAR:2025jut,Komiske:2022enw} all show good agreement with theoretical predictions and indicate a remarkably universal  transition region of $p_{\rm T}R_{\rm L}\sim$ 2-3 GeV across wide kinematic regions, from the lower energies of the Relativistic Heavy Ion Collider (RHIC) to the Large Hadron Collider (LHC). In addition, the result from the CMS Collaboration~\cite{CMS:2024mlf} was used to perform an experimental extraction of the value of $\alpha_{s}$, which resulted in the most precise extraction of this quantity from jet substructure to date. 

ENCs have also been recently applied to both theoretical and experimental studies of larger systems, such as proton-lead~\cite{Andres:2024xvk}, electron-nucleus~\cite{Devereaux:2023vjz}, and lead-lead collisions~\cite{CMS:2025ydi,Andres:2023xwr,Andres:2024ksi,Barata:2023bhh,Yang:2023dwc,Bossi:2024qho,Andres:2024hdd,Andres:2024xvk} where they are used to probe both cold and hot nuclear matter effects, respectively. Additionally ENCs in these large systems still retain their connection to the underlying theory~\cite{Andres:2024xvk} and have the potential to study the energy loss of jets in the quark-gluon plasma (QGP) mitigating in part potential selection biases~\cite{Andres:2024hdd}. 

In this work a measurement of the projected two point energy correlator (E2C) using ALEPH archived data from LEP 1 taken at $\sqrt{s} = 91.2$ GeV, is reported. These results reexamine existing measurements of the E2C in $e^{+}e^{-}$ collision data using all charged particles in the event in order to improve the angular resolution at very small and large angles. Due to recent theoretical improvements including the development of the track function formalism~\cite{Chang:2013rca,Chang:2013iba}, charged-particle-based observables are now theoretically calculable in QCD. The measurement is corrected for smearing in the angle and energy of the correlated pair using a two-dimensional unfolding procedure. In this analysis note, the details of the measurement will be described in the following sections. In Section \ref{sec:Sample} the experimental setup, including details of the data and Monte Carlo (MC) samples, will be described. In Section \ref{sec:EEC} the observable definition used in this analysis will be provided. The bulk of the analysis procedure, including the subsequent matching and unfolding procedures, is described in Section \ref{sec:analysis}. Specifically, the various corrections applied to the E2C distribution are described in Section \ref{sec:effCorr}. In Section \ref{sec:Syst} the systematic uncertainties associated with this analysis procedure are described. The fully-corrected results with the corresponding uncertainties are presented in Section \ref{sec:results}. In Section \ref{sec:opalComparisons} the results are compared to previous experimental measurements. Finally, the results are summarized in Section \ref{sec:summary} and an outlook is provided for future work. In Appendix \ref{app:analysiscode}, the analysis code used is documented. Appendices \ref{app:evtseleff}, \ref{app:CorrectionClosure}, and \ref{app:crosscheck} describe various crosschecks of the final result.

\section{Experimental Setup}
\label{sec:Sample}

This analysis is performed with archived data recorded by the ALEPH detector, described in Ref.~\cite{Decamp:1990jra}. The central part of the detector is designed for the efficient reconstruction of charged particles. Their trajectories, referred to as tracks, are measured by a two-layer silicon strip vertex detector, a cylindrical drift chamber, and a large time projection chamber (TPC). These tracking detectors are situated inside a 1.5 T axial magnetic field generated by a superconducting solenoidal coil. The transverse momenta of the charged particles are reconstructed with a resolution of \(\delta p_{\rm T}/p_{\rm T} = 6 \times 10^{-4} p_{\rm T} \oplus 0.005\) ($p_{\rm T}$ in GeV/$c$)~\cite{ALEPH:1996qht}.

Electrons and photons are identified in the electromagnetic calorimeter (ECAL), which is located between the TPC and the superconducting coil. The ECAL is a sampling calorimeter, comprised of lead plates and proportional wire chambers segmented into \(0.9^\circ \times 0.9^\circ\) projective towers. These are read out in three depth sections and have a total thickness of approximately 22 radiation lengths. Isolated photons are reconstructed with a relative energy resolution of \(0.18/\sqrt{E} + 0.009\) ($E$ in GeV). The iron return yoke, constructed with 23 layers of streamer tubes, also serves as the hadron calorimeter (HCAL) for the detection of charged and neutral hadrons. The relative energy resolution for hadrons is \(0.85/\sqrt{E}\) ($E$ in GeV). Muons are identified based on their patterns in the HCAL and by the muon chambers, which are made of two double layers of streamer tubes located outside the HCAL. The information from the trackers and calorimeters is integrated using an energy-flow algorithm~\cite{ALEPH:1994ayc}. This algorithm generates a set of charged and neutral particles, called energy-flow objects. Only the charged particles reconstructed in the energy-flow algorithm are used for the E2C analysis. 

\subsection{Data and Monte Carlo Samples}\label{sec:dataMC}
For this study an ALEPH archived data sample of $e^{+}e^{-}$ collision data at a center of mass energy of $\sqrt{s} = 91$ GeV in the year 1994 of LEP was used. Note that although the LEP1 data-taking period at this energy ranged from 1992-1995, only the 1994 sample is used because this is the only dataset for which Monte Carlo (MC) with a detector simulation is available, which will be discussed below. The archived data is provided in an MIT Open Data format~\cite{Tripathee:2017ybi} that has been created and verified for numerous other applications~\cite{Badea:2019vey,Chen:2021uws,Chen:2023njr}.

For the study of reconstruction effects and the correction of the data, archived $\textsc{pythia}$ 6.1~\cite{Sjostrand:2000wi} MC events originally simulated by the ALEPH collaboration are used. The MC samples available at the time of this analysis are produced with the 1994 and 1997-2000 run detector conditions. As only the LEP1 data sample is utilized in this analysis, the only MC sample utilized is the 1994 archived MC in order to remain consistent between data and MC. In the future, if more MC samples with a detector simulation become available, additional datasets may be utilized.

\subsection{Event and Charged particle selections}
\label{sec:selections}
There are a number of selections that must be applied to the data and MC to ensure that they are suitable for further analysis. These selections are summarized below in Table \ref{tab:SelectionSummary}. 

\begin{table}[ht]
\caption{Summary table for particle and event selections.}
\begin{center}
\begin{tabularx}{\textwidth}{l|l}
\hline\hline
\multicolumn{2}{l}{Event selection}  \\
\hline
Acceptance              & $7\pi/36 \le \theta_{\rm sphericity} \le 29\pi/36$\\
Hadronic events         & at least five good tracks \\
                        & total reconstructed charged-particle energy $\ge 15$~GeV \\
Non-calibration runs    & $E_{\rm vis} < $ 200 GeV \\
\hline\hline
\multicolumn{2}{l}{Charged particles}  \\
\hline
Acceptance              & $|\cos\theta|<0.94$ \\
High quality tracks     & $p_{\rm T} \ge 0.2$ GeV \\
                        & at least 4 TPC hits \\
Impact parameter        & $d_0<2$~cm, $z_0<10$~cm \\
\hline\hline
\end{tabularx} 
\label{tab:SelectionSummary}
\end{center}
\end{table}

As listed in Table \ref{tab:SelectionSummary}, a standard hadronic event selection is applied as employed in previous LEP 1 ALEPH analyses~\cite{ALEPH:1996oqp, Chen:2021uws,Badea:2019vey}. Events are required to fall within the acceptance of the detector, meaning that they have a sphericity angle of $7\pi/36 \le \theta_{\rm sphericity} \le 29\pi/36$. To suppress electromagnetic interactions, at least five good tracks are required having a minimum total reconstructed charged-particle energy of 15 GeV. In order to remove laser calibration runs, an additional selection requiring that the total visible energy, $E_{\rm vis}$\footnote{$E_{\rm vis}$ is defined as the sum of the energy of all reconstructed particles}, to be less than 200 GeV is applied. Note that laser calibration events account for $\sim 0.006\%$ of the 1994 dataset. After these selections the total contamination from processes such as $e^{+}e^{-} \rightarrow \tau^{+}\tau^{-}$ is estimated to be 0.26\% based on studies from Ref.~\cite{ALEPH:1996oqp}. 

The charged particle selections are summarized in Table~\ref{tab:SelectionSummary}. Charged particles are required to fall within the detector acceptance, such that the absolute value of the cosine of the polar angle is smaller than 0.94. Track candidates are also required to be of so-called ``high-quality". To meet this requirement they must consist of at least four TPC hits and must have a transverse momentum greater than 0.2~GeV. An additional requirement is placed on the impact parameter to have a radial displacement from the interaction point of $d_{0} <2$ cm and a longitudinal displacement of $z_{0}<10$ cm. The sensitivity of the final result to the TPC hit requirement is accounted for in a dedicated systematic uncertainty (see Section \ref{sec:Syst} for more details).

In order to compare the experimental distribution to theory, correction factors to account for the effect of these selections are applied to the distribution following the unfolding procedure. These corrections will be described in Section \ref{sec:effCorr}.

\subsection{Data and Monte Carlo Comparisons}\label{sec:DataMCComp}

\begin{figure}[ht!]
    \centering
    \includegraphics[width = 0.49\textwidth]{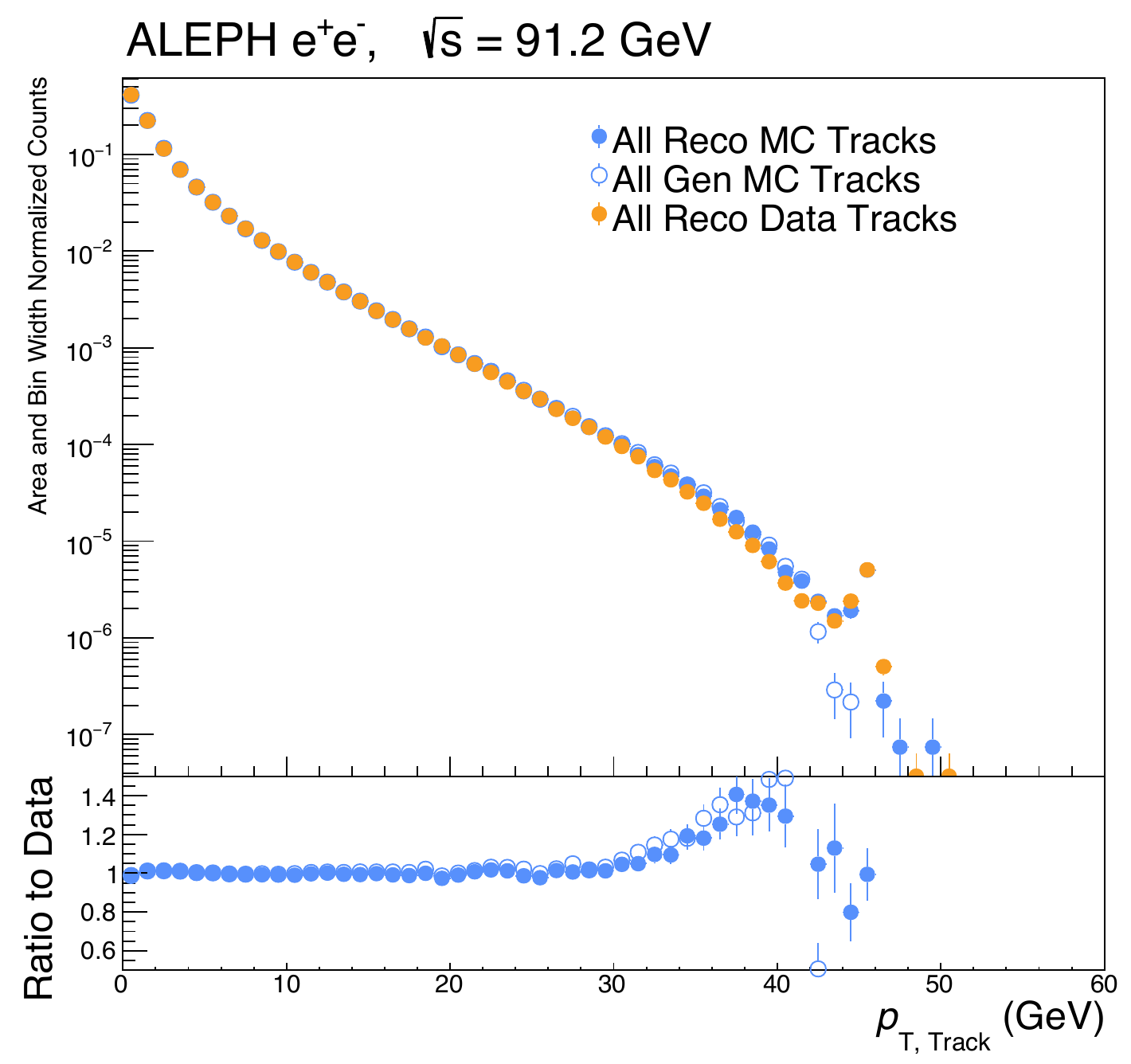}
    \includegraphics[width = 0.49\textwidth]{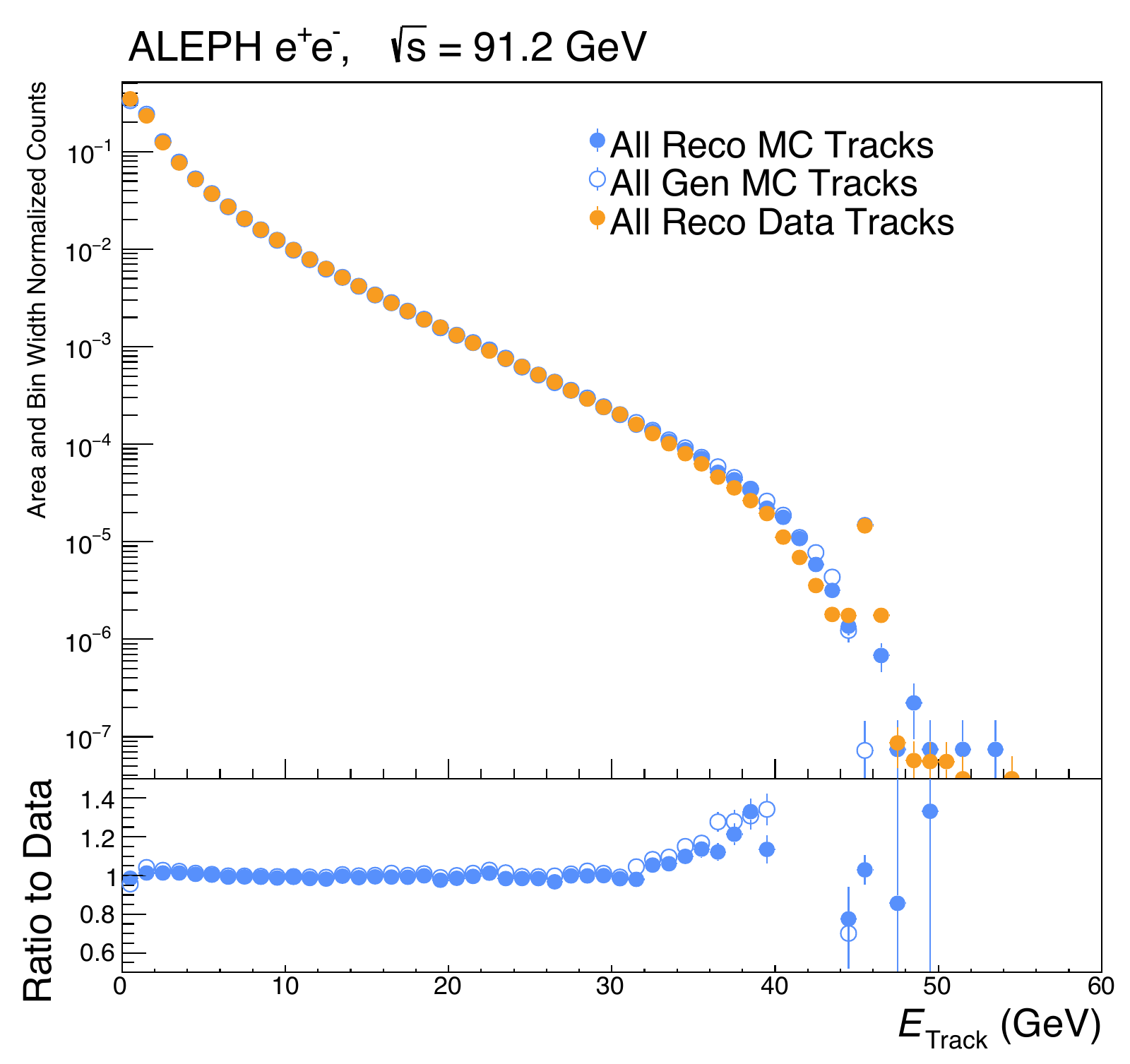}
    \caption{Comparisons between the track $p_{\rm T}$ (left) and the track energy (right) in data, reconstructed MC, and truth MC. These distributions are normalized to their integral, so are purely shape comparisons. Both the reconstructed-level data and MC have a tracking efficiency correction applied.}
    \label{fig:MCdatacomp1}
\end{figure}

To explore the level of agreement between data and the archived MC samples a number of fundamental kinematic variables are compared. For example, the track $p_{\rm T}$ and energy distributions are compared in the left and right panel of Figure \ref{fig:MCdatacomp1}, respectively. Here, the reconstructed level quantities in both data and MC are corrected for the tracking efficiency (see Section \ref{sec:trackingeff} for more details). The distributions shown an excellent agreement up to $\sim 35 $ GeV as expected due to the fact that the tracking efficiency correction itself is derived as a function of $p_{\rm T}$. In both panels of Figure \ref{fig:MCdatacomp1}, there is a spike occurring around $\sim$ 45 GeV that occurs at the reconstructed level, but not the generator level. The impact on the overall E2C distribution with and without this contribution was found to be negligible, and therefore no explicit cut was applied for this contribution. 

\begin{figure}[ht!]
    \centering
    \includegraphics[width = 0.49\textwidth]{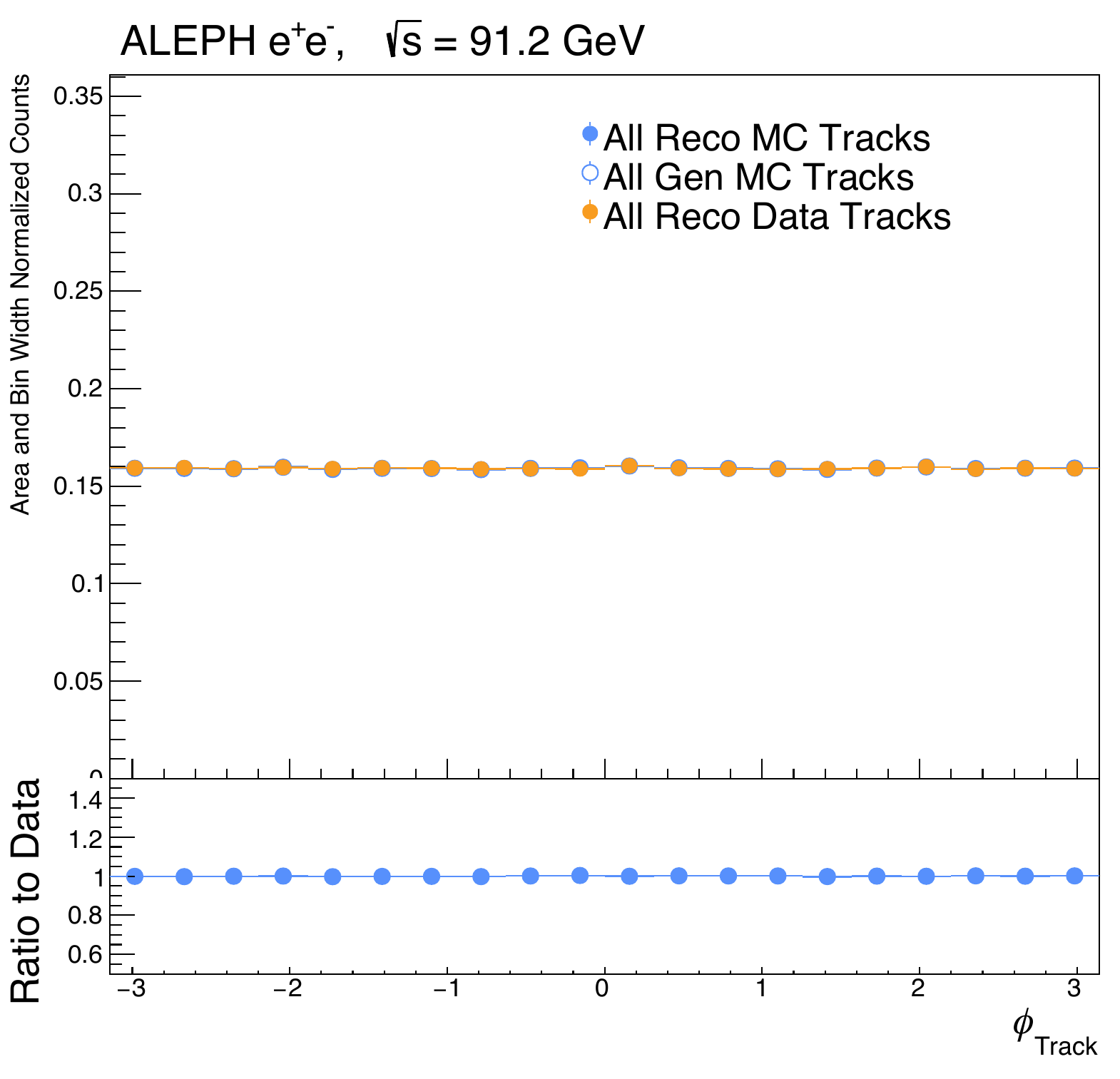}
    \includegraphics[width = 0.49\textwidth]{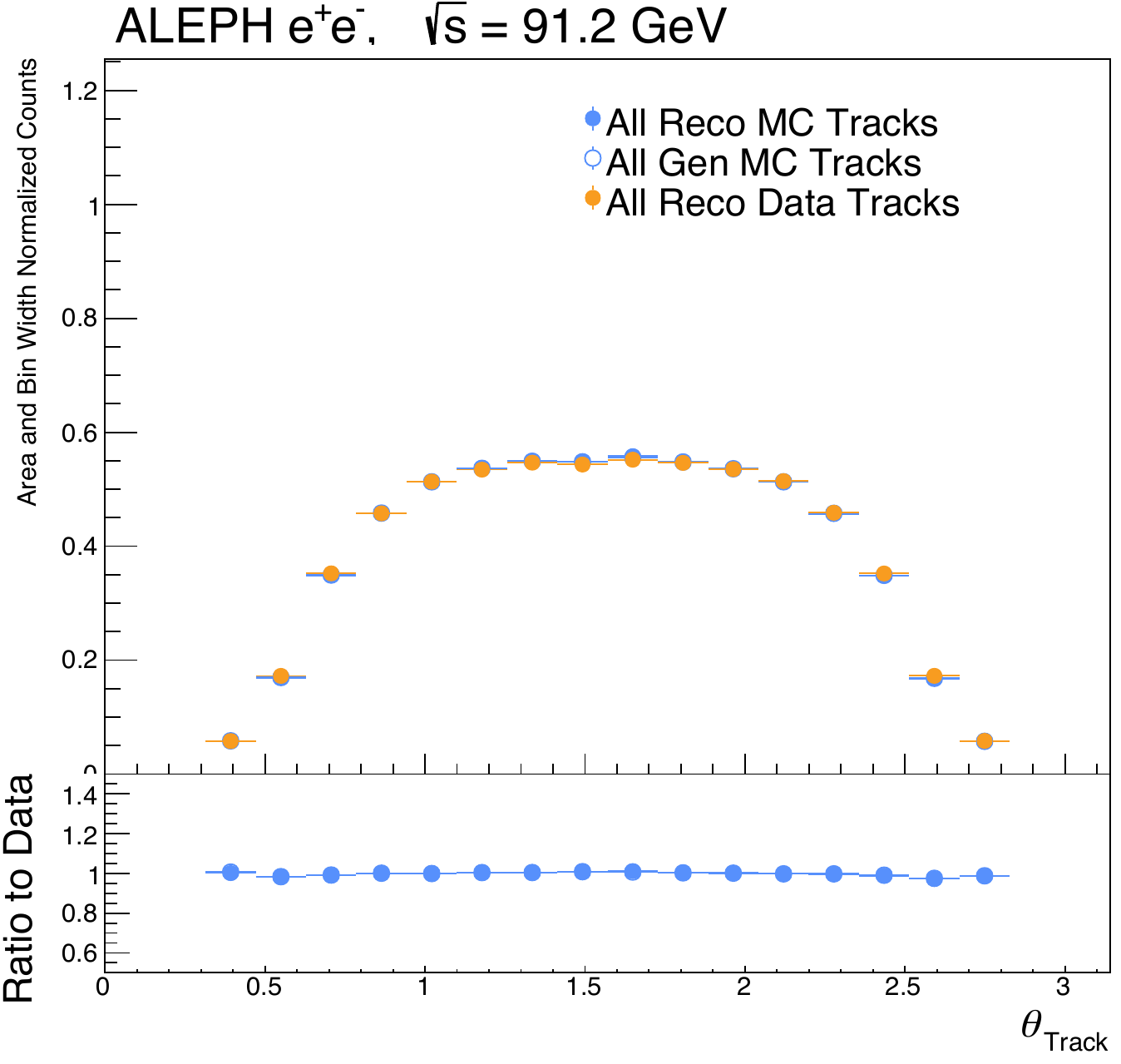}
     \includegraphics[width = 0.49\textwidth]{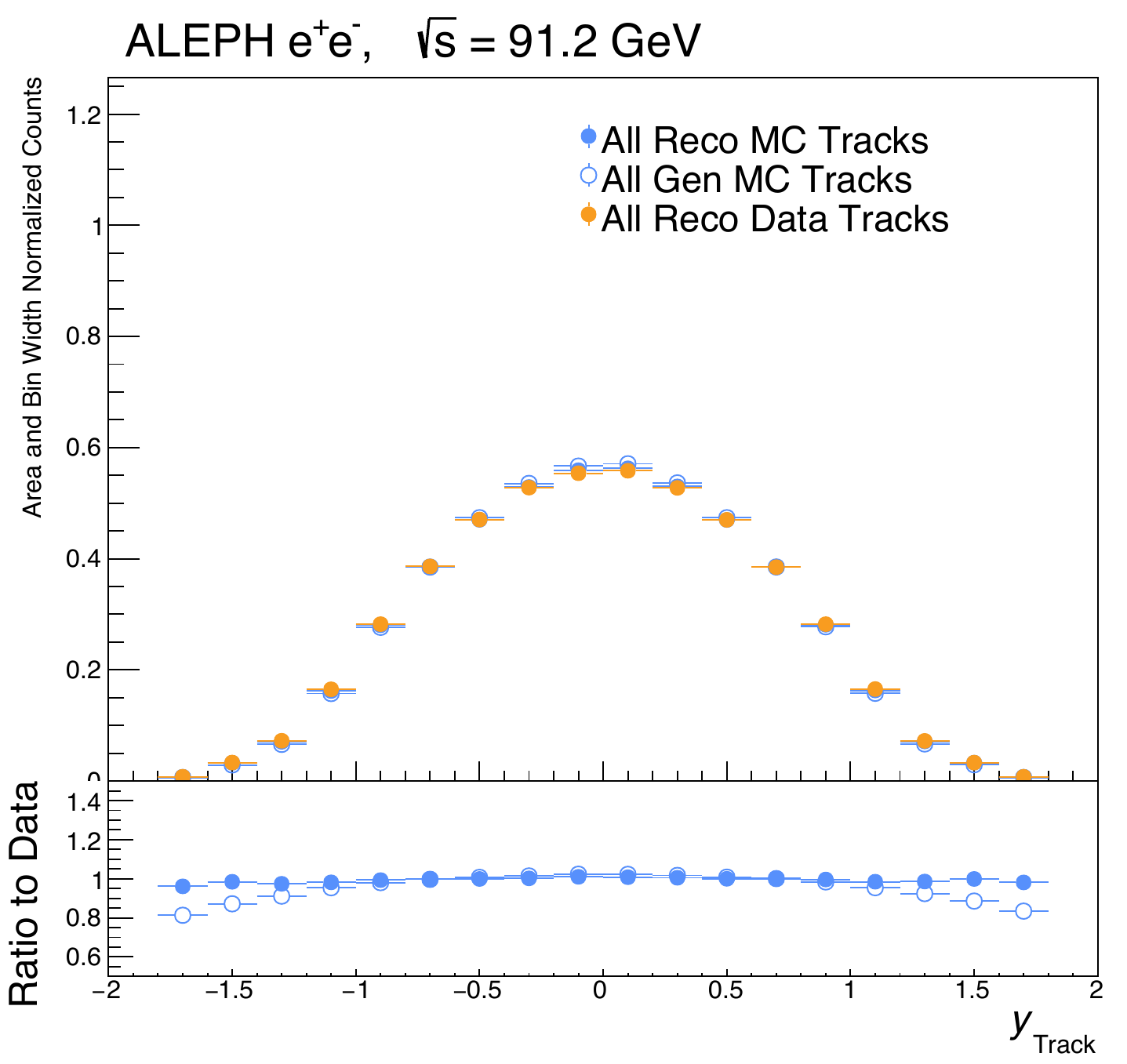}
    \caption{Comparisons between the track $\phi$ (left), the track $\theta$ (middle) and $y$ (right) in data, reconstructed MC, and truth MC.These distributions are normalized to their integral, so are purely shape comparisons. Both the reconstructed-level data and MC have a tracking efficiency correction applied.}
    \label{fig:MCdatacomp2}
\end{figure}

In Figure \ref{fig:MCdatacomp2} similar comparisons are made for the variables of azimuthal angle ($\phi$), polar angle ($\theta$), and particle rapidity ($y$). These distributions generally show good agreement, with some deviations in the forward rapidity region between reconstructed data/MC and the generator-level MC. Most likely, these deviations are due to the fact that the tracking efficiency is not given as a function of $y$ and that an assumption of the pion mass is made when calculating $y$. Overall, these figures demonstrate a reasonable level of agreement between the MC and data distributions for the key charged particle kinematic observables investigated.

\section{Energy-energy correlator}\label{sec:EEC}
Event-shape observables are central to the understanding of the energy flow and radiative patterns of high-energy collider data. The N-point energy correlator observables are one example of such a variable. As mentioned in Section \ref{sec:Introduction}, the EEC was first introduced in order to study energy flux in $e^{+}e^{-}$ collisions~\cite{PhysRevLett.41.1585,Basham:1978zq, Basham:1979gh,Basham:1977iq}, but recent theoretical developments as well as the development of new experimental techniques, such as jet finding algorithms, have led to a renewed interest in energy correlators in a variety of collision systems.  

Different measurements of the energy correlator may have slightly different definitions depending on the application itself. The definition of the ENC with $N = 2, 3$ employed in this work is expressed as an integral over the angle in Equation \ref{eq:ENC} where $\theta_{\rm i j}$ denote the opening angle between charged particles~\footnote{In previous work on EECs in $e^{+}e^{-}$ the angle $\theta_{\rm L}$ is sometimes written as $\chi$.} with indices $i$ and $j$, and the summation loops through all non-identical particle combinations.
\begin{align}\label{eq:ENC}
    \text{E2C}(\theta_{\rm L}) &= \sum^{\rm n}_{\rm i, j} \int d\sigma \frac{E_{\rm i}E_{\rm j}}{E^{2}}\delta(\theta_{\rm L} - \theta_{\rm i j})\nonumber\\
    \text{E3C}(\theta_{\rm L}) &= \sum^{\rm n}_{\rm i, j, k} \int d\sigma \frac{E_{\rm i}E_{\rm j}E_{\rm k}}{E^3}\delta(\theta_{\rm L} - \max(\theta_{\rm i j}, \theta_{\rm ik}, \theta_{\rm j k})),
\end{align}
The energy of particle $i$ is noted as $E_{\rm i}$, while $E$ denotes the total energy in the collision, or $\sqrt{s}$ of the $e^{+}e^{-}$ system. Note that the definition of the ENC provided here differs from some other approaches in a few notable ways. Firstly, the ENC in hadronic collider environments is typically is expressed as a function of the distance variable $\Delta R = \sqrt{\Delta y^{2} + \Delta\phi^{2}}$ instead of the angle $\theta_{\rm L}$. Secondly, all charged particles in a given event are used in this definition as opposed to many other approaches that only utilize particles within reconstructed jets. As will be discussed in detail later, this attribute of the analysis allows not only the collinear limit of QCD to be studied but also the back-to-back region. Thirdly, in this definition, correlations between two particles are only counted once, i.e. $ij$ correlations and $ji$ correlations are not double counted as in previous measurements such as that in Ref.~\cite{OPAL:1990reb}. Lastly, only charged particle correlations are considered in this analysis in order to fully exploit the excellent angular resolution for tracks in ALEPH~\footnote{See Section \ref{sec:Sample} for more details.}. The improvement in the angular resolution compared to previous results can be predominantly attributed to this modification. In the past, prior to the development of the track function formalism~\cite{Chang:2013iba, Chang:2013rca}, track-based observables were not theoretically calculable, so this direction was not favored. 

The ENCs can also be reported as a function of the variable $z$, which is defined in terms of the angle $\theta_{\rm ij}$ as
\begin{equation}\label{eq:z}
    z = \frac{1 - \cos{\theta_{\rm ij}}}{2}.
\end{equation}
Here, the EEC definition can be analogously written in the form given in Equation \ref{eq:ENC_z}, where all the same notational details as specified above for Equation \ref{eq:ENC} apply.

\begin{align}\label{eq:ENC_z}
    \text{E2C}(z) &= \sum^{\rm n}_{\rm i, j} \int d\sigma \frac{E_{\rm i}E_{\rm j}}{E^{2}}\delta(z - z_{\rm i j})\nonumber\\
    \text{E3C}(z) &= \sum^{\rm n}_{\rm i, j, k} \int d\sigma \frac{E_{\rm i}E_{\rm j}E_{\rm k}}{E^3}\delta(z - \max(z_{\rm i j}, z_{\rm ik}, z_{\rm j k})).
\end{align}

\noindent The optimal choice for the angular variable used in the ENC depends on the physics of interest one would like to probe. For example, if one would like to focus on the transition region between the parton shower and hadronization, then writing as a function of $\theta_{\rm ij}$ may be preferable. However, if one would like to probe, for example, the universality of the hadronization regime, then writing the ENC in terms of $z$ may be preferable. For the purposes of this analysis, we report both distributions. 

As mentioned previously, ENCs are commonly constructed using particles within jets and using the jet $p_{\rm T}$ to relate to the hard scattering scale. They are designed to probe both perturbative and non-perturbative QCD in the collinear limit. Due to the clean environment and unique topology of $e^{+}e^{-}$ collisions at $\sqrt{s} =$ 91.2~GeV, where the dominant process is $Z\rightarrow q\bar{q}$, one can also compute the ENCs using all particles in the event probing both the collinear and back-to-back or Sudakov limits of QCD. The back-to-back limit of QCD is relatively unexplored in experiments, and ENCs provide an opportunity to study this interesting region with unprecedented precision. The ENCs in the back-to-back limit also represent the event shape observable under the best theoretical control. Therefore, this measurement provides the most precise test of QCD possible in experiments. Such measurements are also useful for a future extraction of $\alpha_{\rm s}$. Furthermore, an additional benefit of measuring EECs in $e^+e^-$ collisions is that there is no hadronic initial state, which allows for the measurement of the full 3D structure instead of limiting to the transverse EEC that was proposed for hadronic collisions~\cite{Gao:2019ojf,Gao:2023ivm}. Measurements of the ratios of ENCs at different orders is also of interest as they can be useful for the extraction of $\alpha_{\mathrm s}$. This has been seen recently at the CMS experiment, which used the ENC ratios in $pp$ collisions in order to perform the most precise experimental extraction of $\alpha_{s}$ to date~\cite{CMS:2024mlf} using jet substructure observables. Performing such a measurement in $e^{+}e^{-}$ will also provide useful input to the world average of $\alpha_{\rm s}$. Recently, extractions of $\alpha_{\rm s}$ from $e^{+}e^{-}$ event shapes and analytic hadronization characterization were removed from the calculation of the world average~\cite{ParticleDataGroup:2020ssz,ParticleDataGroup:2022pth}, making experimental measurements timely.

\begin{figure}[ht]
    \centering
    \includegraphics[width=0.75\linewidth]{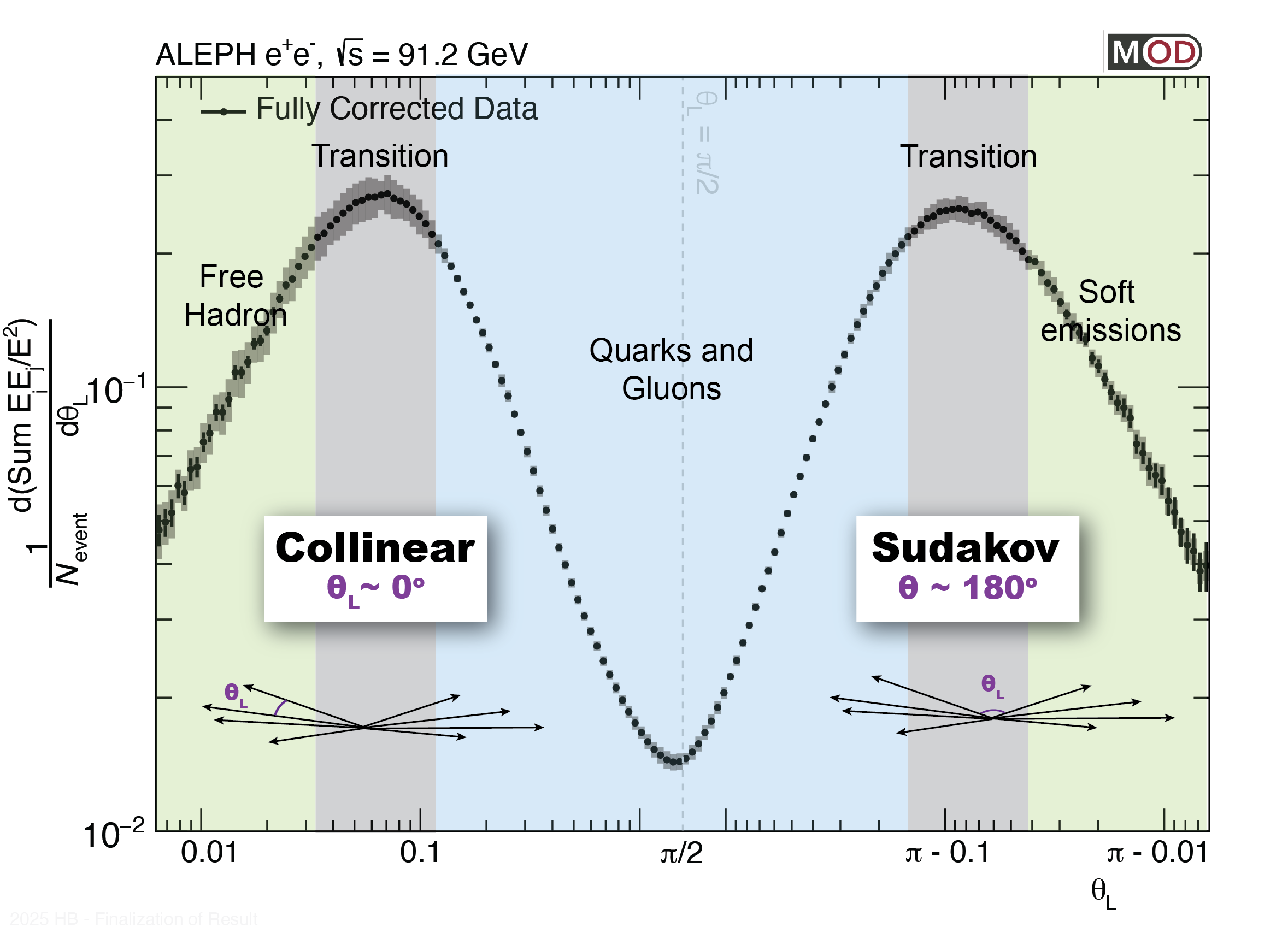}
    \includegraphics[width=0.75\linewidth]{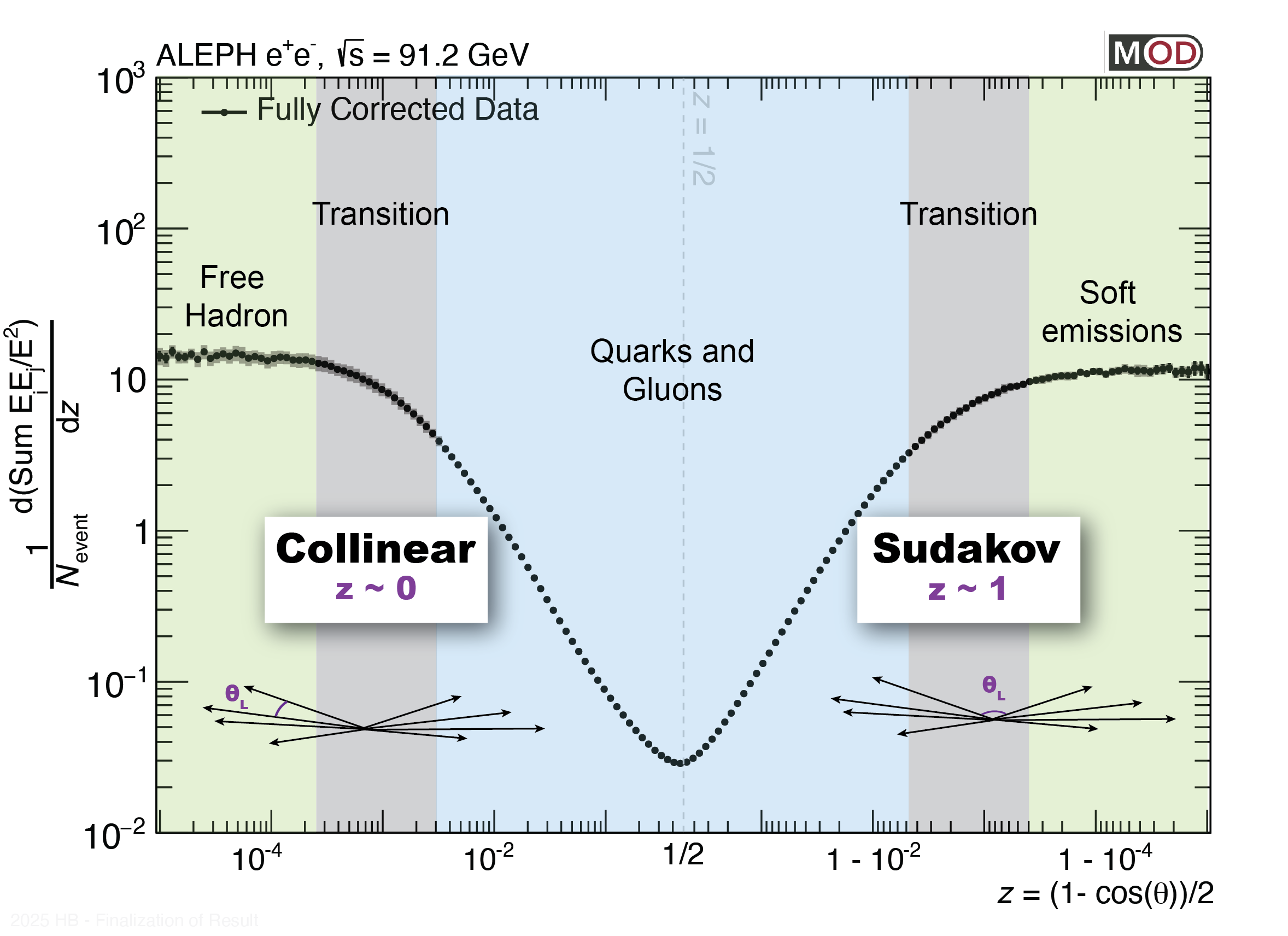}
    \caption{Fully-corrected distributions of the E2C as a function of $\theta_{\rm L}$ (top) and $z$ (bottom) using a ``flipped double-log style" as described in the main text. The various regions of the distribution are marked for illustration purposes, where the green represents the free hadron (or soft emission) region, the blue represents the quark/gluon region and the gray represents the transition between these two regions. The collinear region (where $\theta_{\rm L}\sim 0$ and $z \sim 0$) and the back-to-back or Sudakov region (where $\theta_{\rm L}\sim\pi$ and $z \sim 1$) is also marked.}
    \label{fig:ENCcollinearsudakov}
\end{figure}

To illustrate the shape of the ENC definition used here, and to point out the relevant regions, the fully-corrected E2C distributions that will be reported in Section \ref{sec:results} are shown as a function of $\theta_{\rm L}$ and $z$ in the left and right panels of Figure \ref{fig:ENCcollinearsudakov}, respectively~\footnote{The discussion in this section of the fully-corrected results is largely qualitative, focusing on the shape. For a more in-depth discussion of the results, refer to Section \ref{sec:results}.}. The distributions here are shown using a so-called ``flipped double-log" style. As the name suggests, there are two log axes, one which increases approaching $\theta_{\rm L} = \pi/2$ ($z = 1/2$) and another ``flipped" log axis decreasing from $\theta_{\rm L} = \pi/2$ ($z = 1/2$) to $\theta_{\rm L} = \pi$ ($z = 1$). This is done in order to highlight three regions of interest. In each distribution, the left hand side, corresponding to smaller angles (and therefore smaller values of $z$) represents the collinear limit. This would be analogous to what is probed by studying the ENC inside of jets, where the ENC exhibits a number of different scaling behaviors cleanly broken up into three regions corresponding to the free hadron region, the quark and gluon region, and the transition between these two regions. Similarly the right-hand side, corresponding to larger angles (and therefore larger values of $z$), represents the back-to-back . Analogous to the collinear limit, the Sudakov limit also has three characteristic regions including the quark and gluon region, a region of soft emissions and the transition between these two regions. However in the case of the Sudakov limit, the ordering of these distributions is flipped as compared to the case of the collinear limit, causing the regions to be roughly symmetric when shown in the double log scale.

\begin{figure}[ht]
    \centering
    \includegraphics[width=0.75\linewidth]{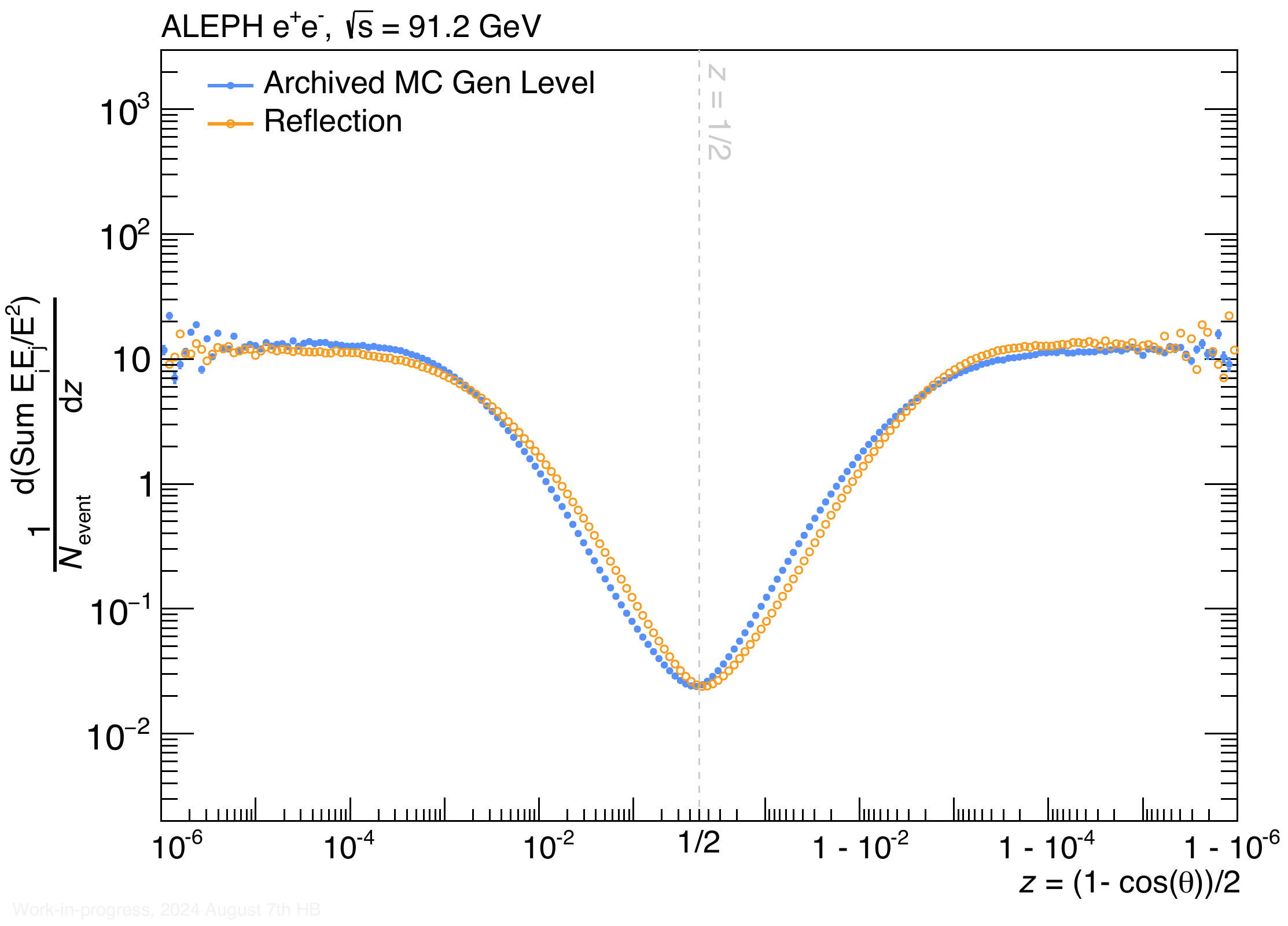}
    \caption{E2C written as a function of $z$ in the generator-level archived \textsc{pythia} 6 MC. The reflected points are also shown in order to indicate the approximate symmetry present in the distribution.}
    \label{fig:reflection}
\end{figure}

Note that these distributions are not symmetric, nor is there any reason to expect  them to be~\footnote{In other words, this is just a coincidental effect that is likely to not exist in samples - for example samples at other center of mass energies.}. However, what may be of interest is the universality of the behavior in the various regions, such as whether or not small and large angle regions exhibit a universal scaling behavior. For such a study the $z$ variable, which is roughly flat in the small and large angle tails of the distribution, is most useful. In order to perform a preliminary investigation of this, one can ``reflect" the E2C distribution as a function of $z$ in an archived MC sample overlaying the original and the reflected distribution. This is shown for the case of archived \textsc{pythia} MC in Figure~\ref{fig:reflection}. From this, one can see that the tails of the distribution are flat and this flattening occurs at approximately the same magnitude, hinting at some degree of universality that will be explored quantitatively in future work. At the very small or very large $z$ regions, the results from the archived MC sample fluctuate beyond the statistical error due to the limited numerical precision of the text-format archived data.

\section{Analysis}
\label{sec:analysis}

\begin{figure}[ht!]
    \centering
    \includegraphics[width=\linewidth]{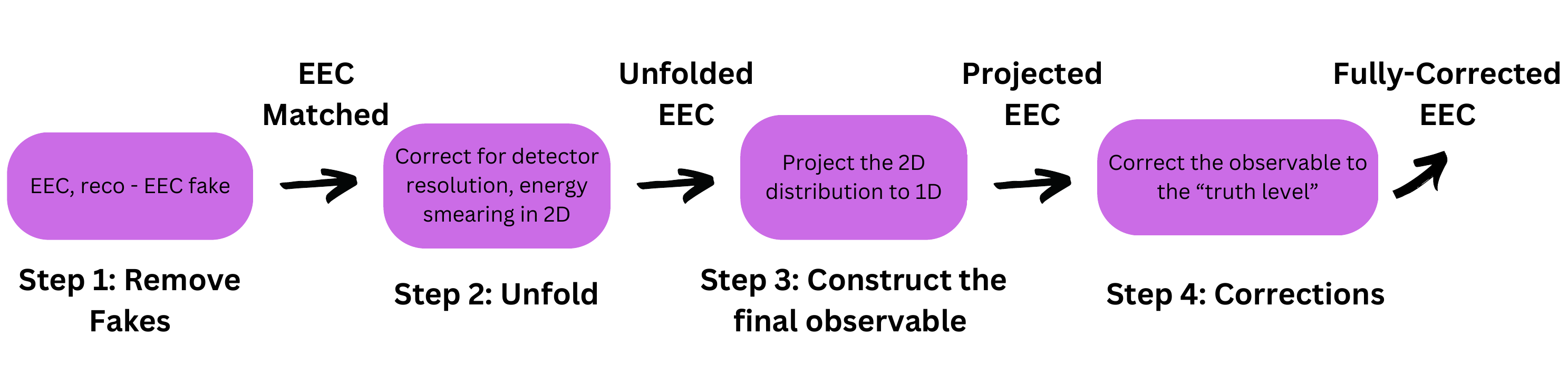}
    \caption{Overview of the various steps of the analysis in their sequential order.}
    \label{fig:analysisOutline}
\end{figure}

An outline of the analysis flow is shown in Figure \ref{fig:analysisOutline}. This outline covers all parts of the procedure following event selection (discussed in Section \ref{sec:Sample}) and prior to the systematic uncertainty determination (discussed in Section \ref{sec:Syst}). Once all event selections are made, the first step is to remove the contribution to the EEC resulting from reconstructed particle pairs lacking a matched generator-level pair (herein referred to as "fake"). The matching procedure will be described in Section \ref{sec:matching}. Then the EEC is unfolded using a two-dimensional unfolding procedure to unfold smearing in the angle (here $\theta_{\rm L}$ or $z$ depending on the choice of variable) and the energy product ($E_{\rm 1}E_{\rm 2}$). The unfolding procedure will be discussed in Section \ref{sec:unfolding}. The output of the unfolding procedure is a two-dimensional histogram in angle and energy that then must be projected onto the angular axis in order to construct the final observable. This process is discussed in Section \ref{sec:projections}. Then, a variety of corrections on the resulting distribution will be performed in order to correct the observable to the so-called ``truth-level" so that it can be compared to generator-level MC distributions and theoretical calculations. These corrections will be discussed in Section \ref{sec:effCorr}. Following these different steps the final observable is constructed and the systematic uncertainties are evaluated, which will be discussed in Section \ref{sec:Syst}.

\subsection{Matching Procedure}\label{sec:matching}
\begin{figure}[ht!]
    \centering
    \includegraphics[width=0.5\linewidth]{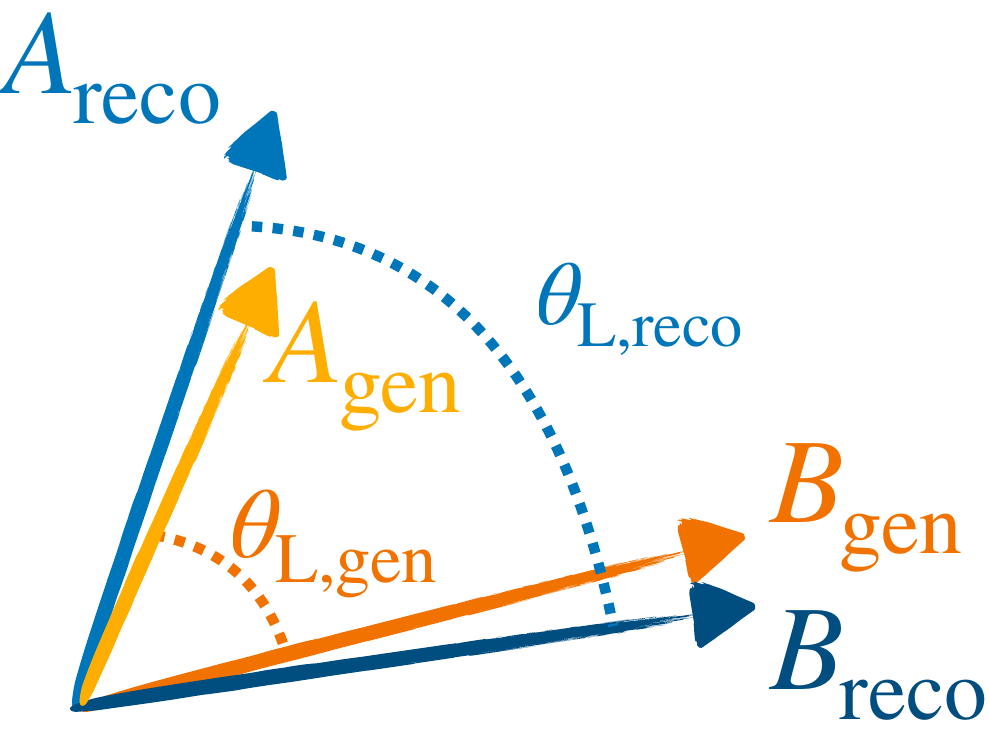}
    \caption{Cartoon visualization of single track and pairwise matching. Tracks shown in blue are considered to be at the reconstructed (reco) level and tracks shown in orange are considered to be at the generator (gen) level.}
    \label{fig:matchedPairCartoon}
\end{figure}

In the matching procedure a correspondence is built between the reconstructed and generator-level information that will then be used to correct the measured distribution from data. For the E2C analysis one must perform track matching at the single-track level and also at the pair level. The conceptual distinction between single track and pair matching can be found in Figure \ref{fig:matchedPairCartoon}. In this Figure, two tracks $A$ and $B$ are considered. These tracks can be matched at the single track level (i.e. matching $A_{\rm reco}$ with $A_{\rm gen}$ and $B_{\rm reco}$ with $B_{\rm gen}$) or matched as a pair where you consider the angle between $A$ and $B$ at the reconstructed level ($\theta_{\rm L, reco}$) and its corresponding gen-level match ($\theta_{\rm L, gen}$).

The matching is performed via the Hungarian Method~\cite{hungarianMatching}, which is a standard algorithm for solving assignment problems in polynomial time. Hungarian maximization is most easily explained using a matrix formulation where the $i$th row and the $j$th column represents the cost of assigning $j$ to $i$ and the algorithm uses this structure to make a unique assignment that minimizes the total cost. Therefore, a central item to the Hungarian method is the metric for determining the cost of a given entry, herein referred to as the matching metric. There are a variety of different choices of metrics that may be applied. In this case the matching metric used takes into account the energy and the angular components as shown in Equation \ref{eq:metric},
\begin{equation}\label{eq:metric}
    \chi^{2}(\theta, \phi, E) = \chi^{2}(\theta) + \chi^{2}(\phi) + \chi^{2}(E) = (\frac{\Delta\theta}{\sigma \theta})^{2} + (\frac{\Delta\phi}{\sigma\phi})^{2} + (\frac{\Delta E}{\sigma E})^{2},
\end{equation}
where $\Delta$ represents the difference between that variable for the two matching candidates (i.e., a reconstructed-level and generator-level particle) and $\sigma$ represents the resolution in that variable. The matching scheme used by default in this analysis is one in which the $\sigma$ values for the various variables follow an energy-dependent description based on the ALEPH tracker performance~\cite{ALEPH:1994ayc}. 
The energy resolution can be inferred from the transverse momentum resolution, which is given by $\frac{\sigma_{p_{\rm T}}}{p_{\rm T}} = 0.0006 \times p_{\rm T} \oplus 0.005$ ($p_{\rm T}$ in GeV/$c$), for the combined tracking system.
The position resolution is related to the reported impact parameter resolution, which is $\sigma(\delta) = 25\mu m + \frac{95\mu m}{p}$ ($p$ in GeV/$c$). The tracking position resolution is modeled in terms of ($\theta,\phi$) with the same momentum dependence, using an empirical form determined by evaluating the correspondence between generator-level and reconstructed-level track pair matching.
The dependence of the results on this particular choice of matching scheme is covered in a corresponding systematic uncertainty. See Section \ref{subsec:matchingsyst} for more details. 

\begin{figure}[ht!]
    \centering
    \includegraphics[width=\linewidth]{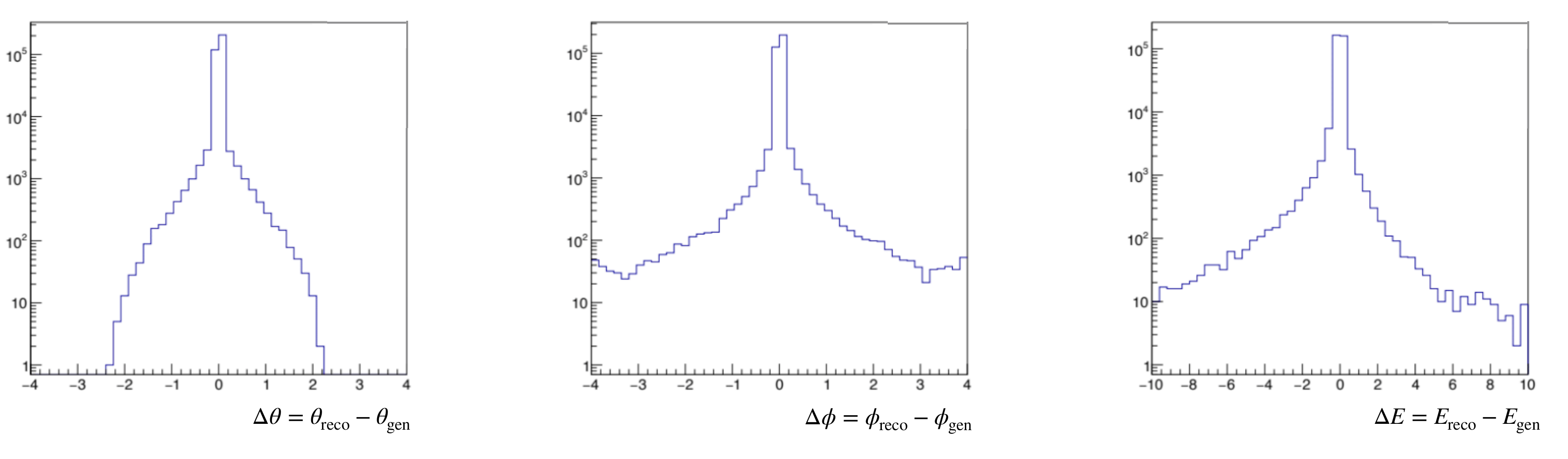}
    \caption{Summary of the single track matching performance via the difference between the reconstructed quantity and the corresponding gen-level quantity for $\theta$ (left), $\phi$ (middle), and energy (right).}
    \label{fig:singletrackperformance}
\end{figure}

To determine the performance of the matching procedure, it is important to evaluate both the single track and the pair-wise matching performance, as both should exhibit good performance for the matching to be effective. The performance of the single-track matching using this scheme is shown in Figure \ref{fig:singletrackperformance}. Here, all of the distributions are strongly peaked (on a log scale) around zero and are largely symmetric, indicating good single track matching performance. The un-normalized width of these distributions corresponds to the relative resolution in that variable. 

\begin{figure}[ht!]
    \centering
    \includegraphics[width=0.49\linewidth]{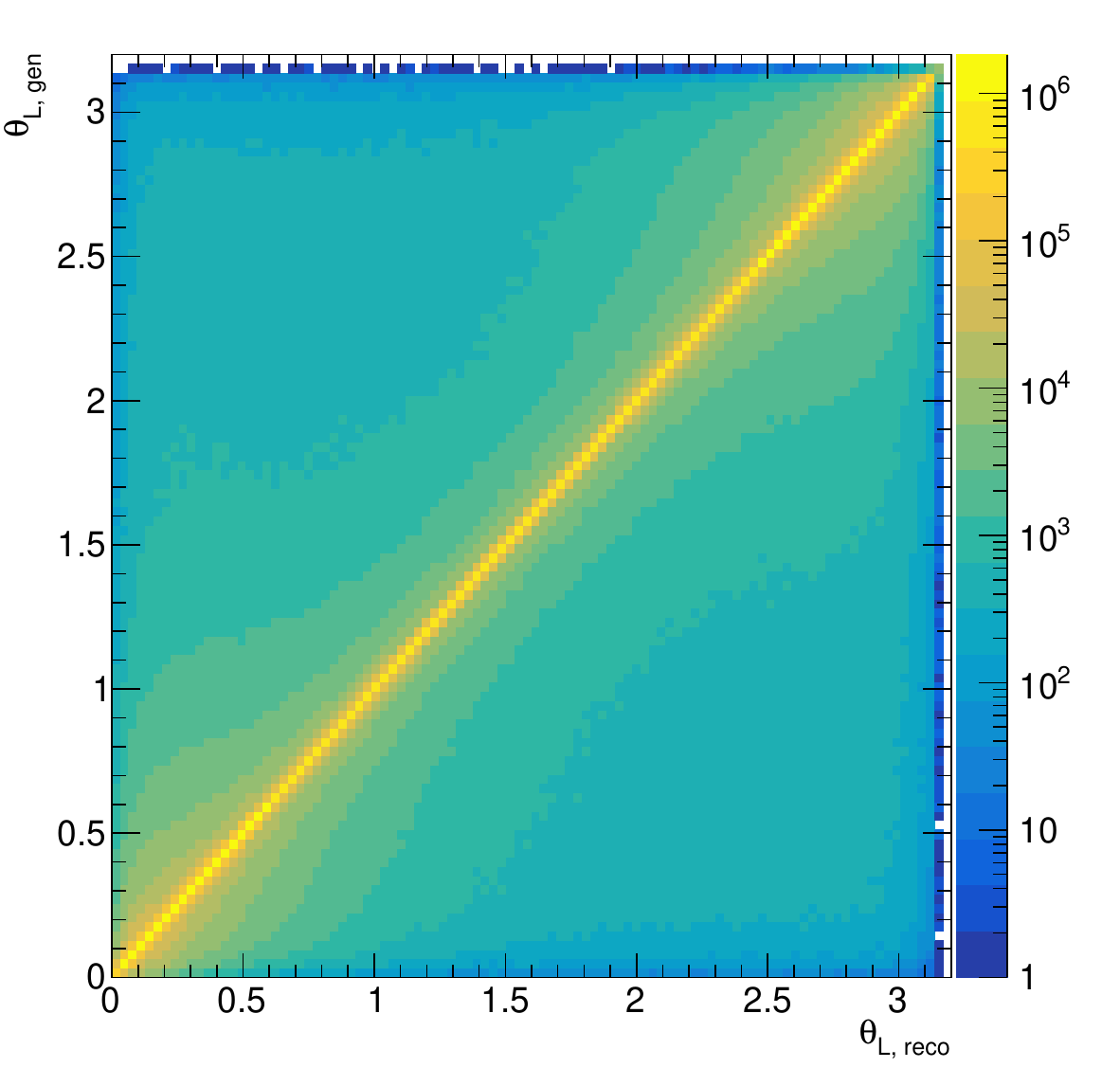}
    \includegraphics[width=0.49\linewidth]{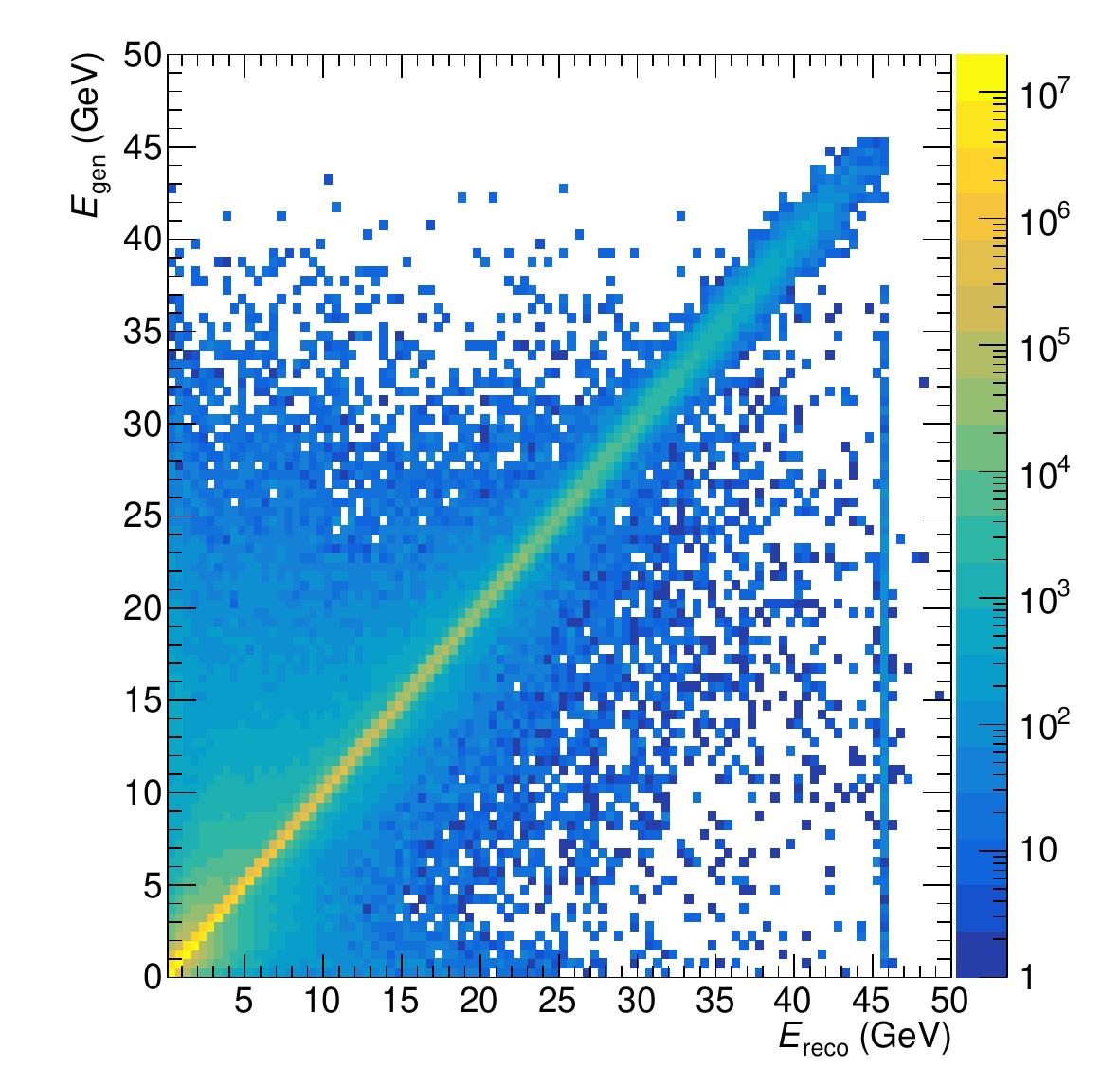}
    \caption{Left: $\theta_{\rm L, gen}$ as a function of the $\theta_{\rm L, reco}$ for the matched track pairs. Right: Energy at the generator level for single tracks as a function of the reconstructed energy of the matched track. }
    \label{fig:matchingPerformanceScheme2}
\end{figure}

The overall performance of the matching method is shown in Figure \ref{fig:matchingPerformanceScheme2}. The left panel of Figure \ref{fig:matchingPerformanceScheme2} shows the angular correspondence for matched reco-gen  pairs by plotting $\theta_{\rm L, gen}$ as a function of $\theta_{\rm L, reco}$. The right panel of Figure \ref{fig:matchingPerformanceScheme2} shows the energy correspondence between single track matches by showing $E_{\rm gen}$ as a function of $E_{\rm reco}$~\footnote{The streak present in this figure at $E_{\rm reco} \sim 45$ GeV was previously discussed in Section \ref{sec:DataMCComp}.}. Both of these metrics are very diagonal, validating that the matching procedure is effective and indicative of the excellent resolution provided by ALEPH. 

The use of this matching procedure results in a high matching efficiency, shown in Figure \ref{fig:matchingEff}, of $\sim 90\%$ and is roughly independent of the angle $\theta_{\rm L, gen}$. This efficiency will later be corrected for in a procedure described in Section \ref{sec:matchingeffcorr}. The corresponding fake rate for this matching scheme is found to be roughly $\sim 5\%$ and is also roughly independent of the angle $\theta_{\rm L, reco}$. 

\subsection{Unfolding}\label{sec:unfolding}
Inherently in any experiment, the measured information is ``smeared", meaning that it deviates from the value that would be measured with a perfect detector. To correct for the finite detector resolution, an unfolding procedure is applied. In this analysis there are two quantities that can be smeared, the angle $\theta_{\rm L}$ and the energy weighting of the particles $E_{\rm 1}E_{\rm 2}$. Therefore, a 2-dimensional unfolding procedure was used to correct the measured E2C distribution. Note that one large advantage of measuring the ENC distributions in $e^{+}e^{-}$  collisions is that the exact momentum transferred is known, due to the colliding objects being fundamental particles. In the case of measuring the ENC in hadronic collision systems, one must also independently correct for the smearing in the momentum transferred, increasing the difficulty of the unfolding problem.  Bayesian Unfolding as implemented in the \texttt{RooUnfold}~\cite{Brenner:2019lmf} package version \texttt{2.0.0} was employed for this task, as this is the last stable release for 2D unfolding~\footnote{See \href{https://gitlab.cern.ch/RooUnfold/RooUnfold/-/issues/16}{https://gitlab.cern.ch/RooUnfold/RooUnfold/-/issues/16} for more information.}. The remainder of this section is organized as follows: In Section \ref{sec:response}, the binning and procedure used to form the response matrix is described. In Section \ref{sec:unfoldingcheck} various checks of the unfolding procedure are described. 

\subsubsection{Forming the Response Matrix}\label{sec:response}
At its core, the unfolding procedure relies on a correspondence between the measured- and truth-level quantities. The response matrix is what captures this correspondence, created by utilizing a mapping between generator and reconstructed-level MC that is then inverted to correct the measured data. Here, the response matrix is formed with archived $\textsc{pythia}$ 6 MC where the correspondence between the measured and generator-level distributions is formed using the matching procedure described in Section \ref{sec:matching}. When filling the response matrix, pairs in a given event are matched and then the bin that matches the reconstructed and generator-level energy and angle is filled. The energy binning utilized in the $E_{i}E_{j}/E^{2}$ axis of the response matrix is provided in Table \ref{tab:energyBinning}. Note that this binning later becomes an important choice when constructing the final distribution after unfolding, which will be discussed in Section \ref{sec:projections}. The binning utilized for the angular axis (either $z$ or $\theta_{\rm L}$) uses a ``double log" style where 200 variable bins are chosen to be evenly spaced on a log scale are used where 100 log bins ranging from $\theta_{\rm L} = 0.002$ to $\theta_{\rm L} = \pi/2$ and 100 ``flipped" log-style bins ranging from $\theta_{\rm L} = \pi - 0.002$ to $\theta_{\rm L} = \pi/2$. The binning used for $\theta_{\rm L}$ and $z$ results is the exact same where the boundaries used are simply converted between $\theta_{\rm L}$ and $z$ by Equation \ref{eq:z}. The fine-binning used here results in a measurement that is more differential than previous results, particularly in the regions of interest. 

\begin{table}[ht!]
    \centering
    \begin{tabular}{ p{0.8\linewidth}}
     \center{Energy Binning}  \\  
$\_\_\_\_\_\_\_\_\_\_\_\_\_\_\_\_\_\_\_\_\_\_\_\_\_\_\_\_\_\_\_\_\_\_\_\_\_\_\_\_\_\_\_\_\_\_\_\_\_\_\_\_\_\_\_\_\_\_\_\_$ \\
  \texttt{(0.0, 0.0001, 0.0002, 0.0005, 0.00075, 0.001, 0.00125, 0.0015, 0.00175, 0.002, 0.00225, 0.0025, 0.00275, 0.003, 0.0035, 0.004, 0.005, 0.007, 0.01, 0.02, 0.03, 0.04, 0.05, 0.07, 0.10, 0.15, 0.20, 0.3)} \\
    \end{tabular}
    \caption{Energy Binning used in the $E_{i}E_{j}/E^{2}$ axis of the unfolding procedure.}
    \label{tab:energyBinning}
\end{table}

\subsubsection{Checks of the Unfolding Procedure}\label{sec:unfoldingcheck}
\begin{figure}[ht!]
    \centering
    \includegraphics[width=0.7\linewidth]{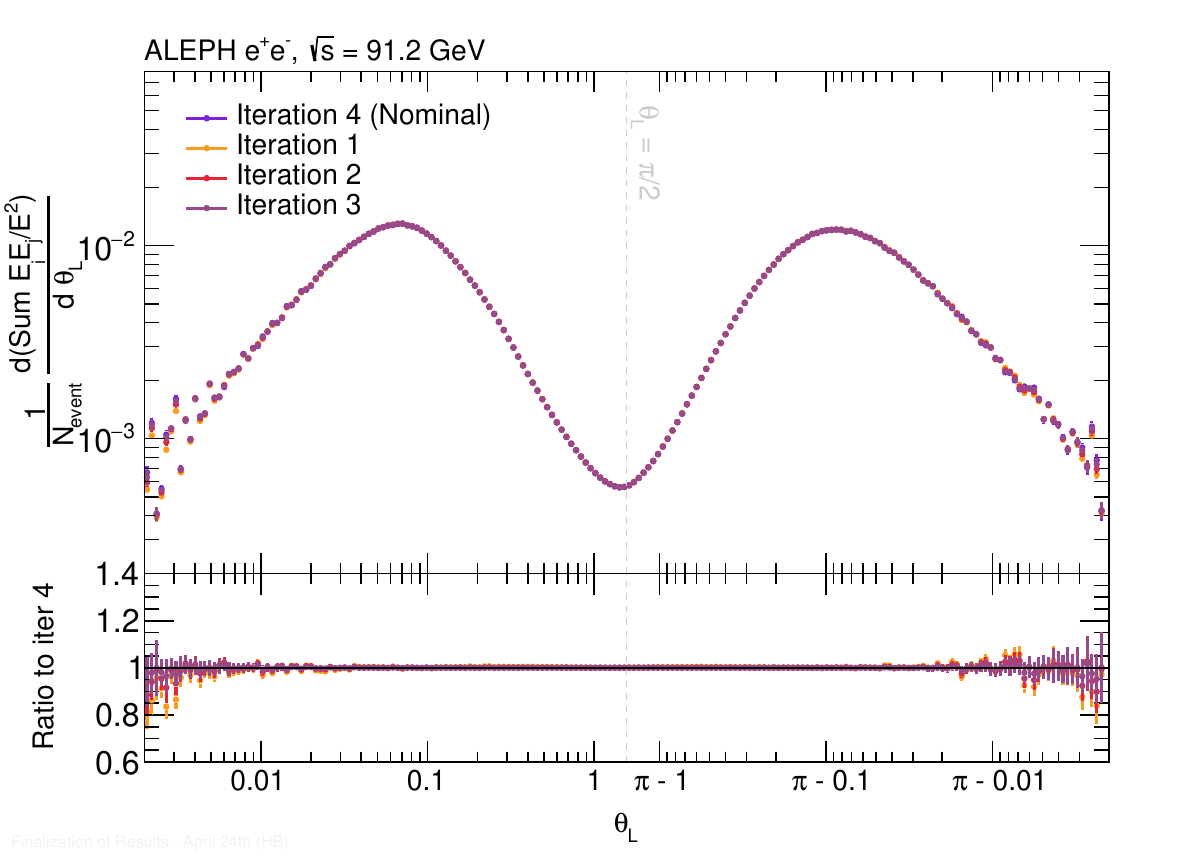}
    \caption{Unfolded E2C distribution as a function of $\theta_{\rm L}$ for a variety of iterations. The ratio to the third iteration is shown in the bottom panel in order to represent the stability of the unfolding procedure. The distribution from archived $\textsc{pythia}$ 6 MC is shown in gray for the purposes of comparison.}
    \label{fig:unfoldingstab}
\end{figure}

At its core, unfolding is a matrix inversion problem, i.e., an inversion of the mapping from the true to reconstructed level quantities. Within statistical uncertainties the reconstructed data
can be explained by the actual physical solution, but also by a large family of unphysical solutions. In a physical solution, a degree of smoothness can be expected, which is imposed in the unfolding procedure through a process called regularization. The final result should have no dependence on the regularization parameter(s) used. In iterative Bayesian unfolding, regularization is imposed by limiting the number of iterations. To test that the result is relatively independent of the regularization parameter, unfolded results for various iteration choices are compared to some nominal value; in this case four iterations is chosen as the nominal value. The stability with the number of iterations for unfolded data is shown for the unfolding in Figure \ref{fig:unfoldingstab}, where good stability is shown after the second iteration with only small deviations with each subsequent iteration. 

\subsection{Constructing the observable}\label{sec:projections}
The result of the procedure described in Section \ref{sec:unfolding} is also a two-dimensional histogram with an angular dimension in bins of $z$ or $\theta_{\rm L}$ and an energy dimension in bins of $E_{\rm i}E_{\rm j}/E^{2}$. However, the final result should be of a one-dimensional histogram of the form given in Equation \ref{eq:ENC}. To convert the two-dimensional histogram in terms of angle and energy weight into the correct form, a projection is performed weighted by the bin center in the $E_{\rm i}E_{\rm j}/E^{2}$ axis to account for the energy weights. Therefore, this will create a difference in the projection and the form given in the Equation \ref{eq:ENC} as the bin center in the $E_{\rm i}E_{\rm j}/E^{2}$ would normally be exact for each pair, but is now approximated by the bin width of the pair that this corresponds to. This difference results in a non-closure which is then corrected, as described in Section \ref{sec:binningcorrection}, and that results a corresponding systematic uncertainty described in Section \ref{subsec:binningsyst}. Note that having finer bins along the energy axis would reduce this systematic uncertainty, but would increase unfolding systematic uncertainties. For this analysis, the energy bins were chosen as reported in Section \ref{sec:response} to balance these effects.

\subsection{Corrections}\label{sec:effCorr}
In this Section, the nature and magnitude of the various corrections in this analysis are presented. In addition, a number of closure checks were performed to make sure that the corrections were applied correctly. These closure checks are presented in Appendix \ref{app:CorrectionClosure}.

\subsubsection{Fake Correction}\label{sec:fakecorrection}
When the matching procedure, as described in Section \ref{sec:matching}, is performed, this implicitly defines the so-called ``fakes". The fake correction accounts for the fact that not all reconstructed tracks have generator-level matches. As indicated in Figure \ref{fig:analysisOutline}, the fake correction is applied prior to the unfolding procedure described in Section \ref{sec:unfolding}. The fake correction is applied to the E2C distribution as function of the reconstructed angle and is given by

\begin{equation}\label{eq:fakefraction}
    \epsilon(\theta_{\rm L, reco} \; \text{or} \; z) = \frac{\text{EEC}_{\rm matched \; pair}(\theta_{\rm L, reco} \; \text{or} \; z)}{\text{EEC}_{\rm all\; pairs}(\theta_{\rm L, reco} \; \text{or} \; z)}. 
\end{equation}

\begin{figure}[ht!]
    \centering
    \includegraphics[width=0.49\linewidth]{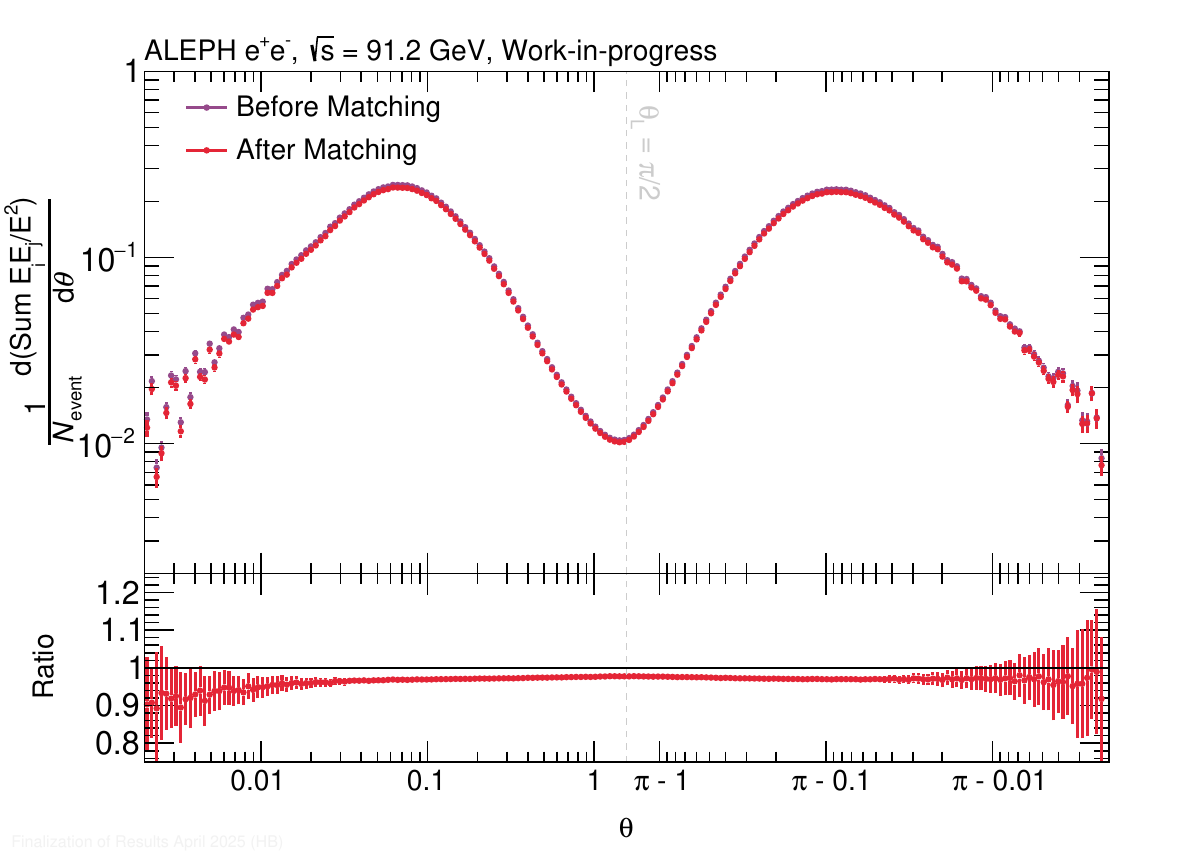}
    \includegraphics[width=0.49\linewidth]{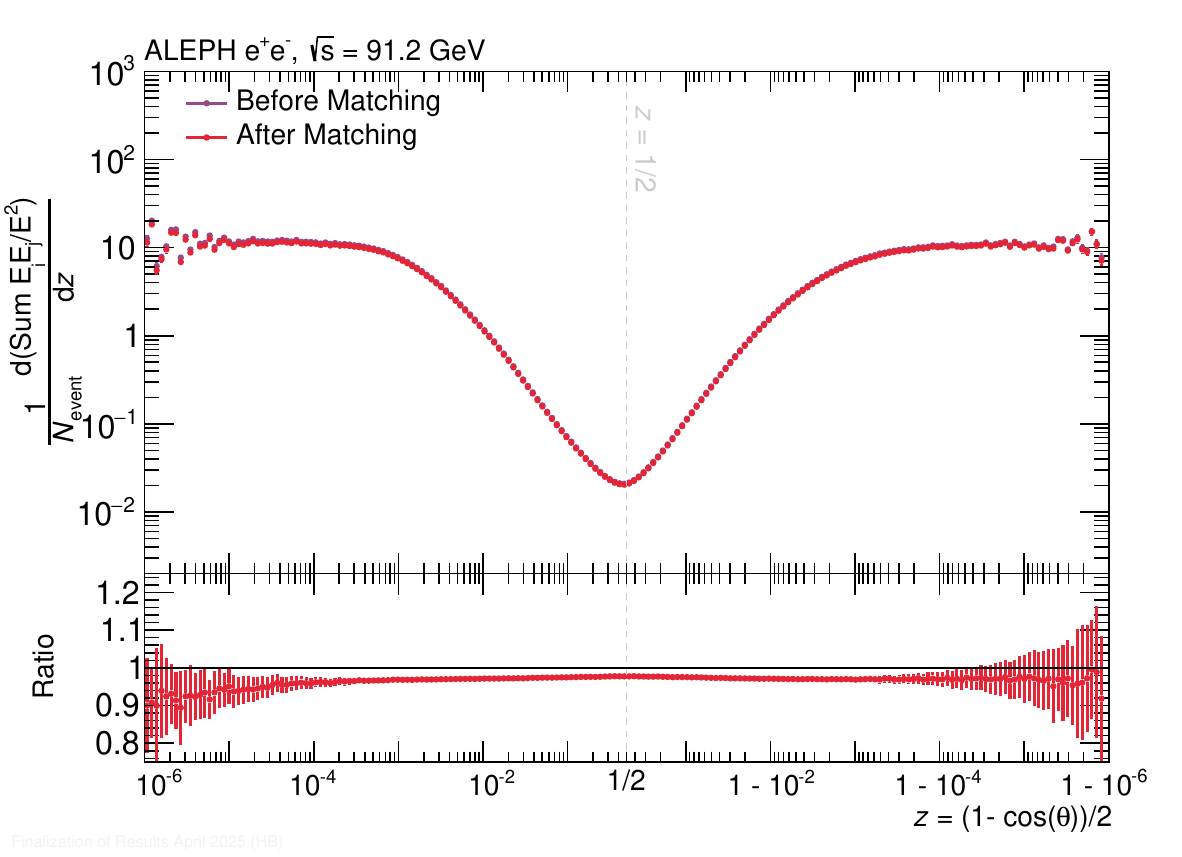}
    \caption{Left: EEC distributions before and after matching as a function of reconstructed-level angle $\theta$. Right: EEC distributions before and after matching as a function of reconstructed level $z$. In both panels the ratio of the two distributions, which represents the fake correction as defined in Equation \ref{eq:fakefraction}, is shown.}
    \label{fig:fakeCorrection}
\end{figure}

\noindent The magnitude of this correction is shown in the bottom panel of both panels of Figure \ref{fig:fakeCorrection}. Note that when it is applied, the correction reduces the overall yield by the factor shown in the bottom panel of Figure \ref{fig:fakeCorrection}. Here it can be seen that the fake fraction is quite low, over all regions of phase space and is relatively flat as a function of $\theta_{\rm L}$ or $z$. Therefore, the fake correction has a relatively small impact on the data. This is consistent with the fake rate estimated from the single track matching procedure of 2.4\%.  

\subsubsection{Binning Correction}\label{sec:binningcorrection}
\begin{figure}[ht!]
    \centering
    \includegraphics[width=0.49\linewidth]{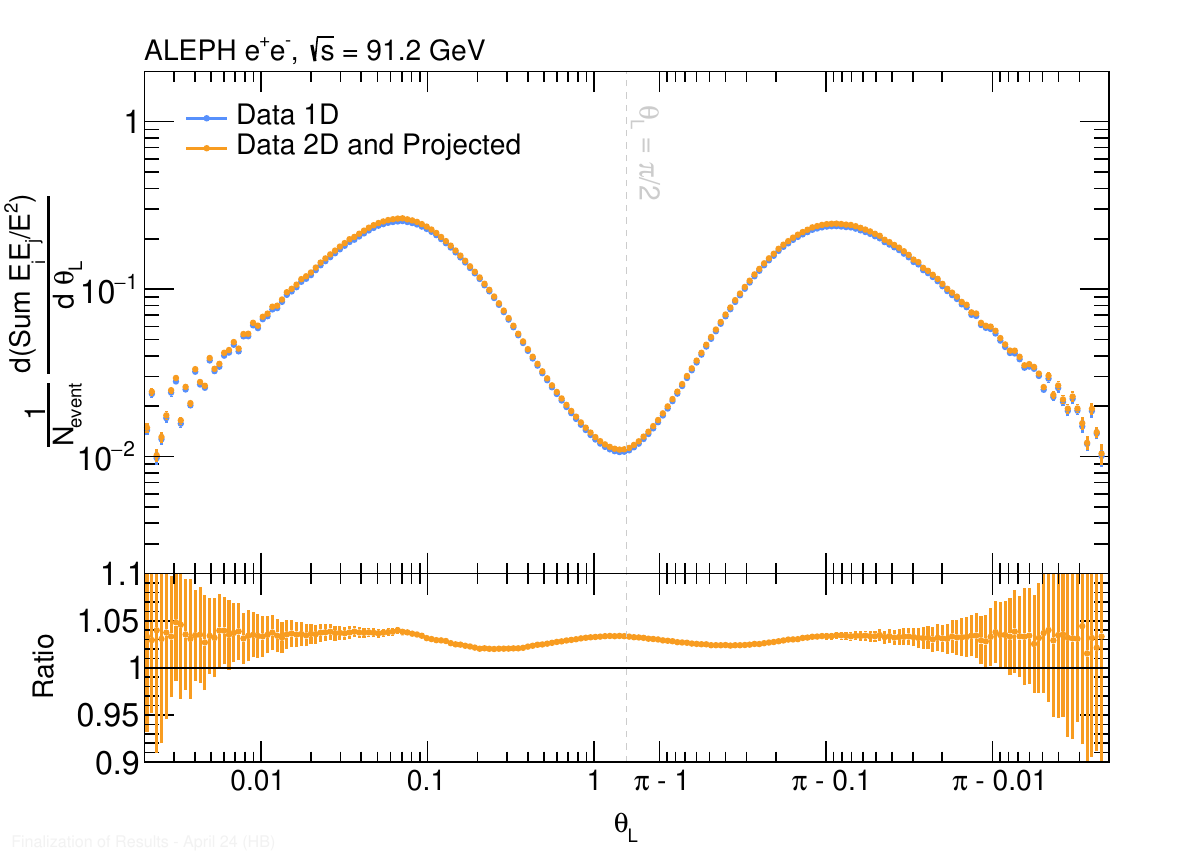}
    \includegraphics[width=0.49\linewidth]{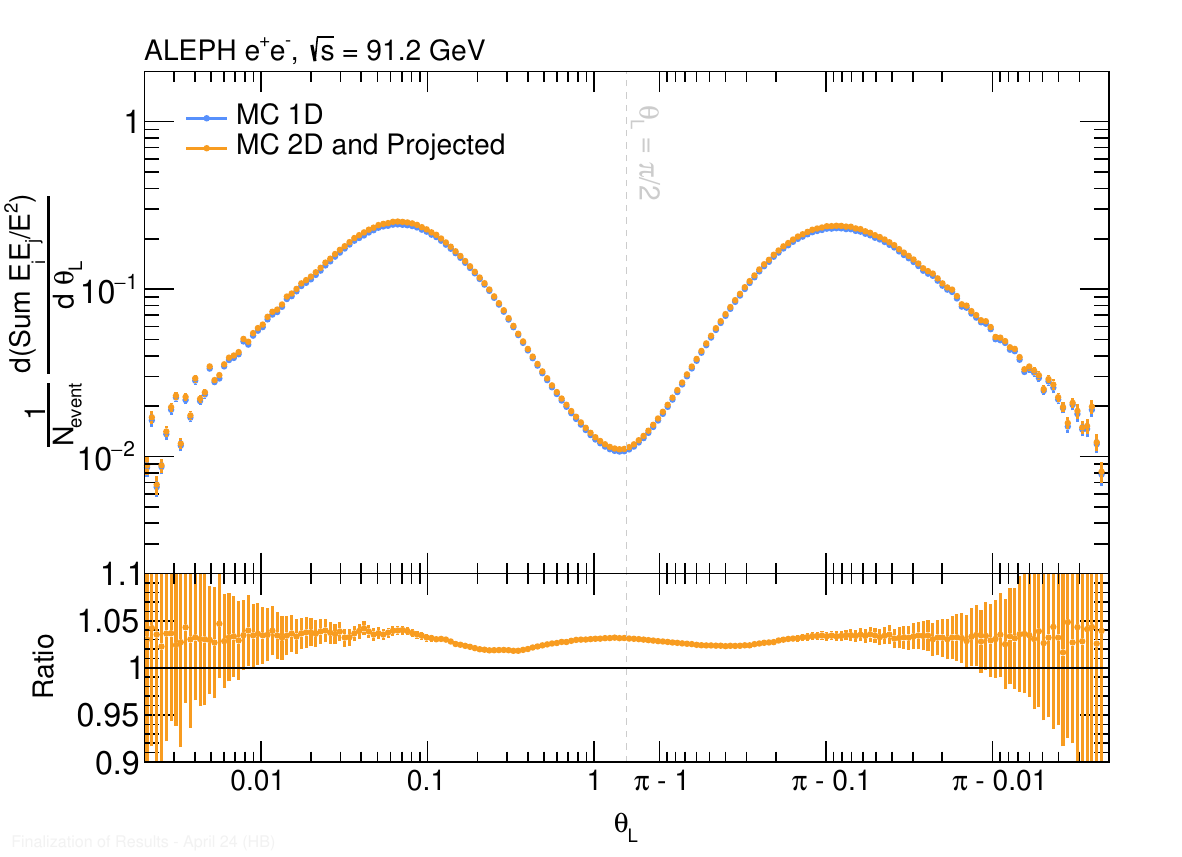}
    \caption{Comparison of the one-dimensional and two-dimensional with projection E2C distributions in data (left) and simulation (right). The ratio of the two distributions represents the binning correction, which is shown in the bottom panel. Note that the ratio in the bottom panel is zoomed in for the purposes of comparison between the two panels.}
    \label{fig:binningSystematic}
\end{figure}

As discussed in Section \ref{sec:projections}, a non-closure exists between the projected form of the two-dimensional histogram and the original form of Equation \ref{eq:ENC}. Such a non-closure can be quantified by filling a 2D distribution and then projecting the distribution using the procedure described in Section \ref{sec:projections} and comparing this to the 1D distribution filled to match the definition in Equation \ref{eq:ENC}. One could derive this correction in either raw uncorrected data or simulation. The correction derived from raw data and simulation are shown separately in the left and right panels of Figure \ref{fig:binningSystematic}, where the bottom panel of each of the figures represents the correction, given by the ratio between the 1D and projected 2D distributions. As shown in Figure \ref{fig:binningSystematic}, the correction is shown to not strongly depend on whether uncorrected data or simulation is used in order to derive the correction. 

\begin{figure}[ht!]
    \centering
    \includegraphics[width=0.49\linewidth]{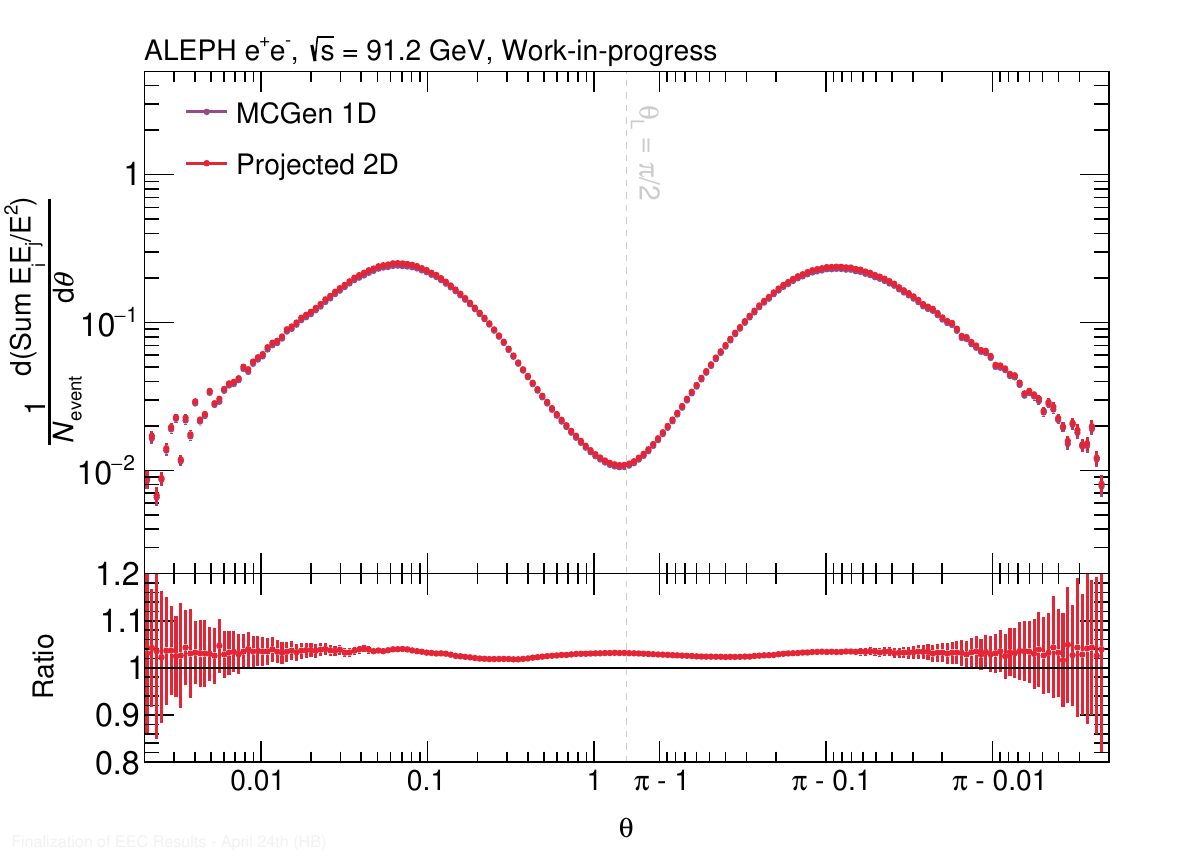}
    \includegraphics[width=0.49\linewidth]{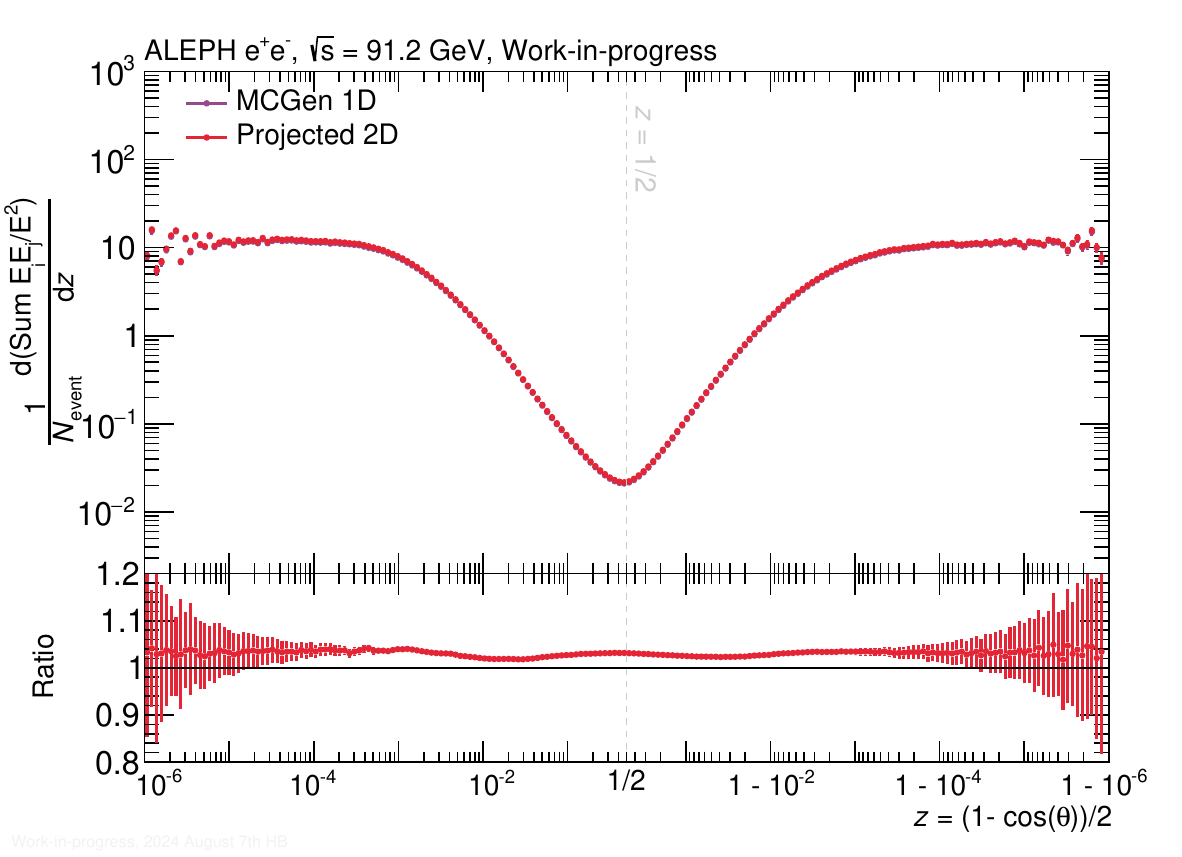}
    \caption{Comparison of the one-dimensional and two-dimensional with projection E2C distributions as a function of $\theta$ (left) and $z$ (right). The ratio of the two distributions represents the binning correction, which is shown in the bottom panel. Note that the ratio in the bottom panel is zoomed out in comparison to the scale used in Figure \ref{fig:binningSystematic}.}
    \label{fig:binningCorrection}
\end{figure}

In the nominal result, simulation was utilized in order to derive the correction. The resulting correction, shown in Figure \ref{fig:binningCorrection} for $\theta_{\rm L}$ (left panel) and $z$ (right panel), is approximately 4\% across the entire phase space. In addition, there is a systematic uncertainty associated with this correction, which will be discussed in Section \ref{subsec:binningsyst}.

\subsubsection{Matching Efficiency Correction}\label{sec:matchingeffcorr}
The procedure for matching individual pairs was first introduced in Section \ref{sec:matching}. The matching efficiency corrects for the fact that not all generator-level pairs have a reconstructed-level match and is defined as written in Equation \ref{eq:matcheff}. Note that equation differs from Equation \ref{eq:fakefraction} in that it is written as a function of $\theta_{\rm L, gen}$.

\begin{equation}\label{eq:matcheff}
    \epsilon(\theta_{\rm L, gen} \; \text{or} \; z) = \frac{\text{EEC}_{\rm matched \; pair}(\theta_{\rm L, gen} \; \text{or} \; z)}{\text{EEC}_{\rm all\; pairs}(\theta_{\rm L, gen} \; \text{or} \; z)}
\end{equation}

\begin{figure}[ht!]
    \centering
    \includegraphics[width=0.49\linewidth]{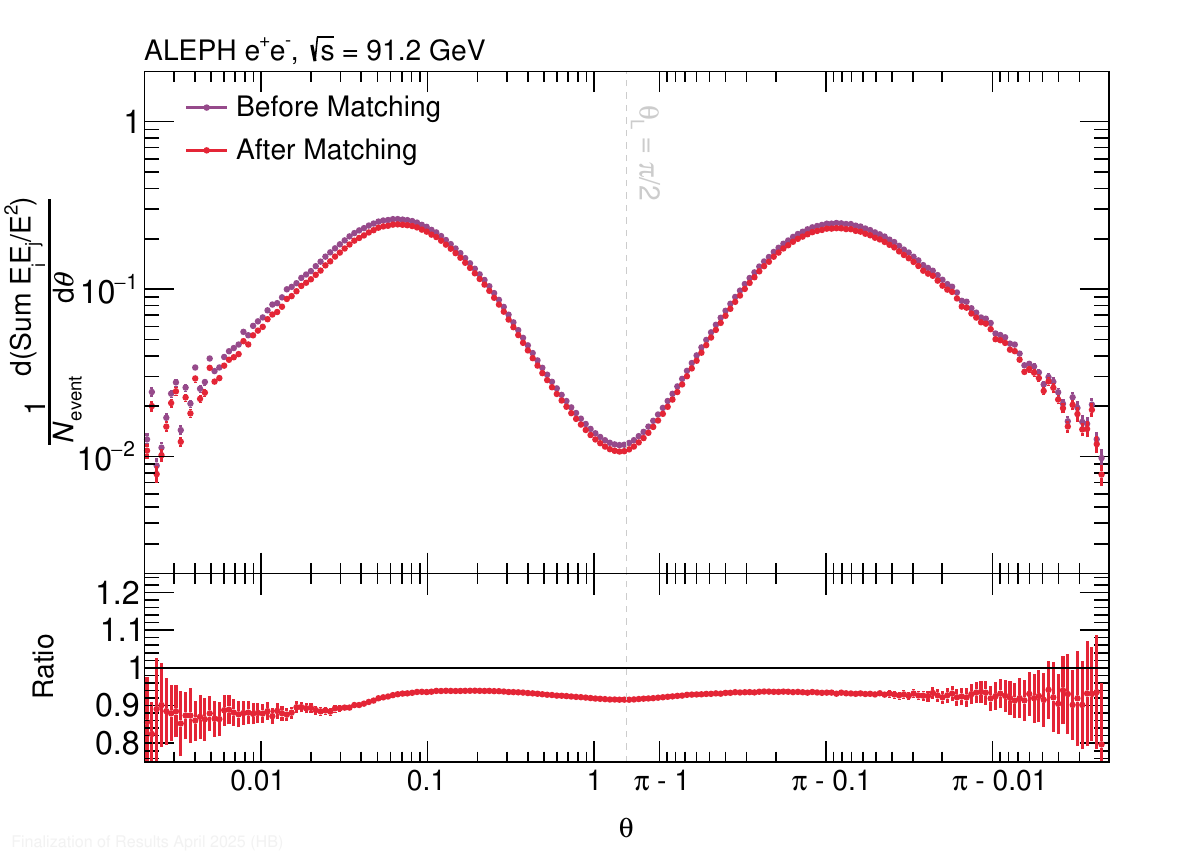}
    \includegraphics[width=0.49\linewidth]{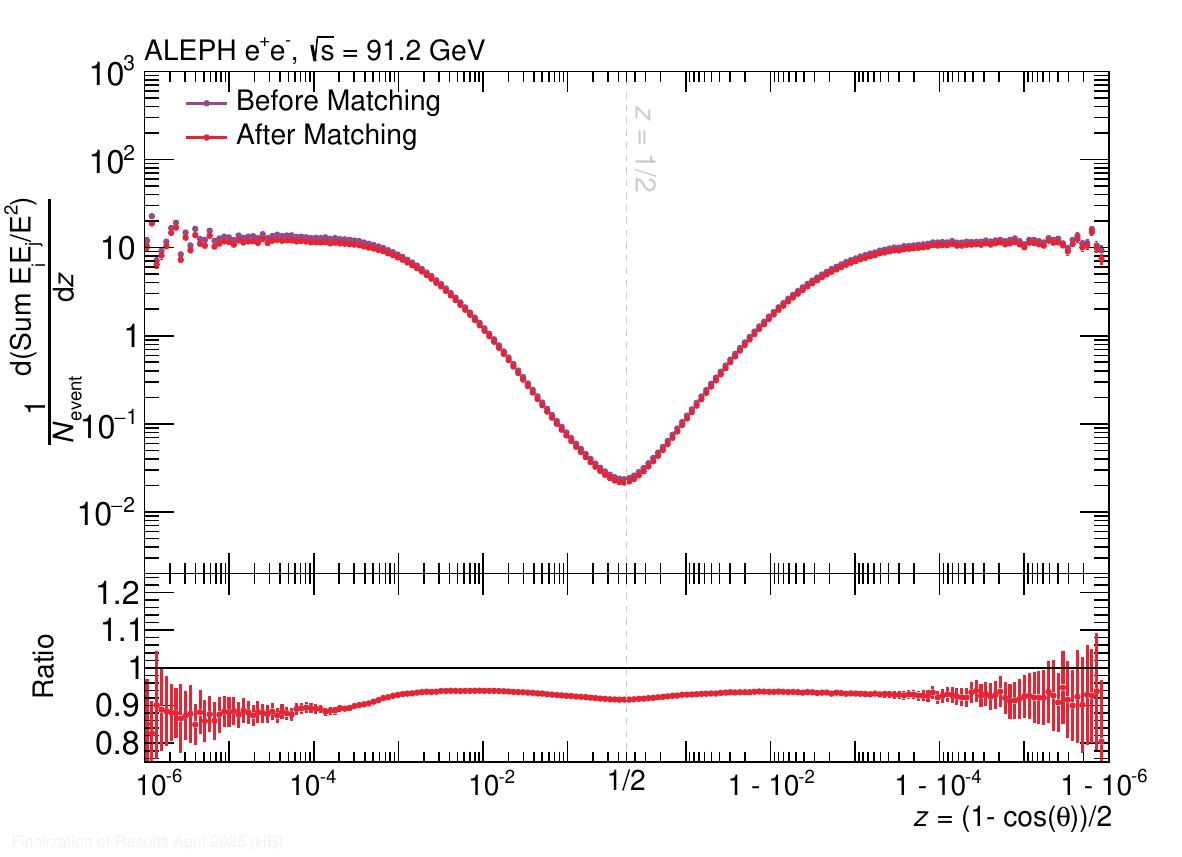}
    \caption{Left: EEC distributions before and after matching as a function of $\theta$. Right: EEC distributions before and after matching as a function of $z$. In both panels the ratio of the two distributions, which represents the matching efficiency as defined in Equation \ref{eq:matcheff}, is shown.}
    \label{fig:matchingEff}
\end{figure}

The matching efficiency, computed as the ratio defined in Equation \ref{eq:matcheff}, is shown in the bottom panel of Figure \ref{fig:matchingEff} as a function of $\theta$ (left panel) and $z$ (right panel). The correction for the matching efficiency is applied following the unfolding procedure, as shown in Figure \ref{fig:analysisOutline}. The uncertainty associated with this correction is incorporated into the systematics by the variation in matching scheme, which also alters the corresponding matching efficiency correction. 

\subsubsection{Event Selection Efficiency Correction}\label{sec:evtselcorr}
The event selection efficiency correction is performed by comparing the EEC distributions before and after the event and charged-particle selections as a function of $z$ and/or $\theta_{\rm L}$. Note that this includes the charged particle selection requirements so that the fully-corrected distribution can be directly compared to theoretical calculations. This is written in Equation \ref{eq:evtselcorr} where $\text{EEC}^{\rm tgen}_{\rm ALL}$ refers to the EEC distribution after the standard event selections written in Table \ref{tab:SelectionSummary} and $\text{EEC}^{\rm tgenBefore}_{\rm ALL}$ refers to the EEC distribution before these standard selections.\footnote{Note that the actual labels here refer to the tree names used in the standard archived ALEPH format.} To apply this correction, the EEC distribution is \textit{divided} by this correction factor such that it properly reflects the distribution prior to event selection that can be compared to MC simulations.

\begin{equation}\label{eq:evtselcorr}
    \text{Event Sel.} = \frac{\text{EEC}^{\rm tgen}_{\rm ALL}}{\text{EEC}^{\rm tgenBefore}_{\rm ALL}}
\end{equation}

\begin{figure}[ht!]
    \centering
    \includegraphics[width=0.49\linewidth]{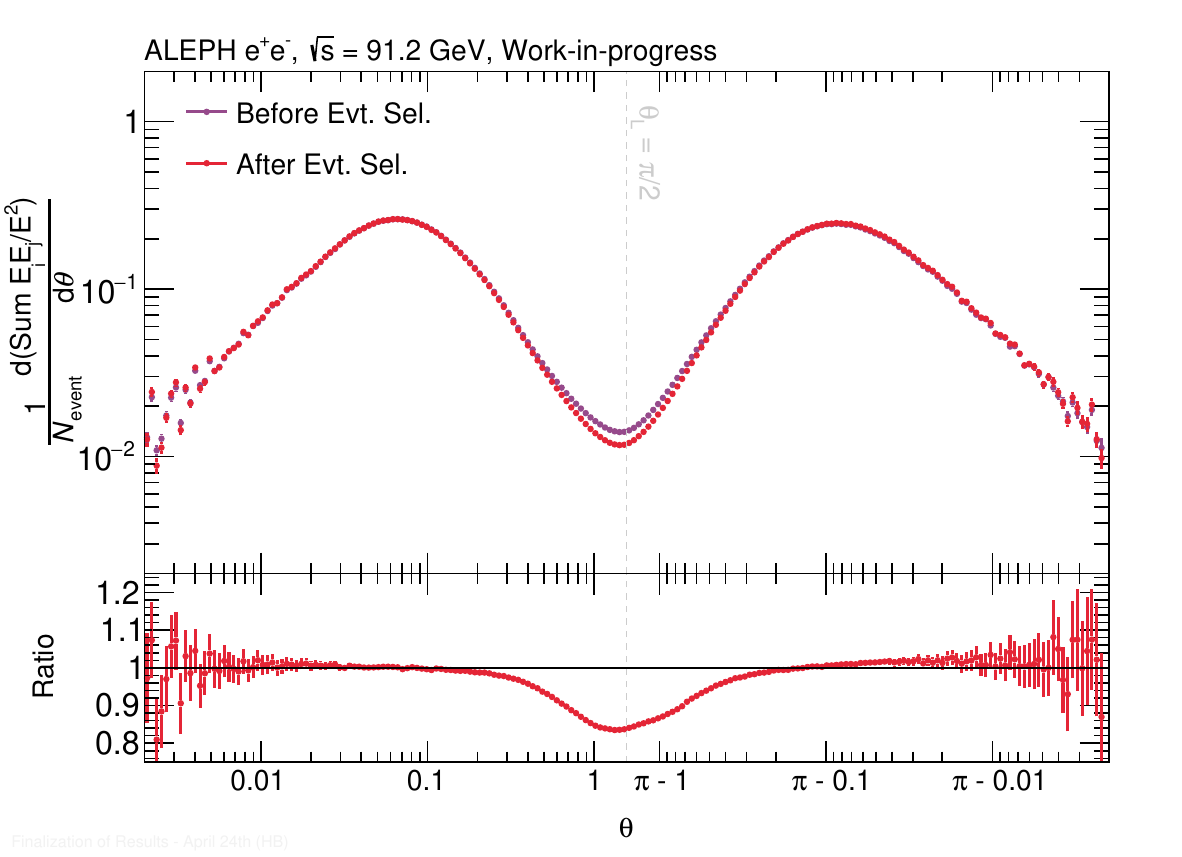}
     \includegraphics[width=0.49\linewidth]{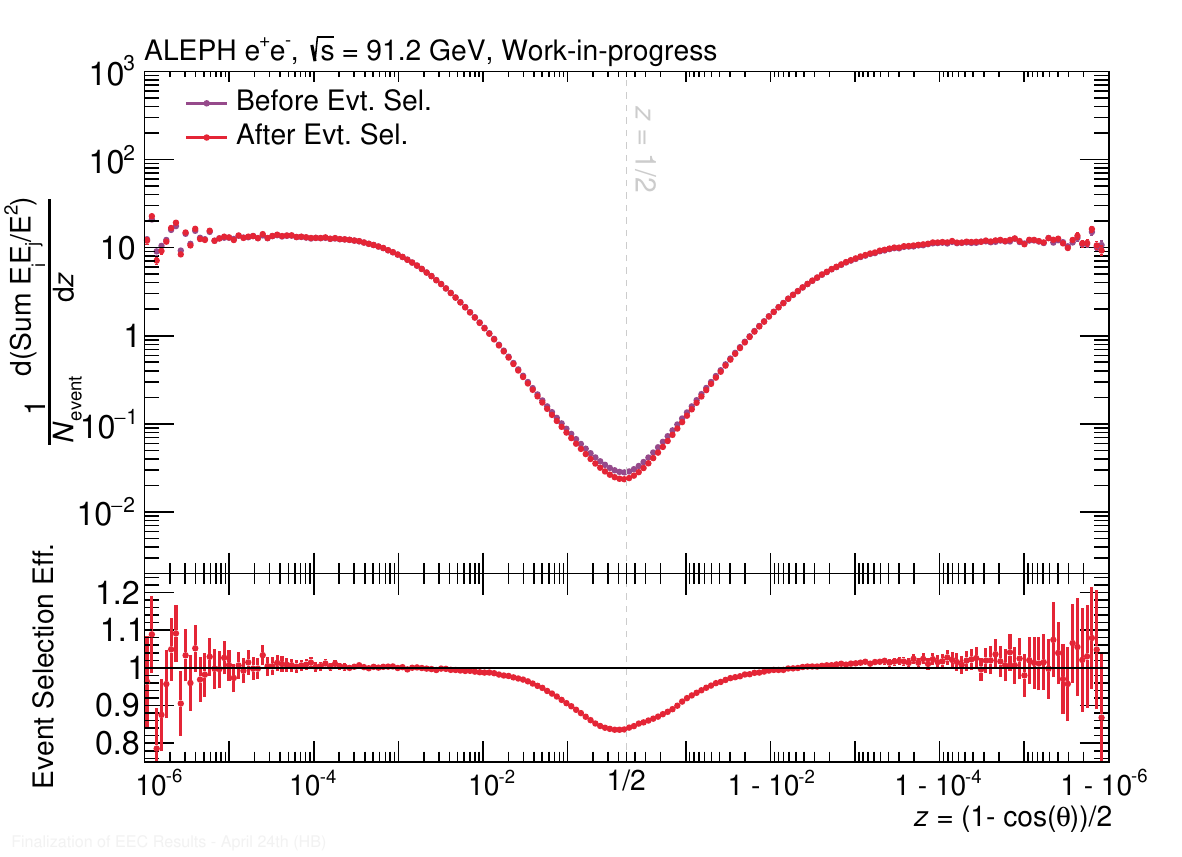}
    \caption{Event selection efficiency corrections as a function of $\theta_{\rm L}$ (left) and $z$ (right). The ratio is shown in the bottom panel that represents the inverse of the event selection efficiency correction.}
    \label{fig:evtSelCorr}
\end{figure}

Examples of this distribution can be found in Figure \ref{fig:evtSelCorr}. Note that the event selection efficiency is at its minimum near $z = 1/2$ and $\theta_{\rm L} = \frac{\pi}{2}$, which is a result of the sphericity cut removing particles at wider angles. Though the choice of the sphericity cut changes the exact magnitude of the event selection efficiency correction, the distribution after the event selection efficiency correction is the same regardless of the sphericity cut used. Alternatively, this could also be phrased as the event selection efficiency correction exhibiting good closure. This closure is demonstrated in Appendix \ref{app:evtseleff}.

\subsubsection{Tracking Efficiency Correction}\label{sec:trackingeff}
Using the MC sample introduced in Section~\ref{sec:Sample}, the reconstruction efficiency of single tracks caused by nonuniform detection efficiency and mis-reconstruction bias can be studied. From this study, one can form a description of the tracking efficiency for single tracks. 

 The efficiency is given by
\begin{equation}
\varepsilon(p_{\rm T}, \theta, \phi, {\rm N}_{\rm Trk}^{\rm Offline}) = \left[\frac{d^3 {\rm N}^{\rm reco}}{dp_{\rm T} d\theta d\phi}/\frac{d^3 {\rm N}^{\rm gen}}{dp_{\rm T} d\theta d\phi}\right]_{{\rm N}_{\rm Trk}^{\rm Offline}},
\end{equation}
where \( {\rm N}^{\rm reco} \) denotes the number of charged particles at the reconstruction level, and \( {\rm N}^{\rm gen} \) denotes the same at the generator level. The efficiency correction factor establishes a correspondence between the reconstruction level and the generator level across the \( p_{\rm T}, \theta, \) and \( \phi \) spectra.

In this analysis, the tracking efficiency correction is implicitly also included in the matching efficiency correction (discussed in Section \ref{sec:matchingeffcorr}). Therefore, no explicit additional correction for the tracking efficiency is applied in the main analysis procedure. However, the tracking efficiency is still included in this description as it is used for a number of different studies, for example the initial studies shown in Section \ref{sec:DataMCComp}.

\section{Systematic Uncertainty Evaluation}\label{sec:Syst}
In the following section the details of the systematic uncertainty evaluation will be provided. For clarity these will be broken up into rough categories as shown in the various subsections. 

\subsection{TPC Hits}
As described in Table \ref{tab:SelectionSummary}, the charged particle selection requires that there be at least 4 TPC hits. For the systematic uncertainty associated with this selection, this requirement was tightened to at least 7 TPC hits. Following the procedure in Ref.~\cite{Chen:2023nsi}, the uncertainty associated with this variation is 0.5\% and is applied independently of $\theta_{\rm L}$ or $z$. 

\subsection{Matching Uncertainty}\label{subsec:matchingsyst}
The matching procedure used to select corresponding pairs between data and the MC has an impact on the pairs that fill the response matrix, therefore impacting the final result. The nominal matching procedure used is described in Section \ref{sec:matching}. In addition, the corresponding correction for the matching efficiency also has an impact on the analysis, which is described in Section \ref{sec:matchingeffcorr}. This systematic uncertainty is evaluated by taking the difference between the nominal matching scheme (as described in Section \ref{sec:matching}) and an alternative scheme with the metric still based on Equation \ref{eq:metric} in which an energy resolution of 10\% is used by setting $\sigma E = 0.1E$ where $E$ is the energy of the track. This represents a conservative estimate of the energy resolution relative to the nominal prescription. The angular resolution is derived from the impact parameter (position) resolution reported by the ALEPH Collaboration~\cite{Buskulic:272484} and is determined to be $\sigma_{\theta} = 10^{-4}$ and $\sigma_{\phi} = 0.002$. 

\begin{figure}[ht!]
    \centering
    \includegraphics[width=0.49\linewidth]{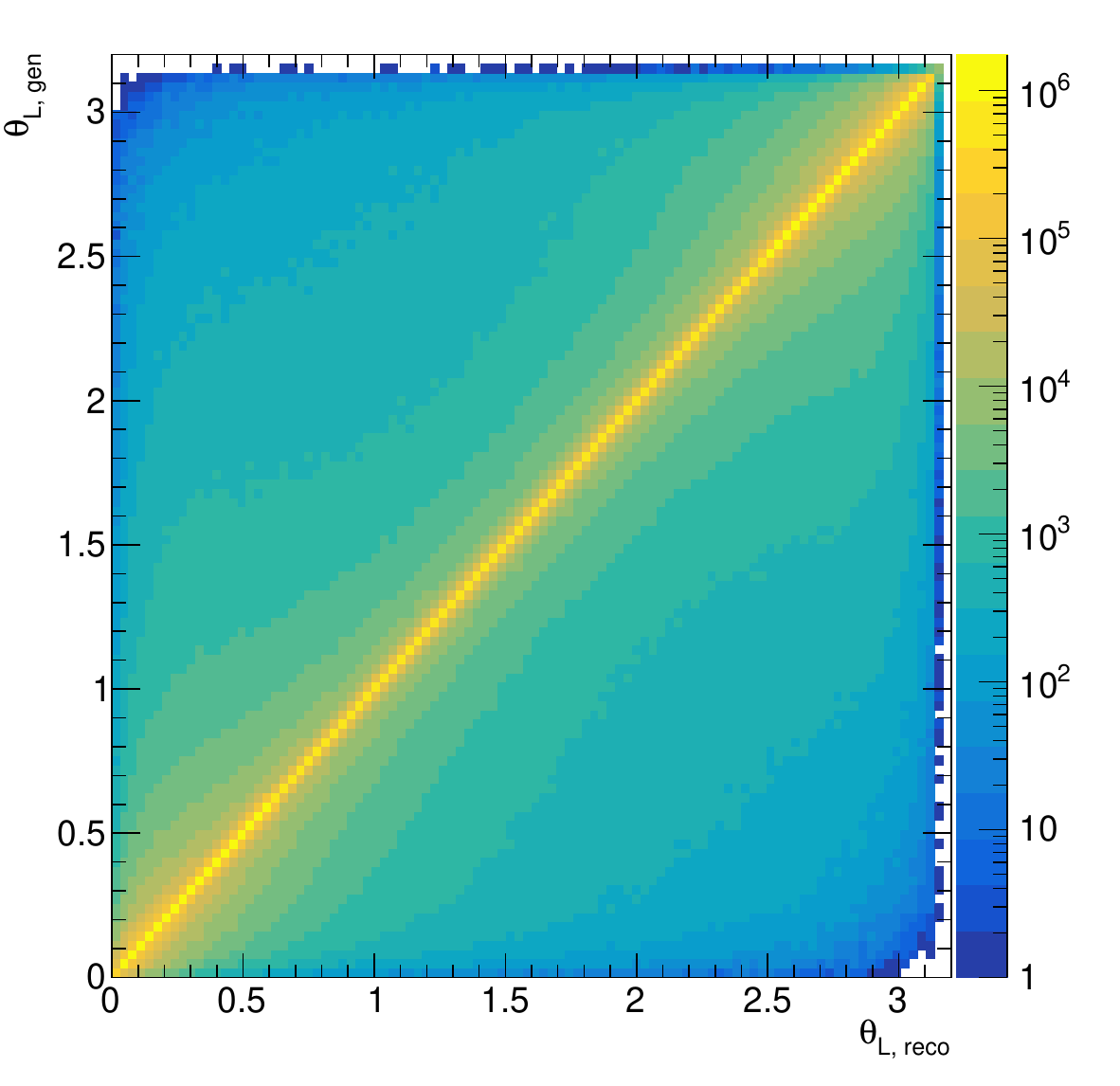}
    \includegraphics[width=0.49\linewidth]{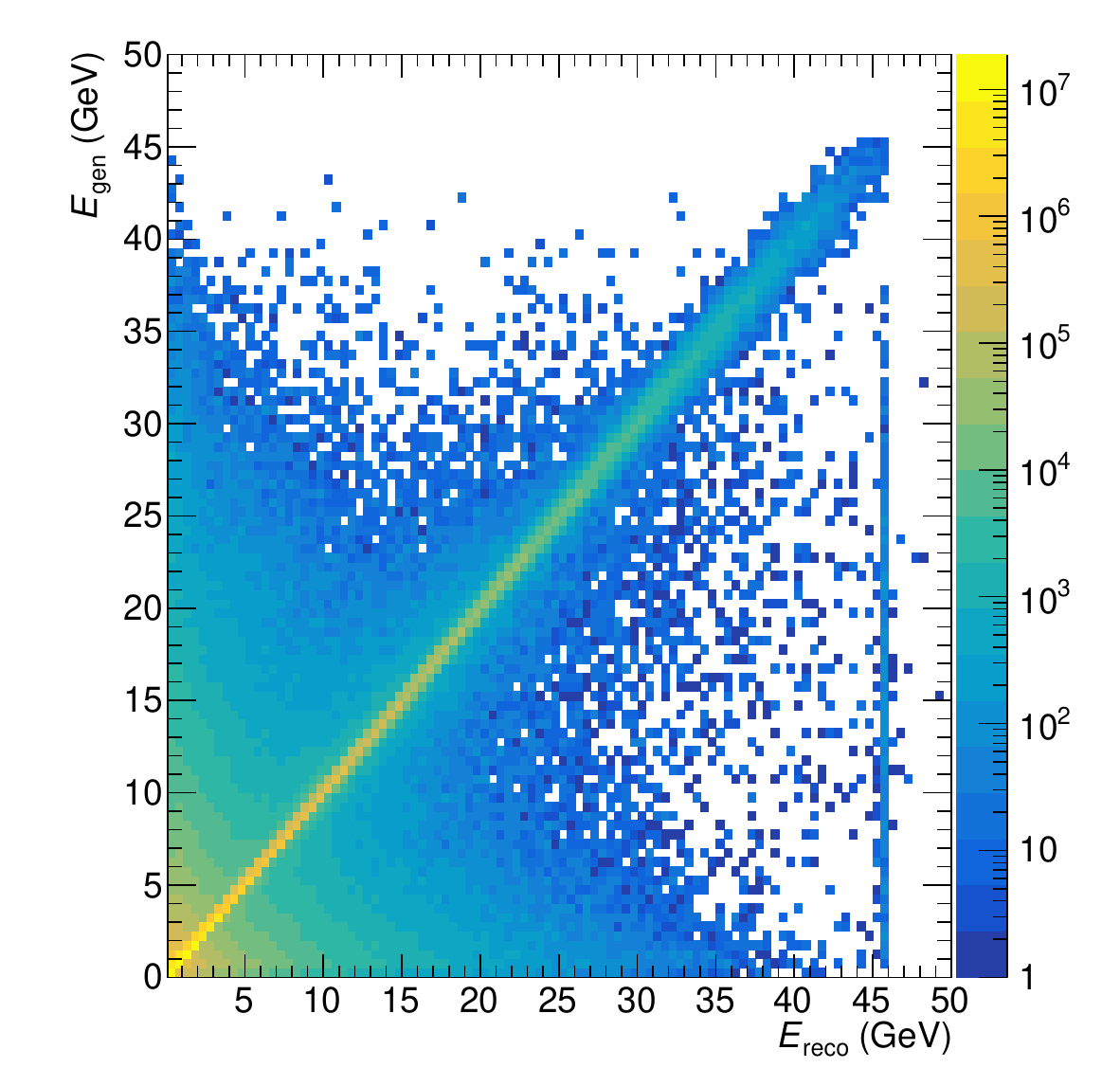}
    \caption{Left: $\theta_{\rm L, gen}$ as a function of the $\theta_{\rm L, reco}$ for the matched track pairs. Right: Energy at the generator level for single tracks as a function of the reconstructed energy of the matched track. }
    \label{fig:matchingPerformance}
\end{figure}

The overall performance of the matching method used for this variation is shown in Figure \ref{fig:matchingPerformance}, where the angular correspondence is shown in the left panel an the correspondence in energy for matched pairs and single tracks, respectively. Both of these metrics are very diagonal, validating that this alternate matching procedure is effective. The difference between these two matching schemes corresponds to a 1\% uncertainty, which covers the difference between the two schemes in all parts of phase space. 

\subsection{Unfolding Uncertainties}\label{sec:UnfSyst}
There are a number of systematic uncertainties associated with the unfolding procedure used in this analysis. The nominal unfolding procedure is described in Section \ref{sec:unfolding}.\\

\noindent\textbf{Regularization parameter:} The first is a systematic uncertainty associated with the regularization parameter of the unfolding procedure. Bayesian unfolding~\cite{D'Agostini:265717} is employed in this analysis, meaning that the regularization comes from the number of iterations. The nominal number of iterations chosen for this analysis is 4. For this systematic variation, the unfolded result from a different iteration was chosen. This alternative iterations choice was chosen to be the optimal number of iterations for the unfolding of the reweighted response, which will be described in the next section. The alternative iteration number used in this analysis is 3. The magnitude of this systematic uncertainty is less than 5\% in all regions of phase space.  \\

\begin{figure}[ht!]
    \centering
    \includegraphics[width=0.7\linewidth]{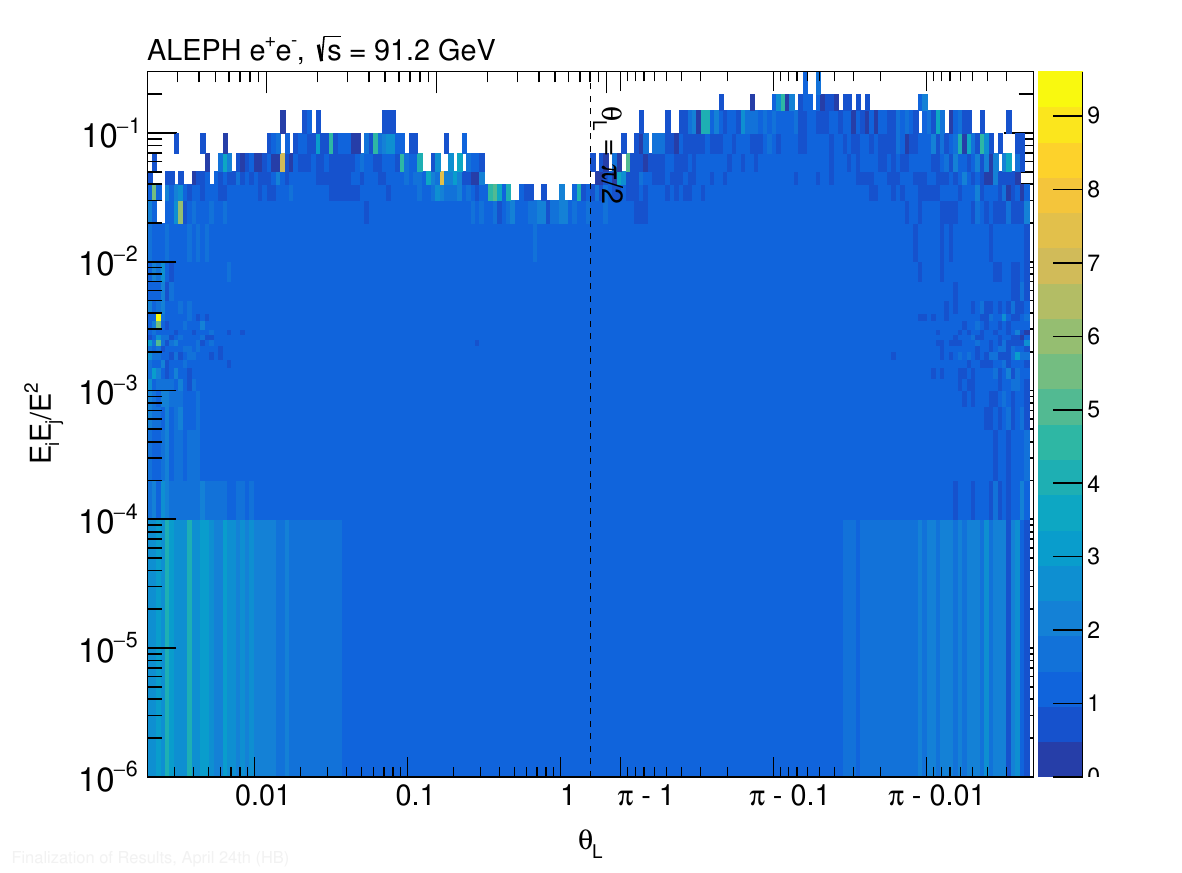}
    \caption{Reweighting factors between data and MC for the systematic uncertainty associated with the prior distribution used in the unfolding procedure.}
    \label{fig:prior}
\end{figure}

\noindent\textbf{Prior Uncertainty:} In this measurement a Bayesian unfolding procedure~\cite{D'Agostini:265717} is applied. Bayesian unfolding seeks to estimate the true distribution using Bayes Theorem given a specific prior distribution, the response matrix, and the measured distribution. Here, the nominal choice for the prior is the archived $\textsc{pythia}$ 6 MC. To account for the uncertainty associated with the choice of prior, it is common to form the response matrix with a different generator, such as HERWIG. However, in this instance such a variation is not possible as the archived $\textsc{pythia}$ 6 MC is the only MC for which a detector simulation available, which is needed for the unfolding procedure. Therefore, in this instance the systematic uncertainty associated with the sensitivity of the result to the prior distribution was evaluated by reweighting the archived $\textsc{pythia}$ 6 MC by the parameterized ratio of the reconstructed-level archived $\textsc{pythia}$6 MC to data, accounting for any shape differences between the two. The reweighting factors prior to parametrization can be found in Figure \ref{fig:prior}. 

\begin{figure}[ht!]
    \centering
    \includegraphics[width=0.7\linewidth]{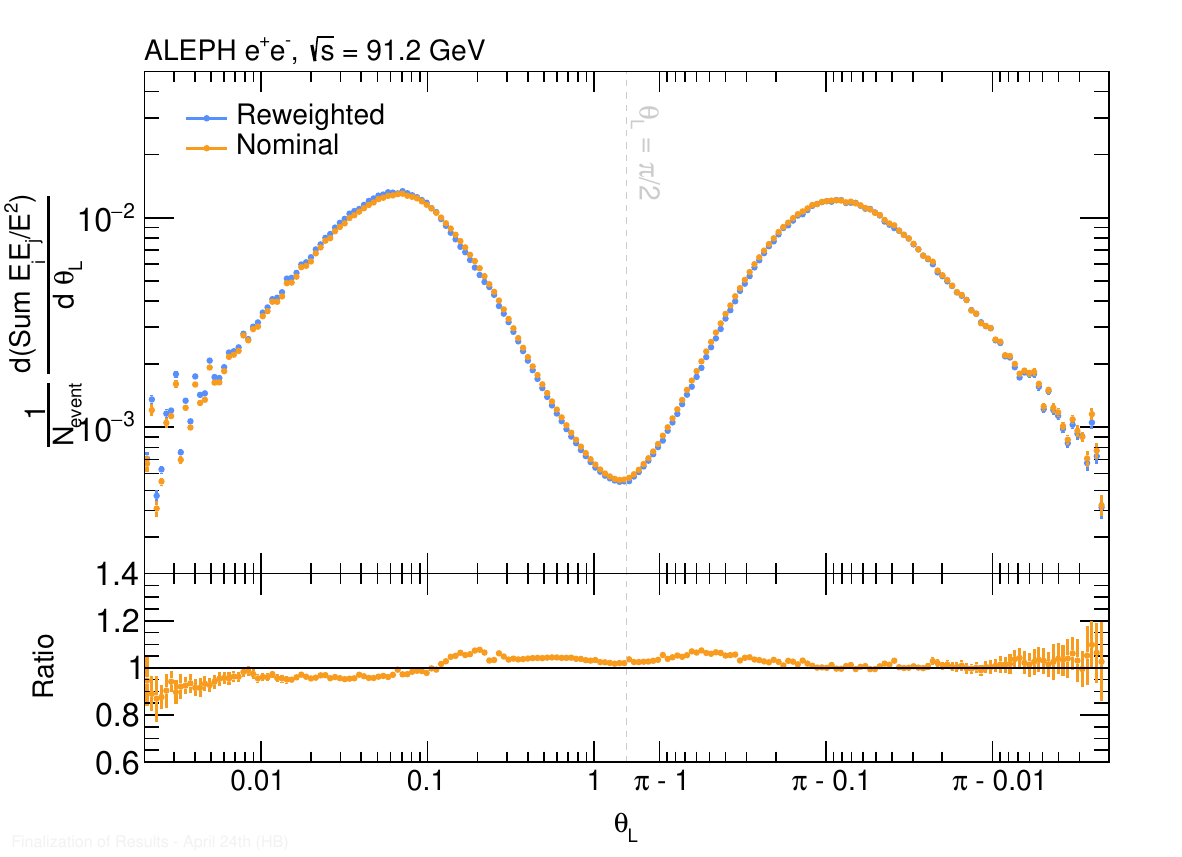}
    \caption{Comparison of the unfolded E2C distribution prior to corrections for the nominal (non-reweighted) case compared to the reweighted variation that is utilized in the prior uncertainty. The ratio between these distributions that will then be used as a systematic uncertainty can be found in the bottom panel. }
    \label{fig:reweightingsyst}
\end{figure}

To parameterize this ratio, each bin in the $x$ (or $\theta_{L}$/$z$) axis is projected onto the $E_{i}E_{j}/E^{2}$ axis and is fit with a constant function. The resulting reweighting factor is then achieved by sampling the corresponding fit. This ensures that the resulting reweighting is not driven solely by fluctuations. The reweighting factor is then applied when filling the response matrix, effectively changing the shape of the mapping between the generator-level and measured distributions to more closely match the shape of data. The resulting distributions are then unfolded and compared, as shown in Figure \ref{fig:reweightingsyst} where the ratio of the two distributions is shown in the bottom panel. 

The difference between the nominal and reweighted unfolded distributions is taken as the systematic uncertainty. Overall, this uncertainty is minimal throughout most of the distribution, but grows to be larger ($\sim 11\%$) in the tails of the distribution. The prior uncertainty is a dominant uncertainty in this analysis. 

\subsection{Binning Systematic}\label{subsec:binningsyst}
As mentioned in Section \ref{sec:projections} the procedure to construct the final observable introduces a non-closure that is corrected for in a corresponding correction, described in Section \ref{sec:binningcorrection}. As shown in Figure \ref{fig:binningSystematic}, this correction factor can be derived using both data and gen-level MC. For the nominal analysis, the correction derived from gen-level MC was employed. The systematic uncertainty attributed to this correction conservatively approximates the difference to be 5\% of the total magnitude of the correction, accounting for any small differences in shape between uncorrected data and gen-level MC. 

\subsection{Summary}
\begin{figure}[ht!]
    \centering
    \includegraphics[width=0.49\linewidth]{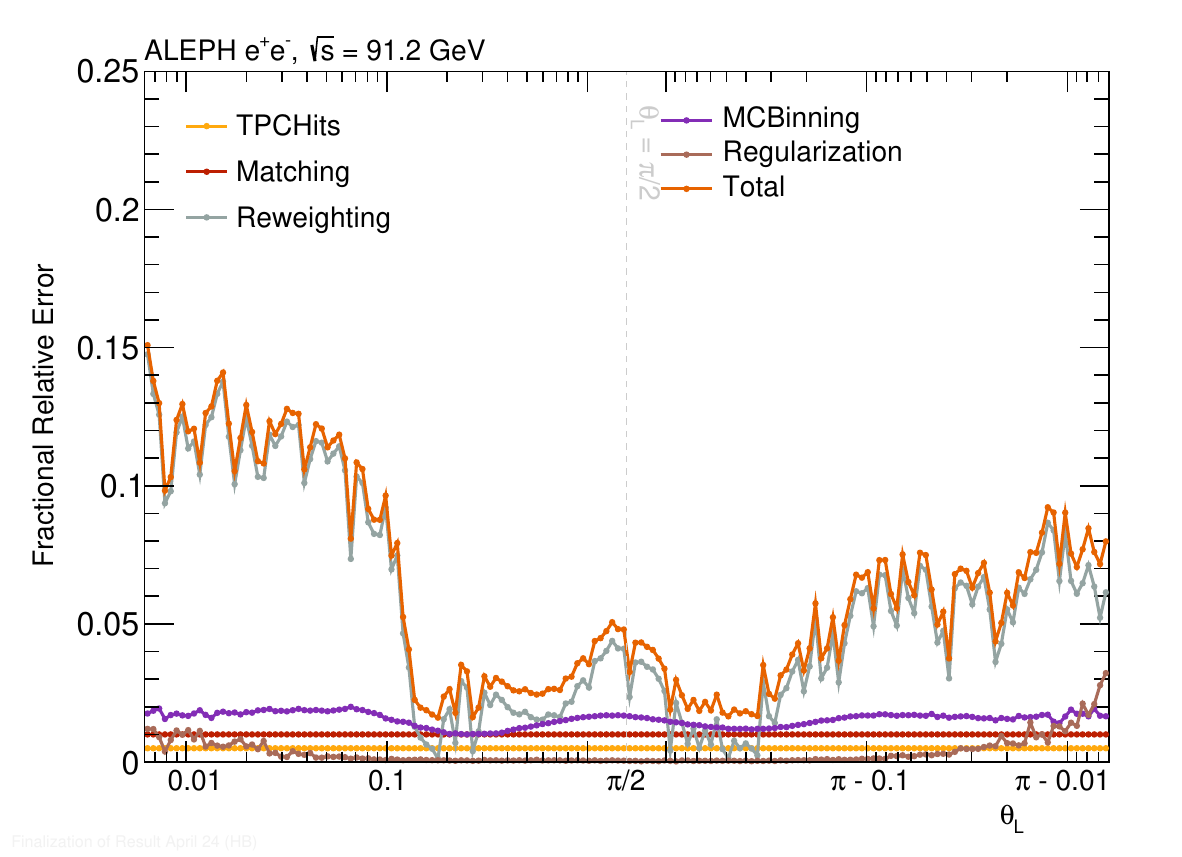}
     \includegraphics[width=0.49\linewidth]{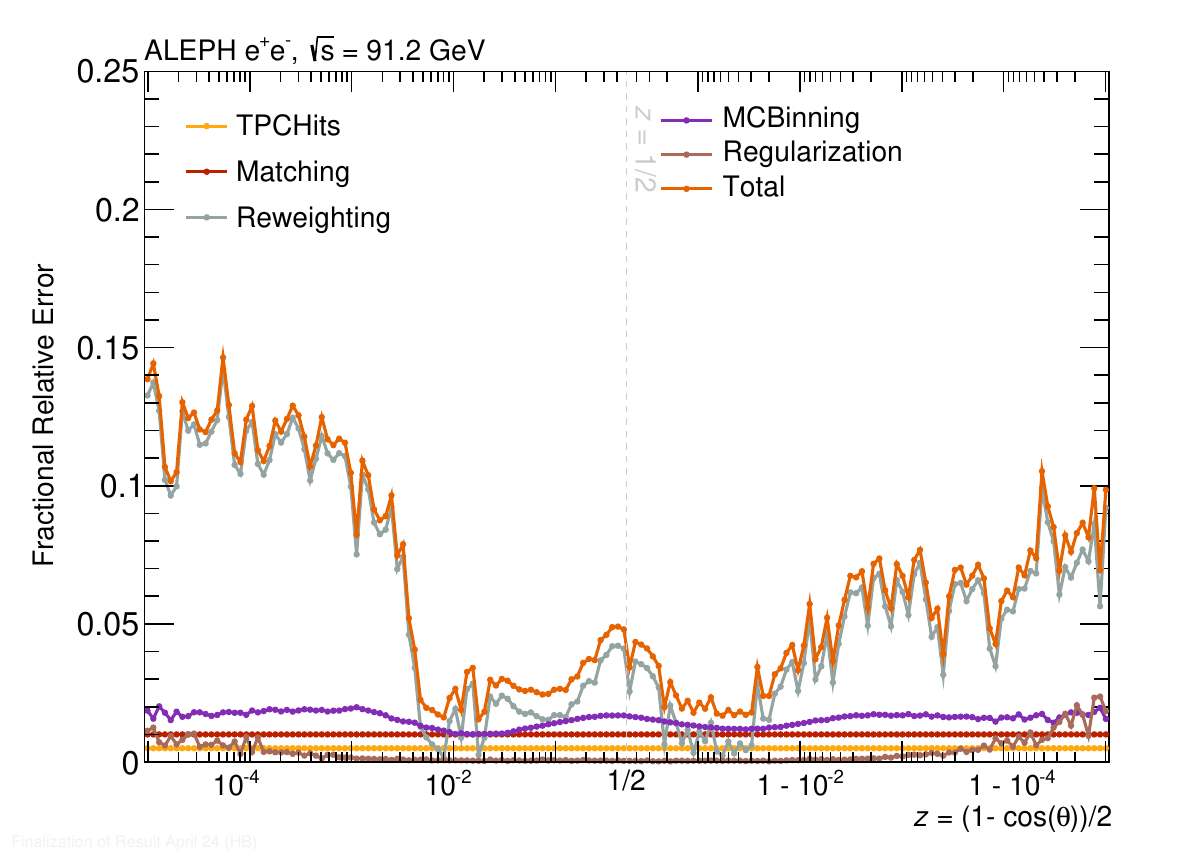}
    \caption{Summary of the systematic uncertainties as a function of  $\theta_{\rm L}$ (left) and $z$ (right).}
    \label{fig:systematicSummary}
\end{figure}

The systematic uncertainties as mentioned above are added in quadrature to produce the final systematic uncertainty applied to the final results. A summary of the relative errors as well as their contribution to the total systematic uncertainty is shown in Figure \ref{fig:systematicSummary}. As demonstrated in Figure \ref{fig:systematicSummary}, the reweighting systematic uncertainty is the dominant uncertainty in this analysis. Though independently derived, the systematic uncertainties are nearly identical for $\theta_{\rm L}$ and $z$ as they are directly related variables. The resulting systematic uncertainties are then added as additional uncertainties on top of the statistical uncertainties to the final results, which will be discussed in the next section of this analysis note.

\section{Results}\label{sec:results}
\begin{figure}
    \centering
    \includegraphics[width=0.49\linewidth]{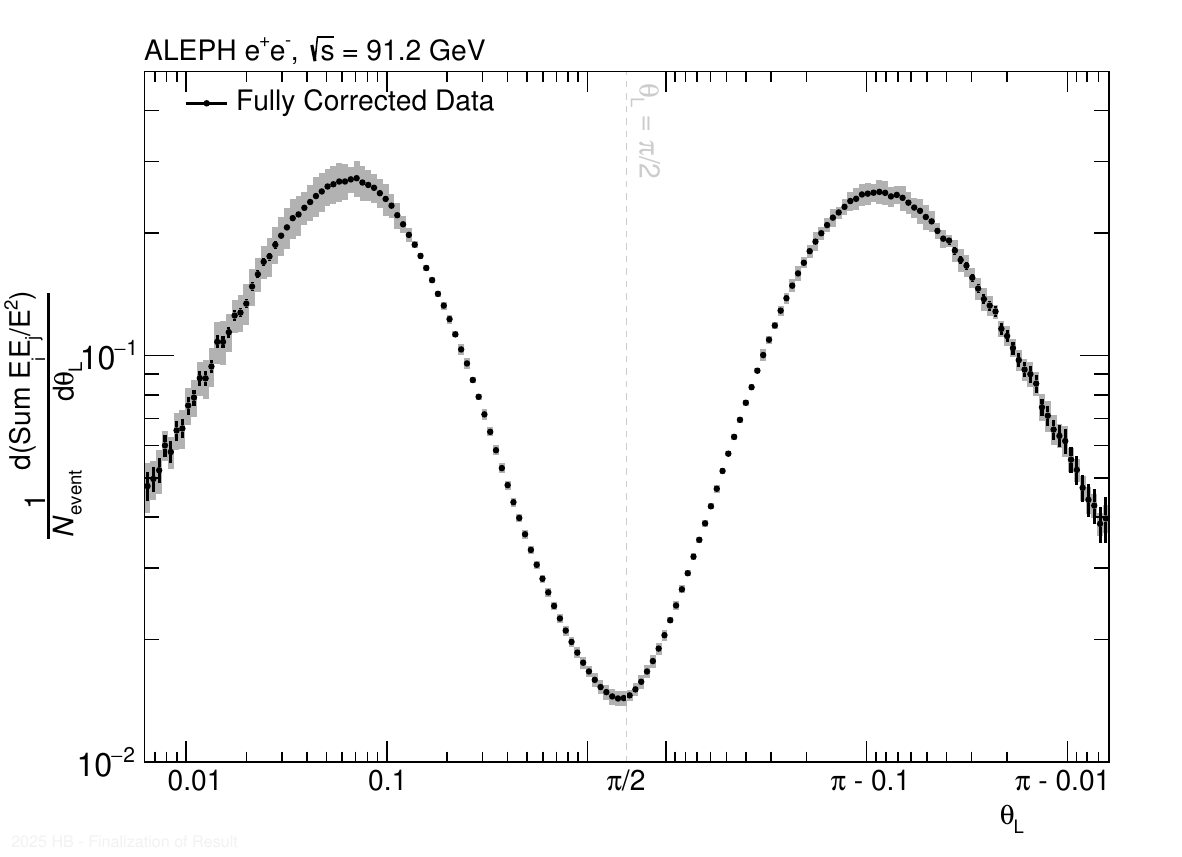}
    \includegraphics[width=0.49\linewidth]{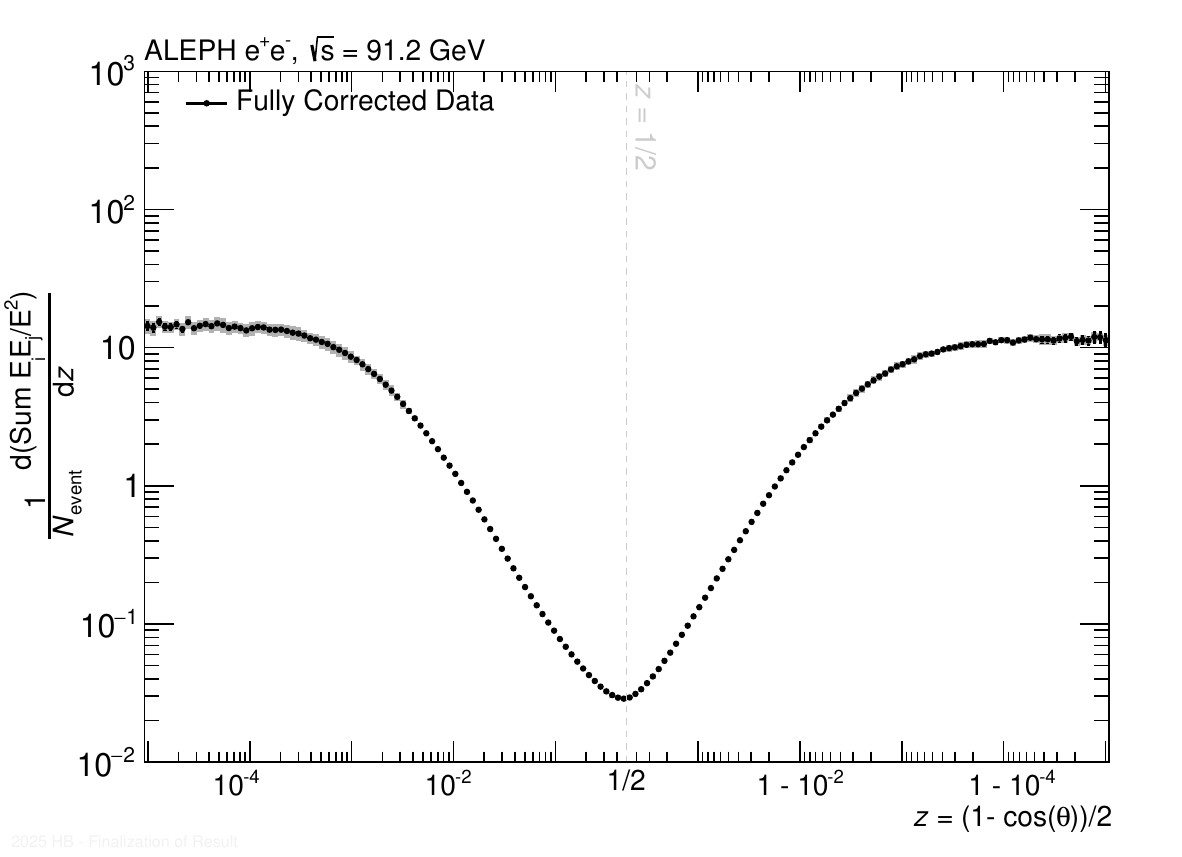}
    \caption{Fully-corrected E2C distributions from ALEPH data written as a function of $\theta_{\rm L}$ (left) and $z$ (right). Statistical error bars are displayed as vertical lines and systematic error bars are shown in the red boxes.}
    \label{fig:ResultsDataOnly}
\end{figure}

\begin{figure}[ht]
    \centering
    \includegraphics[width=0.75\linewidth]{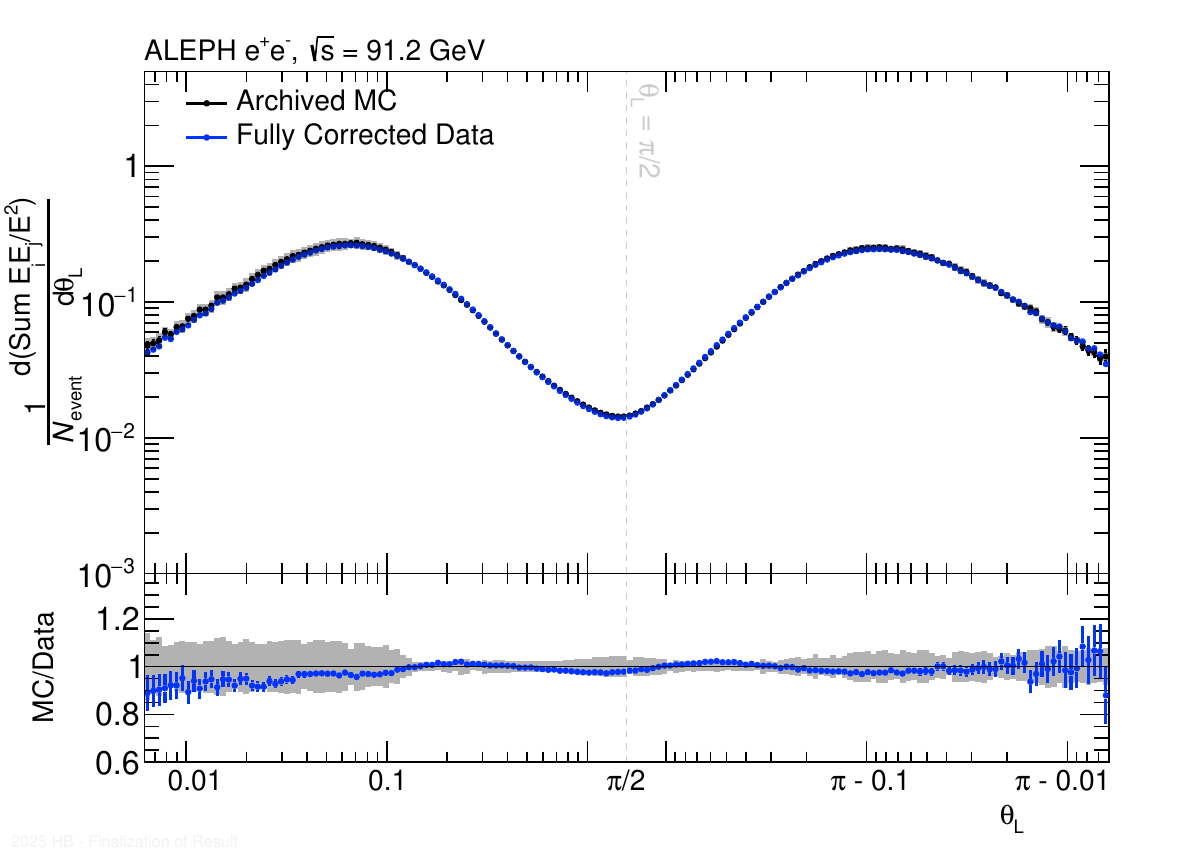}
    \caption{E2C distribution as a function of $\theta_{\rm L}$ for the generator-level archived $\textsc{pythia}$ 6.1 MC distribution before any event selection (blue) and the fully corrected ALEPH data with the corresponding systematic uncertainties (red).}
    \label{fig:resultsMCtheta}
\end{figure}

\begin{figure}[ht]
    \centering
    \includegraphics[width=0.75\linewidth]{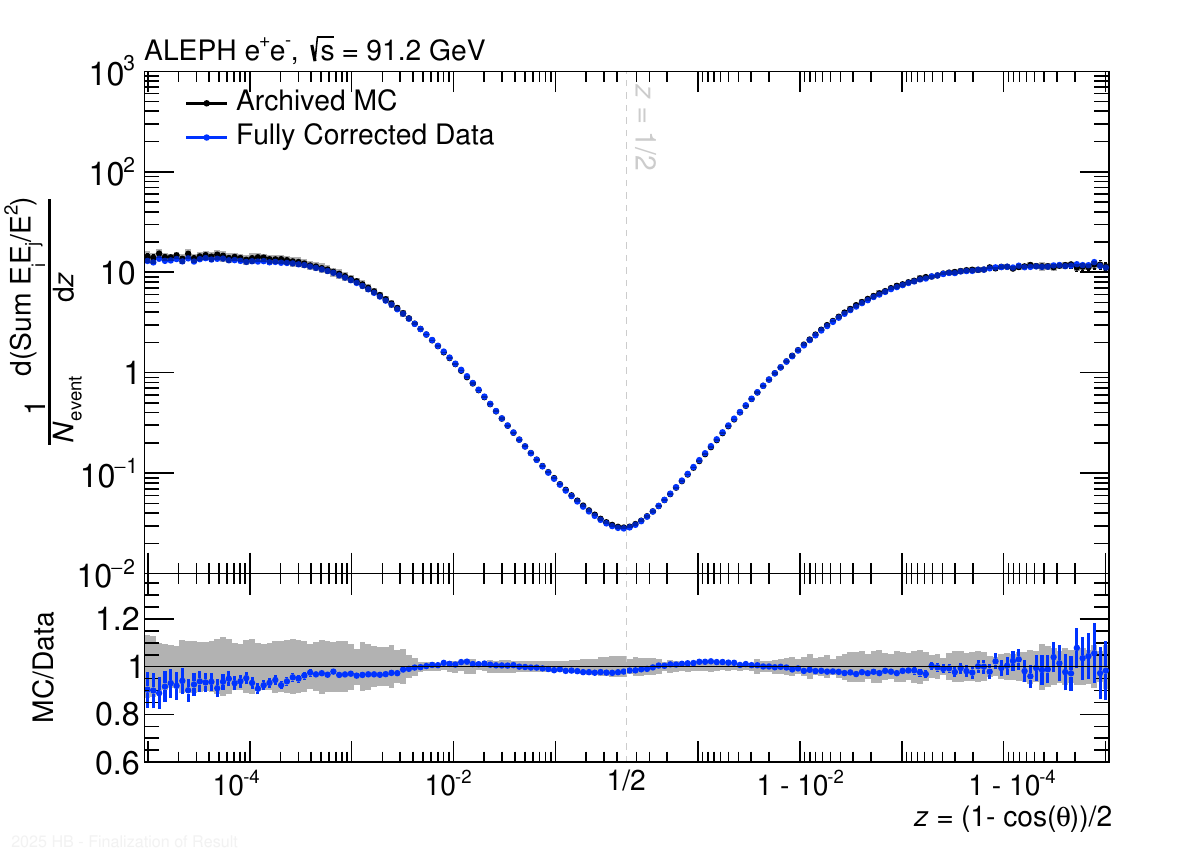}
    \caption{E2C distribution as a function of $z$ for the generator-level archived $\textsc{pythia}$ 6.1 MC distribution before any event selection (blue) and the fully corrected ALEPH data with the corresponding systematic uncertainties (red).}
    \label{fig:resultsMC}
\end{figure}

In Figures \ref{fig:resultsMCtheta} and \ref{fig:resultsMC} the fully-corrected E2C is shown as a function of $\theta_{\rm L}$ and $z$, respectively. These distributions span the angular observable from the collinear region all the way to the back-to-back region. This measurement represents the first fully-corrected measurement of the E2C in $e^{+}e^{-}$ and is the most precise measurement of the of the E2C in $e^{+}e^{-}$ to date,  particularly in the collinear and back-to-back limits. For a comparison of these results to previous OPAL measurements illustrating the improved precision, refer to Section \ref{sec:opalComparisons}. Note that a low and high cutoff for angles of $\theta_{\rm L}\sim 0.006$ and $\theta_{\rm L}\sim \pi - 0.006$, respectively, is placed on the resulting distribution to reflect the 3-digit precision of the original ALEPH text file format. The fully corrected data as a function of $\theta_{\rm L}$ and $z$ is shown in Figure \ref{fig:ResultsDataOnly}. Here, one can see by eye the excellent precision of the data sample spanning the full kinematic phase space. 

In addition, the E2C distributions are also compared to the archived $\textsc{pythia}$ 6.1~\cite{Sjostrand:2000wi} MC (shown in blue in Figures  \ref{fig:resultsMCtheta} and \ref{fig:resultsMC}), with the ratio of the two distributions shown in the bottom panel. 

As mentioned previously, and shown in MC in Figure \ref{fig:reflection}, the universality of the scaling behavior in the free hadron regions, corresponding to the collinear and Sudakov limits, is of great interest. In Figure \ref{fig:resultsMC} one can see that the collinear and Sudakov free hadron regions are roughly compatible with one another.~\footnote{Quantitative comparisons between the two are left for future work.} Due to the compatibility of the resulting distribution between data and MC, the level of agreement between the collinear and Sudakov regions is roughly that of Figure \ref{fig:reflection}.

These results are also compared to a variety of MC simulations. In addition to the archived $\textsc{pythia}$ 6.1 sample, the results are also compared to a $\textsc{pythia}$ 8.3~\cite{Bierlich:2022pfr} sample with a variety of different shower implementations. In the default, so-called ``simple" shower implementation, the parton shower proceeds in a $p_{\rm T}$-ordered fashion. Two variations of this parton shower implemented in  $\textsc{pythia}$ 8.3, the \textsc{DIRE} and \textsc{VINCIA} showers, are also shown. The $\textsc{pythia}$ DIRE~\cite{Hoche:2015sya} shower~\footnote{For more information, see \url{https://pythia.org/latest-manual/DireShowers.html}.} is a dipole-like shower implementation where the ordering is based on the transverse momentum in the soft approximation, therefore presenting a hybrid option in between traditional dipole and parton shower implementations. The $\textsc{pythia}$ VIrtual Numerical Collider with Interleaved Antennae (\textsc{VINCIA}) shower~\footnote{For more information, see \url{https://pythia.org/latest-manual/Vincia.html}.} is a $p_{\rm T}$-ordered showering model based on the antenna formalism~\cite{Giele:2007di}. For the hadronization model both $\textsc{pythia}$ 6.1 and all variations of $\textsc{pythia}$ 8.3 shown here employ a Lund string model of hadronization. 

\begin{figure}[ht!]
    \centering
    \includegraphics[width=0.75\linewidth]{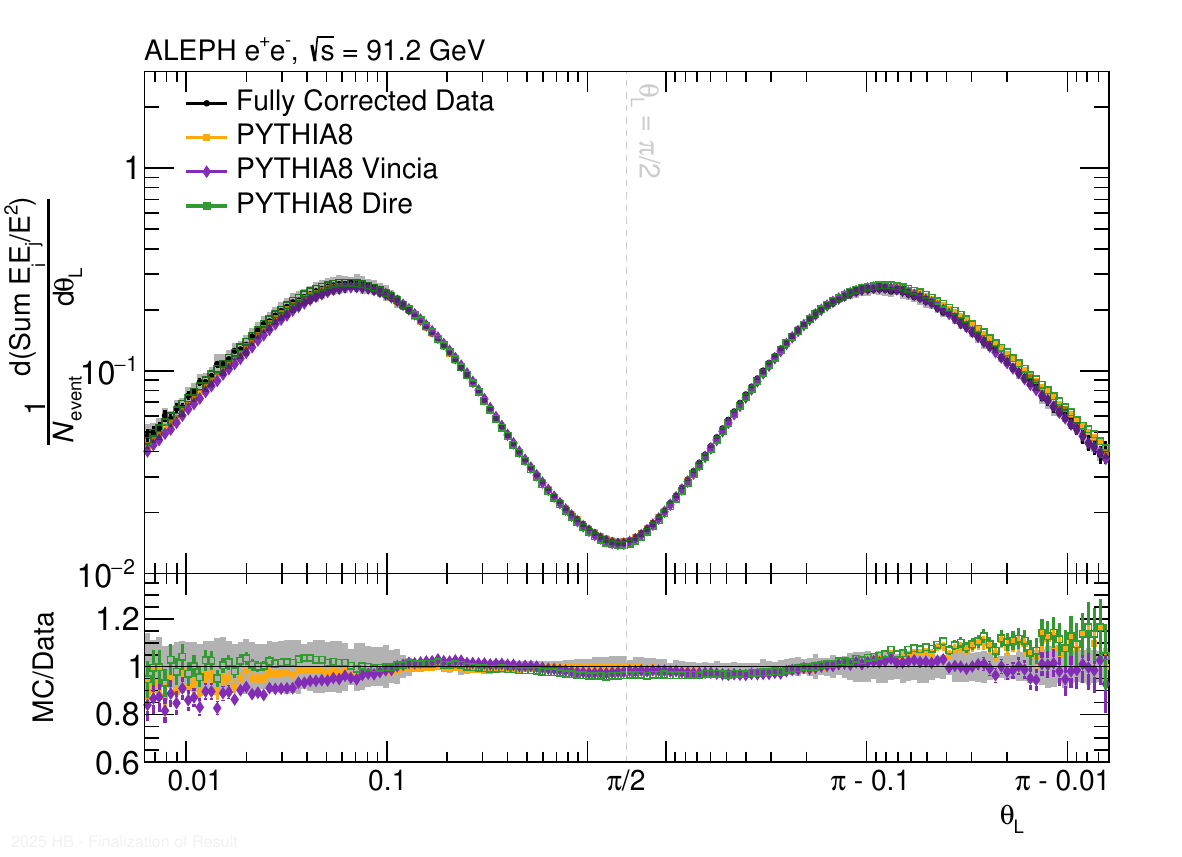}
    \includegraphics[width = 0.75\linewidth]{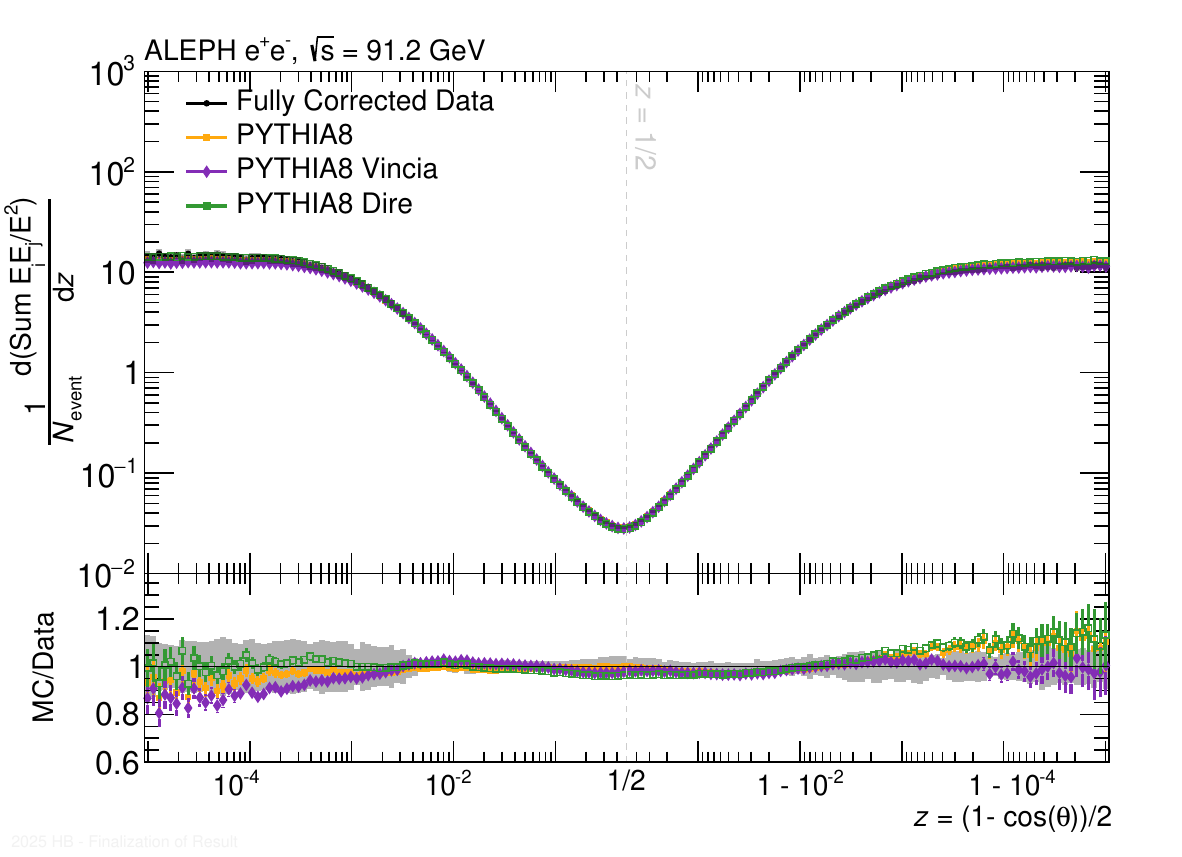}
    \caption{Comparison between the E2C distribution in data and the various PYTHIA8 simulations as a function of $\theta_{\rm L}$ (top) and $z$ (bottom). In the bottom panel of each of the figures the ratio with the data points is taken. The black error band represents the systematic uncertainties in the data points and is propagated to the ratio in the bottom panel.}
    \label{fig:DataMCPYTHIA8}
\end{figure}

A comparison between data and \textsc{pythia} 6.1 is shown in Figures \ref{fig:resultsMCtheta} and \ref{fig:resultsMC}. The comparison between data and \textsc{pythia} 8.3 with the different parton shower variations is shown in Figure \ref{fig:DataMCPYTHIA8}. One can see that in general, the various models span a wide range of predictions, differing by about 10-20\% in some regions from one another. Amongst the various models, the archived $\textsc{pythia}$ 6.1 MC tends to have the best agreement across all parts of phase space with some deviations present at small angles. The $\textsc{pythia}$ 8.3 nominal and \textsc{DIRE} parton showers exhibit a similar shape, deviating from the data in the large angle region, but otherwise providing a good description. The $\textsc{pythia}$ 8.3 \textsc{VINCIA} parton shower shows good agreement with data throughout most of the distribution, but showing ~10-20\% deviations in the collinear free hadron region. 


 The E2C measurement presented in this note allows for differential comparisons to be made from the collinear to the back-to-back regimes. Overall, these comparisons with MC show large deviations, comparable to those seen in previous ALEPH analyses~\cite{Chen:2021uws}. Therefore, measurements such as this one can greatly aid to improve the understanding and modeling of the $e^{+}e^{-}$ system. This also would improve LHC event generators, many of which have input from $e^{+}e^{-}$ collision data.

\section{Comparisons with previous results}
\label{sec:opalComparisons}
As mentioned in Section \ref{sec:Introduction}, N-point energy correlators were originally proposed for the study of $e^{+}e^{-}$ collisions and were measured by the OPAL~\cite{OPAL:1990reb,OPAL:1993pnw}, DELPHI collaboration~\cite{DELPHI:1990sof}, ALEPH~\cite{ALEPH:1990vew} and L3~\cite{L3:1991qlf,L3:1992btq} collaborations at LEP and the SLD experiment at SLAC~\cite{SLD:1994idb}. The OPAL collaboration has provided tabulated results of the data points, making it possible to do exact comparisons here.~\footnote{Note that a separate analysis of DELPHI archived data is in progress and will be utilized in the future for the purposes of comparison.} For OPAL the definition used for the measurement of the E2C is written in Equation \ref{eq:OPAL} where $\chi$ is the angle (in degrees) between two particles $i$ and $j$ with energies $E_{\rm i}$ and $E_{\rm j}$, $E_{\rm vis}$ is the sum of energies of all particles in the event, $\Delta \chi$ is the angular bin width, and $N$ is the total number of events. 
\begin{equation}\label{eq:OPAL}
    EEC(\chi) = \frac{1}{\Delta \chi N}\int^{\chi + \Delta\chi/2}_{\chi - \Delta\chi/2} \sum^{N}_{\text{events}}\sum_{i,j}\frac{E_{\rm i}E_{\rm j}}{E^{2}_{\rm vis}}\delta(\chi^{'}-\chi_{\rm i,j})d\chi^{'}.
\end{equation}
 Note that there are some key differences how OPAL defines the energy correlators and how they are defined in this work as described in Section \ref{sec:EEC}. Firstly, all possible correlations of particles are used in constructing the energy correlator. That is, correlations of a particle with itself (i.e. $i = j$ case) are included as well as a double counting of correlations are included (i.e. $(i,j)$ and $(j,i)$).
In addition, both charged and neutral particles are utilized, whereas in our case, only \textit{charged} particles are included in the correlation. In addition, we normalize by the fixed $E$ determined by the center of mass energy (91.2 GeV) whereas OPAL normalized by the total energy of all particles in the event.\footnote{These two quantities will be similar, but not the same.}

\begin{figure}[ht!]
    \centering
    \includegraphics[width=0.75\linewidth]{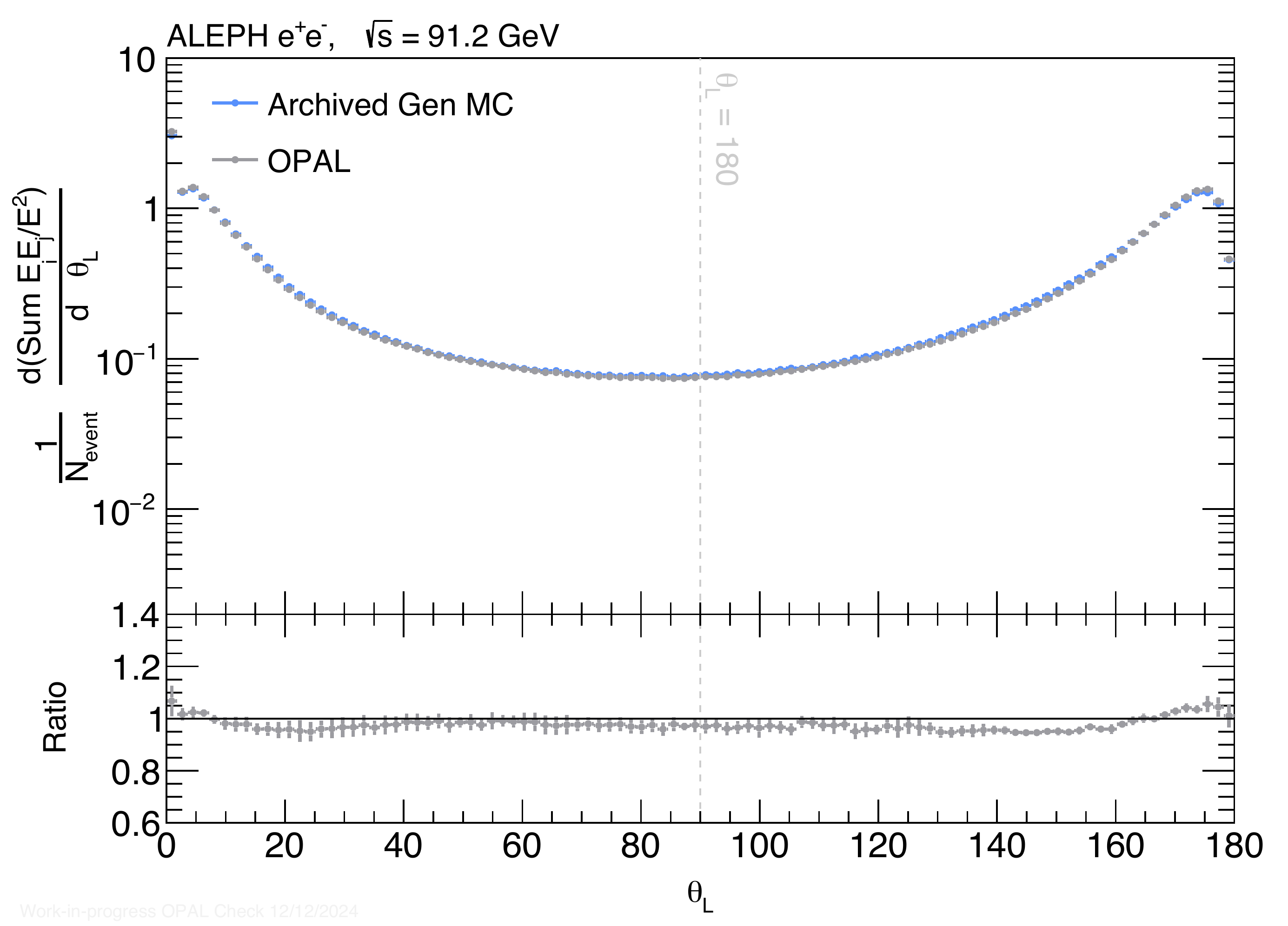}
    \caption{Published OPAL data from Ref.~\cite{OPAL:1993pnw} compared to archived $\textsc{pythia}$ 6 MC with the same selections and binning. The ratio between these curves is included in the bottom panel. Note that the $\theta_{\rm L}$ axis is presented in units of degrees. }
    \label{fig:opalLinear}
\end{figure}

Due to the above-mentioned difference in the definitions, it is difficult to directly compare the results presented in Section \ref{sec:results} to the OPAL data tabulated in Ref.~\cite{OPAL:1993pnw}. Therefore, we have instead compared the OPAL results to archived MC at the generator level using the original OPAL definition, which shows good agreement with the results presented in Section \ref{sec:results}. This comparison is shown in Figure \ref{fig:opalLinear}. Here, the binning of the archived $\textsc{pythia}$ 6 MC curve was modified in order to allow for a direct comparison between the results, which is included in the bottom panel. These two results are consistent within 5\% of one another, indicating a general agreement with the trends seen in $\textsc{pythia}$ 6 archived MC and therefore the ALEPH data. Note that this 5\% value should not be taken as the overall level of agreement between ALEPH data and OPAL, as here archived $\textsc{pythia}$ 6 MC is used in the comparison. However, this exercise gives confidence that the two results are qualitatively consistent. 

\begin{figure}[ht!]
    \centering
    \includegraphics[width=0.75\linewidth]{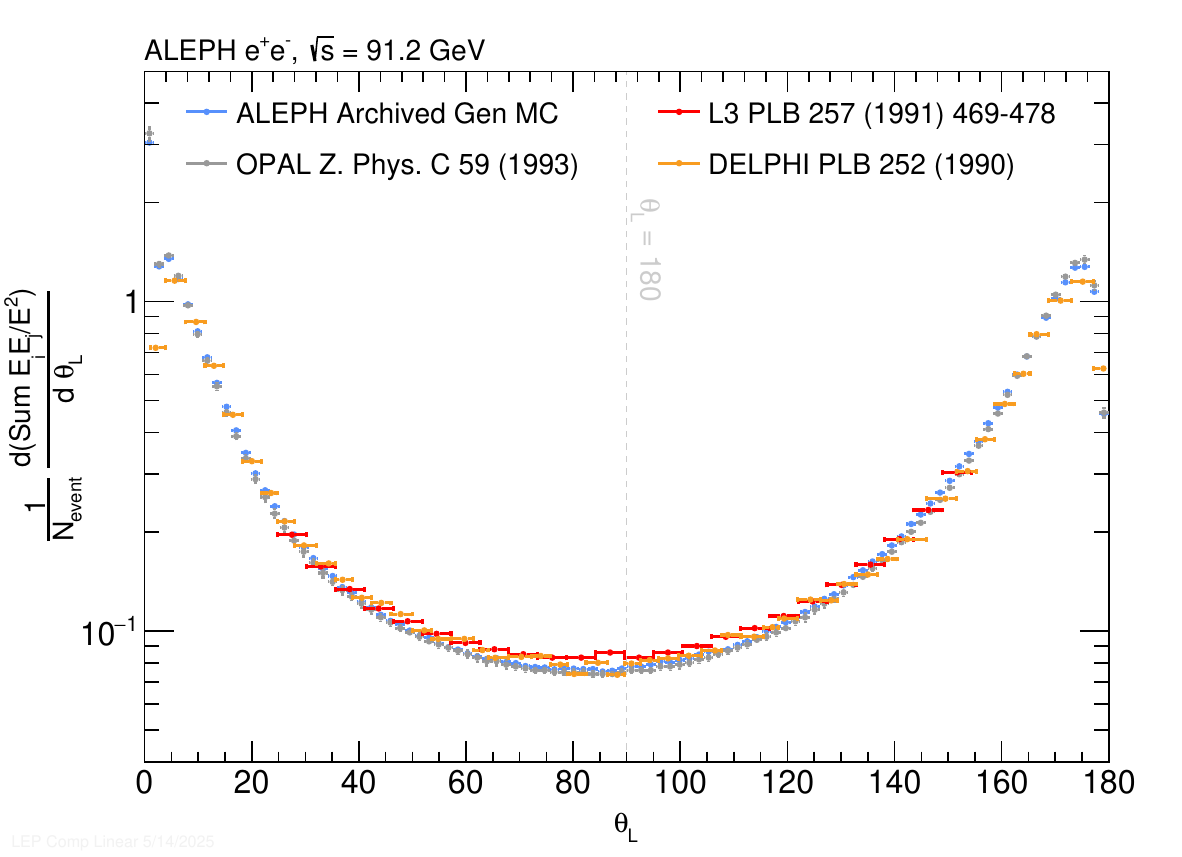}
    \caption{Published LEP dat compared to archived $\textsc{pythia}$ 6 MC with the same selections.}
    \label{fig:LEPLinear}
\end{figure}

\begin{figure}[ht!]
    \centering
    \includegraphics[width=0.75\linewidth]{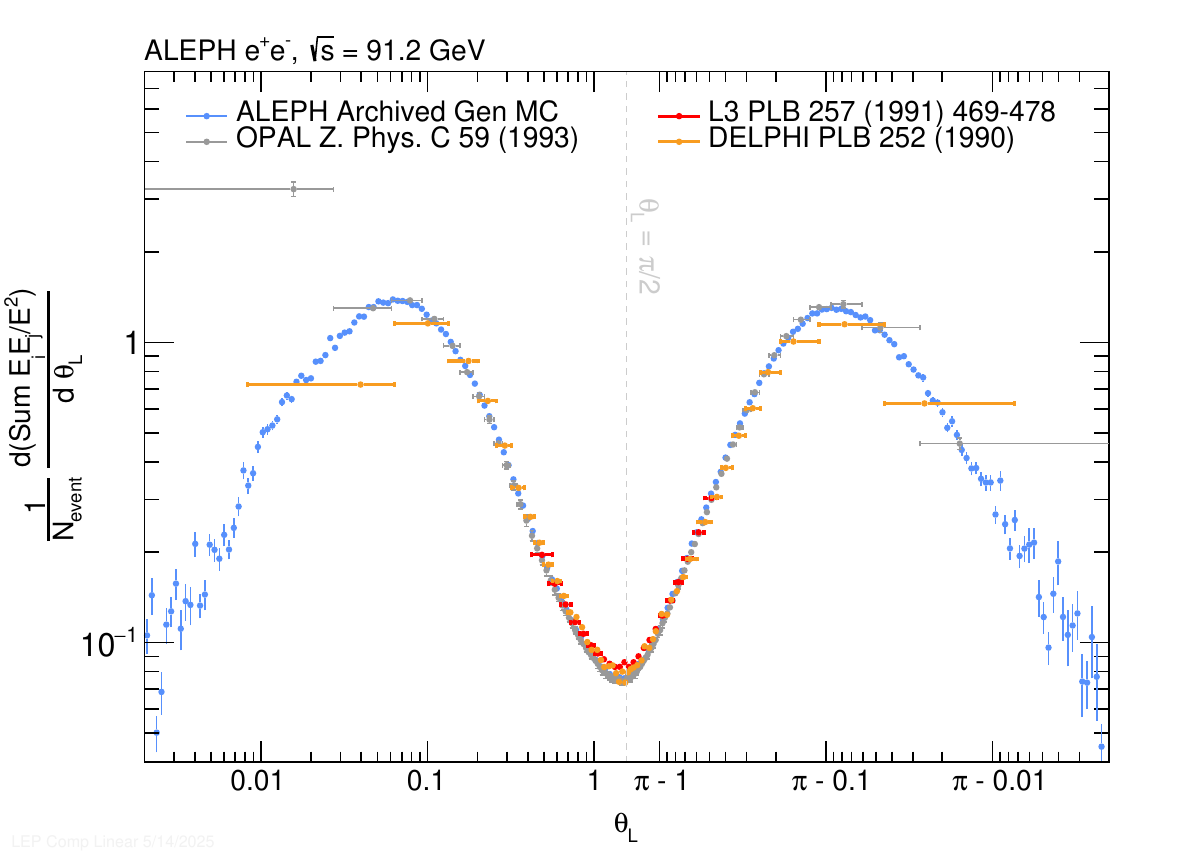}
    \caption{Published OPAL data from Ref.~\cite{OPAL:1993pnw} compared to archived $\textsc{pythia}$ 6 MC with the binning utilized in this analysis. Note that these are plotted using the double log style in order to place emphasis on the angular limits of the distribution. }
    \label{fig:opalLog}
\end{figure}

An additional comparison between the OPAL data from Ref.~\cite{OPAL:1993pnw} and archived $\textsc{pythia}$ 6 MC can be found in Figure \ref{fig:opalLog}.  Here, the results are plotted using the characteristic \textit{double log} style introduced in this work, with the binning of the archived $\textsc{pythia}$ 6 MC binning equivalent to the binning utilized in the results presented in Section \ref{sec:results}. Here, one can see that the results presented in this note provide unprecedented precision, particularly in the small and large angle limits of the distribution. This precision is particularly useful for the investigations of hardonization in the collinear and the back-to-back limit. 

\section{Summary and Future Work}
\label{sec:summary}
As originally envisioned, the N-point energy correlation functions are a useful way in which to study QCD by successfully isolating the scaling behavior across various regions of phase space. Making such measurements in the $e^{+}e^{-}$ collision environment offers a number of advantages due to the fact that the colliding objects are fundamental particles. For example, a precise knowledge of the momentum transferred in the collision allows for a 2D unfolding procedure to be employed vs. a 3D unfolding procedure, reducing the complexity of the unfolding problem. In addition, this is a relatively simple environment for theoretical comparisons where complications such as beam remnants, gluonic initial-state radiation, and parton distribution functions are largely avoided. New experimental and theoretical developments make now the ideal time to re-analyze the ENC in $e^{+}e^{-}$ data while taking advantage of these new tools.  

In this analysis note the details related to the first fully-corrected measurement of the E2C spanning from the collinear to the back-to-back limit of QCD using ALEPH archived data are presented. Here, due to the clean environment of $e^{+}e^{-}$ collisions all charged particles in the event are utilized for forming the E2C as opposed to only those within reconstructed jets, allowing for a large kinematic reach spanning collinear and back-to-back regions. Utilizing only charged particles allows for improved precision in the regions of interest as compared to existing EEC measurements with LEP data. 

When comparing the results of this analysis to ALEPH archived $\textsc{pythia}$ 6 MC and theoretical calculations, these studies show excellent agreement and provide crucial tests for QCD calculations and phenomenological models. This is especially true in the relatively-unexplored Sudakov limit where this measurement provides one of the most precise experimental constraints. 

One natural extension of this work would be to also perform a measurement of the fully-corrected energy correlation asymmetry as measured in Ref.~\cite{OPAL:1990reb}. Measuring such a distribution as a function of both $z$ and $\theta_{\rm L}$ would additionally be useful in order to provide a qualitative measure of how universal the scaling behavior is in the free hadron region. Another natural extension would be to measure higher orders of the ENC distributions. For these efforts, it may also be useful to take advantage of recent work performed in order to reduce the computation time of the higher point correlators~\cite{Budhraja:2024xiq,Alipour-fard:2024szj}. By using a ratio of higher point correlators to the E2C, a value for $\alpha_{\rm s}$ can be extracted from the data. Such a measurement would be timely as the $\alpha_{\rm s}$ fits from $e^{+}e^{-}$ event shapes and analytical hadronization were recently excluded from the calculation of the world average~\cite{ParticleDataGroup:2020ssz,ParticleDataGroup:2022pth}. 

This measurement marks the beginning of a new investigative direction in $e^{+}e^{-}$ collisions, revisiting concepts from the 1970s to address contemporary physics questions. This work also has the potential to shape the future, serving as a catalyst to inspire and inform studies at the proposed $e^{+}e^{-}$ colliders~\cite{Benedikt:2651299}.

\section*{Acknowledgments}
The authors would like to thank Roberto Tenchini and Guenther Dissertori from the ALEPH Collaboration for their useful comments and suggestions on the use of ALEPH data. The authors would also like to thanks Max Jaarsma, Yibei Li, Ian Moult, Wouter Waalewijn, and HuaXing Zhu for providing the theoretical prediction for this result. In particular, the authors would also like to thank Ian Moult for useful discussions related to this analysis. 

\clearpage

\bibliographystyle{JHEP}
\typeout{}
\bibliography{EEC}

\providecommand{\href}[2]{#2}\begingroup\raggedright\begin{thebibliography}{10}

\bibitem{Badea:2019vey}
A.~Badea, A.~Baty, P.~Chang, G.~M. Innocenti, M.~Maggi, C.~Mcginn et~al., \emph{{Measurements of two-particle correlations in $e^+e^-$ collisions at 91 GeV with ALEPH archived data}}, \href{https://doi.org/10.1103/PhysRevLett.123.212002}{\emph{Phys. Rev. Lett.} {\bfseries 123} (2019) 212002}, [\href{https://arxiv.org/abs/1906.00489}{{\ttfamily 1906.00489}}].

\bibitem{Chen:2023njr}
Y.-C. Chen et~al., \emph{{Long-range near-side correlation in $e^+e^-$ collisions at 183-209 GeV with ALEPH archived data}}, \href{https://doi.org/10.1016/j.physletb.2024.138957}{\emph{Phys. Lett. B} {\bfseries 856} (2024) 138957}, [\href{https://arxiv.org/abs/2312.05084}{{\ttfamily 2312.05084}}].

\bibitem{Cacciari:2008gp}
M.~Cacciari, G.~P. Salam and G.~Soyez, \emph{{The anti-$k_t$ jet clustering algorithm}}, \href{https://doi.org/10.1088/1126-6708/2008/04/063}{\emph{JHEP} {\bfseries 04} (2008) 063}, [\href{https://arxiv.org/abs/0802.1189}{{\ttfamily 0802.1189}}].

\bibitem{Chen:2021uws}
Y.~Chen et~al., \emph{{Jet energy spectrum and substructure in $e^+e^-$ collisions at 91.2 GeV with ALEPH Archived Data}}, \href{https://doi.org/10.1007/JHEP06(2022)008}{\emph{JHEP} {\bfseries 06} (2022) 008}, [\href{https://arxiv.org/abs/2111.09914}{{\ttfamily 2111.09914}}].

\bibitem{PhysRevLett.41.1585}
C.~L. Basham, L.~S. Brown, S.~D. Ellis and S.~T. Love, \emph{Energy correlations in electron-positron annihilation: Testing quantum chromodynamics}, \href{https://doi.org/10.1103/PhysRevLett.41.1585}{\emph{Phys. Rev. Lett.} {\bfseries 41} (Dec, 1978) 1585--1588}.

\bibitem{Basham:1978zq}
C.~Basham, L.~Brown, S.~Ellis and S.~Love, \emph{{Energy Correlations in electron-Positron Annihilation in Quantum Chromodynamics: Asymptotically Free Perturbation Theory}}, \href{https://doi.org/10.1103/PhysRevD.19.2018}{\emph{Phys. Rev. D} {\bfseries 19} (1979) 2018}.

\bibitem{Basham:1979gh}
C.~L. Basham, L.~S. Brown, S.~D. Ellis and S.~T. Love, \emph{{Energy Correlations in Perturbative Quantum Chromodynamics: A Conjecture for All Orders}}, \href{https://doi.org/10.1016/0370-2693(79)90601-4}{\emph{Phys. Lett. B} {\bfseries 85} (1979) 297--299}.

\bibitem{Basham:1977iq}
C.~L. Basham, L.~S. Brown, S.~D. Ellis and S.~T. Love, \emph{{Electron - Positron Annihilation Energy Pattern in Quantum Chromodynamics: Asymptotically Free Perturbation Theory}}, \href{https://doi.org/10.1103/PhysRevD.17.2298}{\emph{Phys. Rev. D} {\bfseries 17} (1978) 2298}.

\bibitem{OPAL:1990reb}
{\scshape OPAL} collaboration, M.~Z. Akrawy et~al., \emph{{A Measurement of energy correlations and a determination of alpha-s (M2 (Z0)) in e+ e- annihilations at s**(1/2) = 91-GeV}}, \href{https://doi.org/10.1016/0370-2693(90)91098-V}{\emph{Phys. Lett. B} {\bfseries 252} (1990) 159--169}.

\bibitem{OPAL:1993pnw}
{\scshape OPAL} collaboration, P.~D. Acton et~al., \emph{{A Determination of alpha-s (M (Z0)) at LEP using resummed QCD calculations}}, \href{https://doi.org/10.1007/BF01555834}{\emph{Z. Phys. C} {\bfseries 59} (1993) 1--20}.

\bibitem{DELPHI:1990sof}
{\scshape DELPHI} collaboration, P.~Abreu et~al., \emph{{Energy-energy correlations in hadronic final states from Z0 decays}}, \href{https://doi.org/10.1016/0370-2693(90)91097-U}{\emph{Phys. Lett. B} {\bfseries 252} (1990) 149--158}.

\bibitem{L3:1991qlf}
{\scshape L3} collaboration, B.~Adeva et~al., \emph{{Determination of alpha-s from energy-energy correlations measured on the Z0 resonance.}}, \href{https://doi.org/10.1016/0370-2693(91)91925-L}{\emph{Phys. Lett. B} {\bfseries 257} (1991) 469--478}.

\bibitem{L3:1992btq}
{\scshape L3} collaboration, O.~Adrian et~al., \emph{{Determination of alpha-s from hadronic event shapes measured on the Z0 resonance}}, \href{https://doi.org/10.1016/0370-2693(92)90463-E}{\emph{Phys. Lett. B} {\bfseries 284} (1992) 471--481}.

\bibitem{SLD:1994idb}
{\scshape SLD} collaboration, K.~Abe et~al., \emph{{Measurement of alpha-s (M(Z)**2) from hadronic event observables at the Z0 resonance}}, \href{https://doi.org/10.1103/PhysRevD.51.962}{\emph{Phys. Rev. D} {\bfseries 51} (1995) 962--984}, [\href{https://arxiv.org/abs/hep-ex/9501003}{{\ttfamily hep-ex/9501003}}].

\bibitem{CMS:2024mlf}
{\scshape CMS} collaboration, A.~Hayrapetyan et~al., \emph{{Measurement of energy correlators inside jets and determination of the strong coupling $\alpha_\mathrm{S}(m_\mathrm{Z})$}},  \href{https://arxiv.org/abs/2402.13864}{{\ttfamily 2402.13864}}.

\bibitem{ALICE:2024dfl}
{\scshape ALICE} collaboration, S.~Acharya et~al., \emph{{Exposing the parton-hadron transition within jets with energy-energy correlators in pp collisions at $\sqrt{\textit s}=5.02$ TeV}},  \href{https://arxiv.org/abs/2409.12687}{{\ttfamily 2409.12687}}.

\bibitem{STAR:2025jut}
{\scshape STAR} collaboration, S.~Collaboration, \emph{{Measurement of Two-Point Energy Correlators Within Jets in $p$+$p$ Collisions at $\sqrt{s}$ = 200 GeV}},  \href{https://arxiv.org/abs/2502.15925}{{\ttfamily 2502.15925}}.

\bibitem{Komiske:2022enw}
P.~T. Komiske, I.~Moult, J.~Thaler and H.~X. Zhu, \emph{{Analyzing N-Point Energy Correlators inside Jets with CMS Open Data}}, \href{https://doi.org/10.1103/PhysRevLett.130.051901}{\emph{Phys. Rev. Lett.} {\bfseries 130} (2023) 051901}, [\href{https://arxiv.org/abs/2201.07800}{{\ttfamily 2201.07800}}].

\bibitem{Tkachov:1999py}
F.~V. Tkachov, \emph{{A Theory of jet definition}}, \href{https://doi.org/10.1142/S0217751X02009977}{\emph{Int. J. Mod. Phys. A} {\bfseries 17} (2002) 2783--2884}, [\href{https://arxiv.org/abs/hep-ph/9901444}{{\ttfamily hep-ph/9901444}}].

\bibitem{Kardos:2018kqj}
A.~Kardos, S.~Kluth, G.~Somogyi, Z.~Tulip\'ant and A.~Verbytskyi, \emph{{Precise determination of $\alpha _{S}(M_Z)$ from a global fit of energy\textendash{}energy correlation to NNLO+NNLL predictions}}, \href{https://doi.org/10.1140/epjc/s10052-018-5963-1}{\emph{Eur. Phys. J. C} {\bfseries 78} (2018) 498}, [\href{https://arxiv.org/abs/1804.09146}{{\ttfamily 1804.09146}}].

\bibitem{Andres:2024xvk}
C.~Andres, F.~Dominguez, J.~Holguin, C.~Marquet and I.~Moult, \emph{{Simple Scaling Laws for Energy Correlators in Nuclear Matter}},  \href{https://arxiv.org/abs/2411.15298}{{\ttfamily 2411.15298}}.

\bibitem{Devereaux:2023vjz}
K.~Devereaux, W.~Fan, W.~Ke, K.~Lee and I.~Moult, \emph{{Imaging Cold Nuclear Matter with Energy Correlators}},  \href{https://arxiv.org/abs/2303.08143}{{\ttfamily 2303.08143}}.

\bibitem{CMS:2025ydi}
{\scshape CMS} collaboration, V.~Chekhovsky et~al., \emph{{Observation of nuclear modification of energy-energy correlators inside jets in heavy ion collisions}},  \href{https://arxiv.org/abs/2503.19993}{{\ttfamily 2503.19993}}.

\bibitem{Andres:2023xwr}
C.~Andres, F.~Dominguez, J.~Holguin, C.~Marquet and I.~Moult, \emph{{A coherent view of the quark-gluon plasma from energy correlators}}, \href{https://doi.org/10.1007/JHEP09(2023)088}{\emph{JHEP} {\bfseries 09} (2023) 088}, [\href{https://arxiv.org/abs/2303.03413}{{\ttfamily 2303.03413}}].

\bibitem{Andres:2024ksi}
C.~Andres, F.~Dominguez, J.~Holguin, C.~Marquet and I.~Moult, \emph{{Towards an Interpretation of the First Measurements of Energy Correlators in the Quark-Gluon Plasma}},  \href{https://arxiv.org/abs/2407.07936}{{\ttfamily 2407.07936}}.

\bibitem{Barata:2023bhh}
J.~a. Barata, P.~Caucal, A.~Soto-Ontoso and R.~Szafron, \emph{{Advancing the understanding of energy-energy correlators in heavy-ion collisions}}, \href{https://doi.org/10.1007/JHEP11(2024)060}{\emph{JHEP} {\bfseries 11} (2024) 060}, [\href{https://arxiv.org/abs/2312.12527}{{\ttfamily 2312.12527}}].

\bibitem{Yang:2023dwc}
Z.~Yang, Y.~He, I.~Moult and X.-N. Wang, \emph{{Probing the Short-Distance Structure of the Quark-Gluon Plasma with Energy Correlators}}, \href{https://doi.org/10.1103/PhysRevLett.132.011901}{\emph{Phys. Rev. Lett.} {\bfseries 132} (2024) 011901}, [\href{https://arxiv.org/abs/2310.01500}{{\ttfamily 2310.01500}}].

\bibitem{Bossi:2024qho}
H.~Bossi, A.~S. Kudinoor, I.~Moult, D.~Pablos, A.~Rai and K.~Rajagopal, \emph{{Imaging the wakes of jets with energy-energy-energy correlators}}, \href{https://doi.org/10.1007/JHEP12(2024)073}{\emph{JHEP} {\bfseries 12} (2024) 073}, [\href{https://arxiv.org/abs/2407.13818}{{\ttfamily 2407.13818}}].

\bibitem{Andres:2024hdd}
C.~Andres, J.~Holguin, R.~Kunnawalkam~Elayavalli and J.~Viinikainen, \emph{{Minimizing Selection Bias in Inclusive Jets in Heavy-Ion Collisions with Energy Correlators}},  \href{https://arxiv.org/abs/2409.07514}{{\ttfamily 2409.07514}}.

\bibitem{Chang:2013rca}
H.-M. Chang, M.~Procura, J.~Thaler and W.~J. Waalewijn, \emph{{Calculating Track-Based Observables for the LHC}}, \href{https://doi.org/10.1103/PhysRevLett.111.102002}{\emph{Phys. Rev. Lett.} {\bfseries 111} (2013) 102002}, [\href{https://arxiv.org/abs/1303.6637}{{\ttfamily 1303.6637}}].

\bibitem{Chang:2013iba}
H.-M. Chang, M.~Procura, J.~Thaler and W.~J. Waalewijn, \emph{{Calculating Track Thrust with Track Functions}}, \href{https://doi.org/10.1103/PhysRevD.88.034030}{\emph{Phys. Rev. D} {\bfseries 88} (2013) 034030}, [\href{https://arxiv.org/abs/1306.6630}{{\ttfamily 1306.6630}}].

\bibitem{Decamp:1990jra}
{\scshape ALEPH} collaboration, D.~Decamp et~al., \emph{{ALEPH: A detector for electron-positron annnihilations at LEP}}, \href{https://doi.org/10.1016/0168-9002(90)91831-U}{\emph{Nucl. Instrum. Meth.} {\bfseries A294} (1990) 121--178}.

\bibitem{ALEPH:1996qht}
{\scshape ALEPH} collaboration, D.~Buskulic et~al., \emph{{Production of orbitally excited charm mesons in semileptonic B decays}}, \href{https://doi.org/10.1007/s002880050351}{\emph{Z. Phys. C} {\bfseries 73} (1997) 601--612}.

\bibitem{ALEPH:1994ayc}
{\scshape ALEPH} collaboration, D.~Buskulic et~al., \emph{{Performance of the ALEPH detector at LEP}}, \href{https://doi.org/10.1016/0168-9002(95)00138-7}{\emph{Nucl. Instrum. Meth. A} {\bfseries 360} (1995) 481--506}.

\bibitem{Tripathee:2017ybi}
A.~Tripathee, W.~Xue, A.~Larkoski, S.~Marzani and J.~Thaler, \emph{{Jet Substructure Studies with CMS Open Data}}, \href{https://doi.org/10.1103/PhysRevD.96.074003}{\emph{Phys. Rev.} {\bfseries D96} (2017) 074003}, [\href{https://arxiv.org/abs/1704.05842}{{\ttfamily 1704.05842}}].

\bibitem{Sjostrand:2000wi}
T.~Sjostrand, P.~Eden, C.~Friberg, L.~Lonnblad, G.~Miu, S.~Mrenna et~al., \emph{{High-energy physics event generation with PYTHIA 6.1}}, \href{https://doi.org/10.1016/S0010-4655(00)00236-8}{\emph{Comput. Phys. Commun.} {\bfseries 135} (2001) 238--259}, [\href{https://arxiv.org/abs/hep-ph/0010017}{{\ttfamily hep-ph/0010017}}].

\bibitem{ALEPH:1996oqp}
{\scshape ALEPH} collaboration, R.~Barate et~al., \emph{{Studies of quantum chromodynamics with the ALEPH detector}}, \href{https://doi.org/10.1016/S0370-1573(97)00045-8}{\emph{Phys. Rept.} {\bfseries 294} (1998) 1--165}.

\bibitem{Gao:2019ojf}
A.~Gao, H.~T. Li, I.~Moult and H.~X. Zhu, \emph{{Precision QCD Event Shapes at Hadron Colliders: The Transverse Energy-Energy Correlator in the Back-to-Back Limit}}, \href{https://doi.org/10.1103/PhysRevLett.123.062001}{\emph{Phys. Rev. Lett.} {\bfseries 123} (2019) 062001}, [\href{https://arxiv.org/abs/1901.04497}{{\ttfamily 1901.04497}}].

\bibitem{Gao:2023ivm}
A.~Gao, H.~T. Li, I.~Moult and H.~X. Zhu, \emph{{The transverse energy-energy correlator at next-to-next-to-next-to-leading logarithm}}, \href{https://doi.org/10.1007/JHEP09(2024)072}{\emph{JHEP} {\bfseries 09} (2024) 072}, [\href{https://arxiv.org/abs/2312.16408}{{\ttfamily 2312.16408}}].

\bibitem{ParticleDataGroup:2020ssz}
{\scshape Particle Data Group} collaboration, P.~A. Zyla et~al., \emph{{Review of Particle Physics}}, \href{https://doi.org/10.1093/ptep/ptaa104}{\emph{PTEP} {\bfseries 2020} (2020) 083C01}.

\bibitem{ParticleDataGroup:2022pth}
{\scshape Particle Data Group} collaboration, R.~L. Workman et~al., \emph{{Review of Particle Physics}}, \href{https://doi.org/10.1093/ptep/ptac097}{\emph{PTEP} {\bfseries 2022} (2022) 083C01}.

\bibitem{hungarianMatching}
H.~W. Kuhn, \emph{The hungarian method for the assignment problem}, \href{https://doi.org/https://doi.org/10.1002/nav.3800020109}{\emph{Naval Research Logistics Quarterly} {\bfseries 2} (1955) 83--97}, [\href{https://arxiv.org/abs/https://onlinelibrary.wiley.com/doi/pdf/10.1002/nav.3800020109}{{\ttfamily https://onlinelibrary.wiley.com/doi/pdf/10.1002/nav.3800020109}}].

\bibitem{Brenner:2019lmf}
L.~Brenner, R.~Balasubramanian, C.~Burgard, W.~Verkerke, G.~Cowan, P.~Verschuuren et~al., \emph{{Comparison of unfolding methods using RooFitUnfold}}, \href{https://doi.org/10.1142/S0217751X20501456}{\emph{Int. J. Mod. Phys. A} {\bfseries 35} (2020) 2050145}, [\href{https://arxiv.org/abs/1910.14654}{{\ttfamily 1910.14654}}].

\bibitem{Chen:2023nsi}
Y.-C. Chen, Y.-J. Lee, Y.~Chen, P.~Chang, C.~McGinn, T.-A. Sheng et~al., \emph{{Analysis note: two-particle correlation in $e^+e^-$ collisions at 91-209 GeV with archived ALEPH data}},  \href{https://arxiv.org/abs/2309.09874}{{\ttfamily 2309.09874}}.

\bibitem{Buskulic:272484}
{\scshape ALEPH} collaboration, D.~Buskulic et~al., \emph{{Performance of the ALEPH detector at LEP}}, \href{https://doi.org/10.1016/0168-9002(95)00138-7}{\emph{Nucl. Instrum. Methods Phys. Res., A} {\bfseries 360} (1995) 481--506}.

\bibitem{D'Agostini:265717}
G.~D'Agostini, \emph{{A multidimensional unfolding method based on Bayes' Theorem}},  tech. rep., DESY, Hamburg, 1994.

\bibitem{Bierlich:2022pfr}
C.~Bierlich et~al., \emph{{A comprehensive guide to the physics and usage of PYTHIA 8.3}}, \href{https://doi.org/10.21468/SciPostPhysCodeb.8}{\emph{SciPost Phys. Codeb.} {\bfseries 2022} (2022) 8}, [\href{https://arxiv.org/abs/2203.11601}{{\ttfamily 2203.11601}}].

\bibitem{Hoche:2015sya}
S.~H\"oche and S.~Prestel, \emph{{The midpoint between dipole and parton showers}}, \href{https://doi.org/10.1140/epjc/s10052-015-3684-2}{\emph{Eur. Phys. J. C} {\bfseries 75} (2015) 461}, [\href{https://arxiv.org/abs/1506.05057}{{\ttfamily 1506.05057}}].

\bibitem{Giele:2007di}
W.~T. Giele, D.~A. Kosower and P.~Z. Skands, \emph{{A simple shower and matching algorithm}}, \href{https://doi.org/10.1103/PhysRevD.78.014026}{\emph{Phys. Rev. D} {\bfseries 78} (2008) 014026}, [\href{https://arxiv.org/abs/0707.3652}{{\ttfamily 0707.3652}}].

\bibitem{ALEPH:1990vew}
{\scshape ALEPH} collaboration, D.~Decamp et~al., \emph{{Measurement of alpha-s from the structure of particle clusters produced in hadronic Z decays}}, \href{https://doi.org/10.1016/0370-2693(91)91926-M}{\emph{Phys. Lett. B} {\bfseries 257} (1991) 479--491}.

\bibitem{Budhraja:2024xiq}
A.~Budhraja and W.~J. Waalewijn, \emph{{FastEEC: Fast evaluation of N-point energy correlators}}, \href{https://doi.org/10.1016/j.physletb.2025.139276}{\emph{Phys. Lett. B} {\bfseries 861} (2025) 139276}, [\href{https://arxiv.org/abs/2406.08577}{{\ttfamily 2406.08577}}].

\bibitem{Alipour-fard:2024szj}
S.~Alipour-fard, A.~Budhraja, J.~Thaler and W.~J. Waalewijn, \emph{{New Angles on Energy Correlators}},  \href{https://arxiv.org/abs/2410.16368}{{\ttfamily 2410.16368}}.

\bibitem{Benedikt:2651299}
M.~Benedikt, A.~Blondel, O.~Brunner, M.~Capeans~Garrido, F.~Cerutti, J.~Gutleber et~al., \emph{{FCC-ee: The Lepton Collider: Future Circular Collider Conceptual Design Report Volume 2. Future Circular Collider}},  Tech. Rep.~2, CERN, Geneva, 2019.
\newblock 10.1140/epjst/e2019-900045-4.

\end{thebibliography}\endgroup

\clearpage
\appendix

\section{Analysis Code}\label{app:analysiscode}
The code used in this analysis is available publicly at \newline \url{https://github.com/FHead/PhysicsEEJetEEC/tree/main}. For any information regarding the code or its use for the analysis of ALEPH data, please contact the authors. 

\section{Dependence of Track and Event Selection Efficiency on Sphericity}\label{app:evtseleff}
\begin{figure}[ht!]
    \centering
    \includegraphics[width=0.32\linewidth]{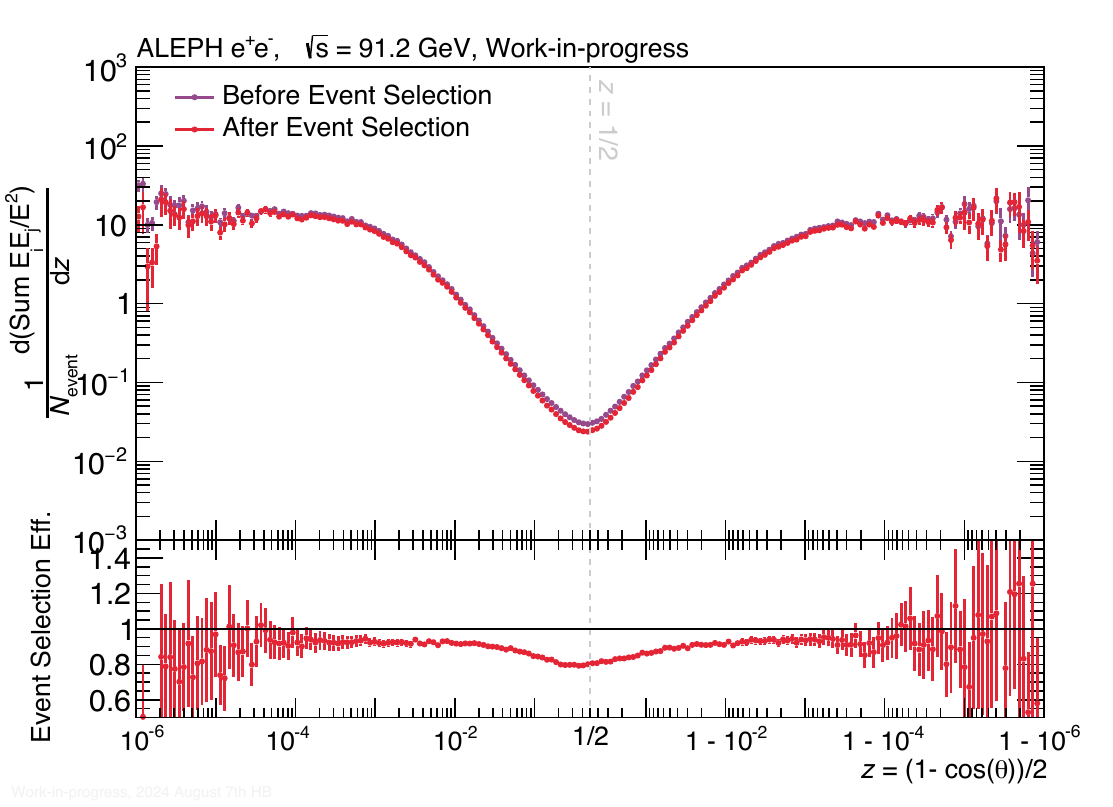}
    \includegraphics[width=0.32\linewidth]{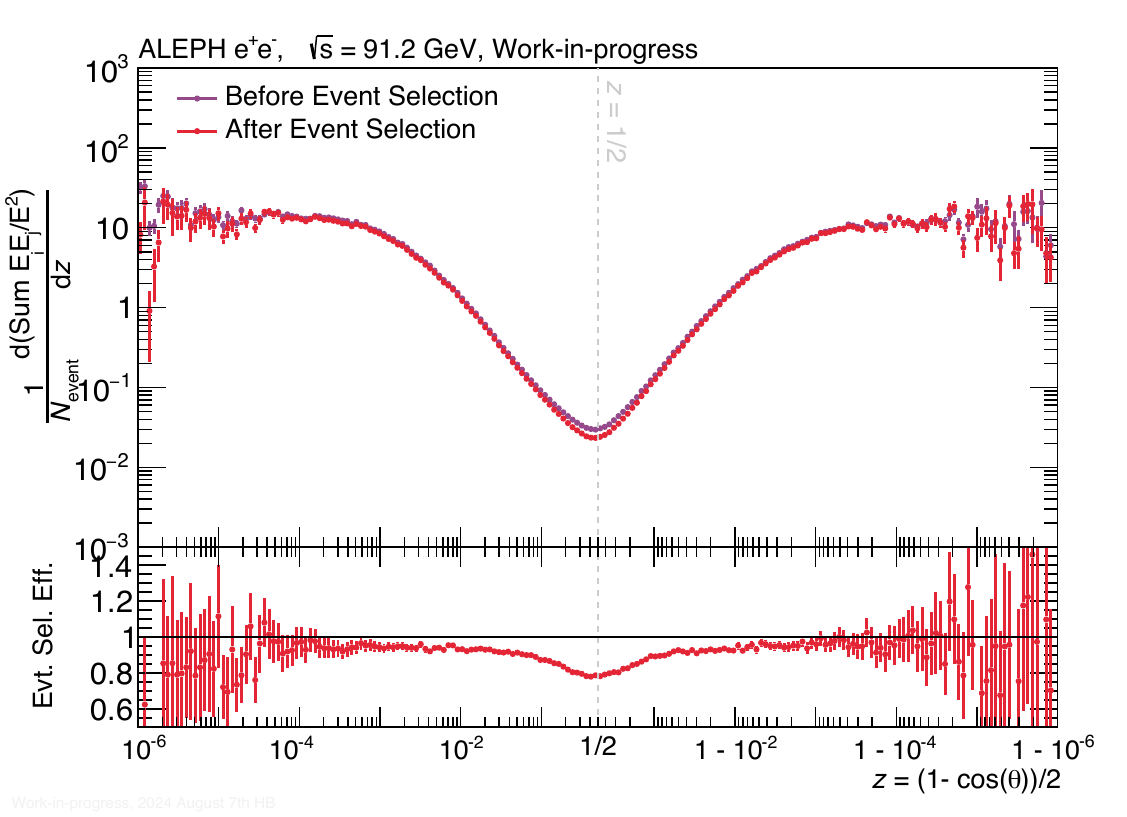}
    \includegraphics[width = 0.32\linewidth]{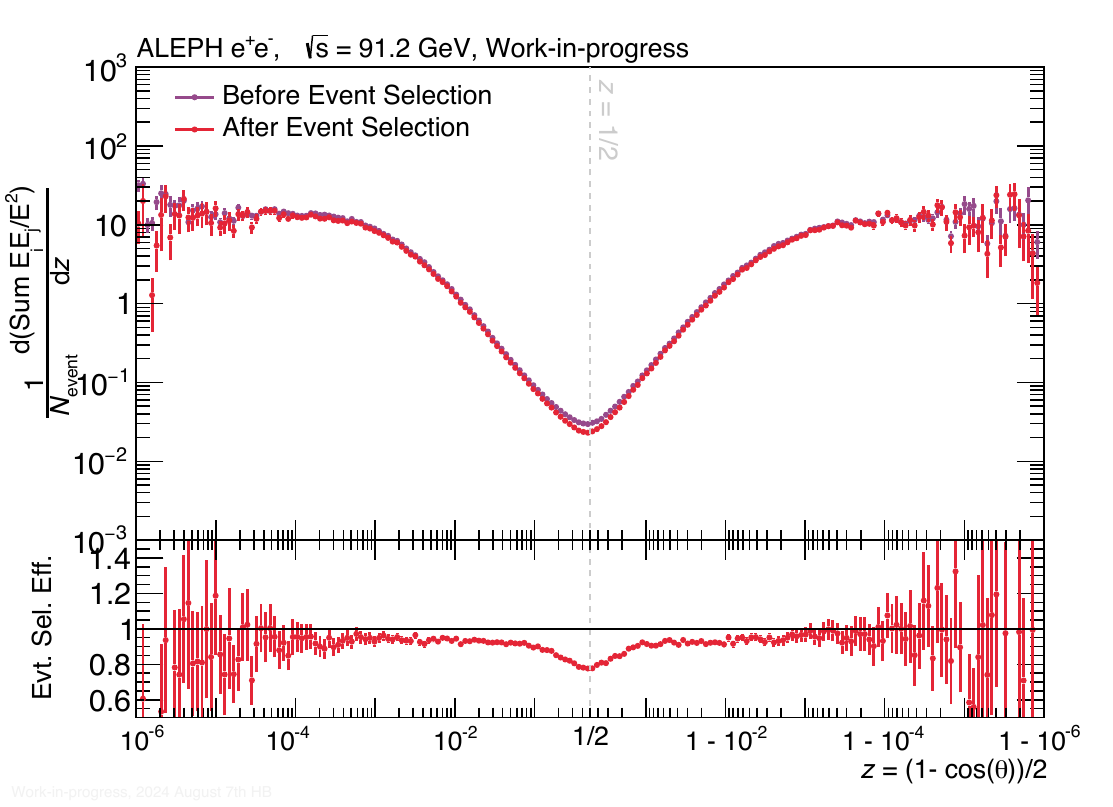}
    \includegraphics[width = 0.32\linewidth]{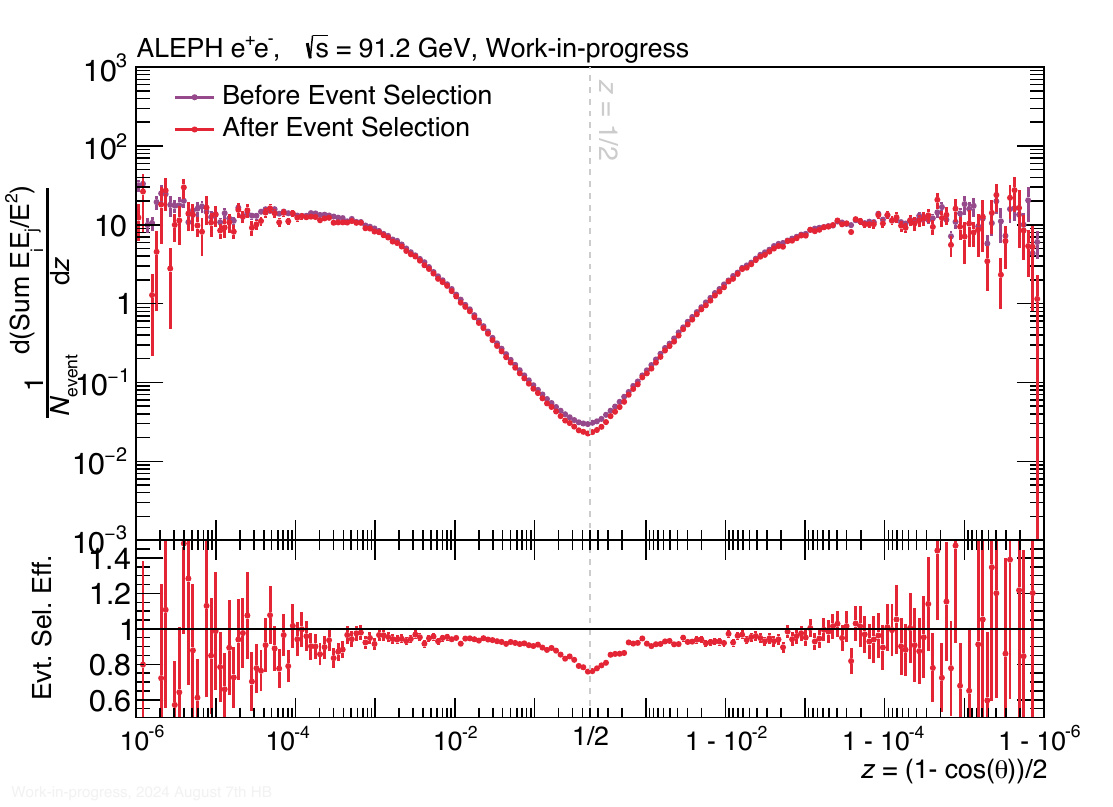}
    \includegraphics[width = 0.32\linewidth]{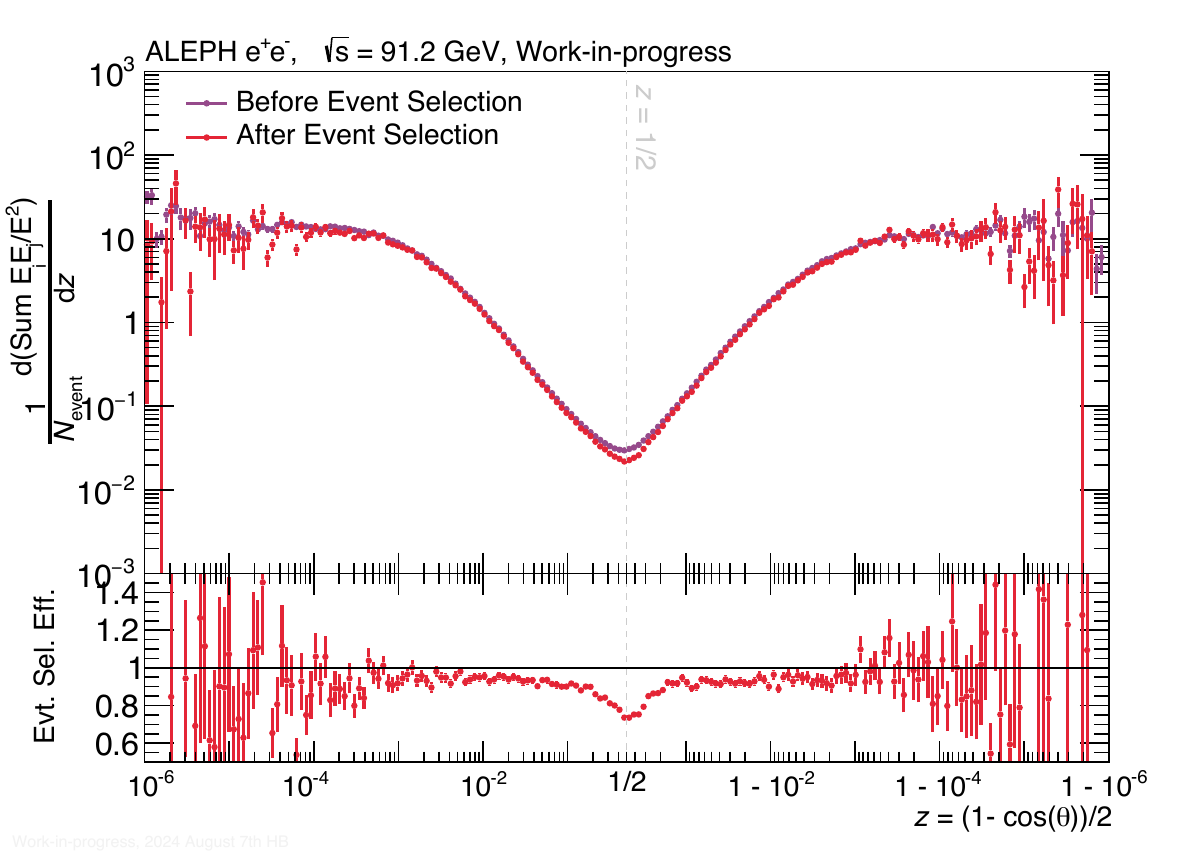}
    \includegraphics[width = 0.32\linewidth]{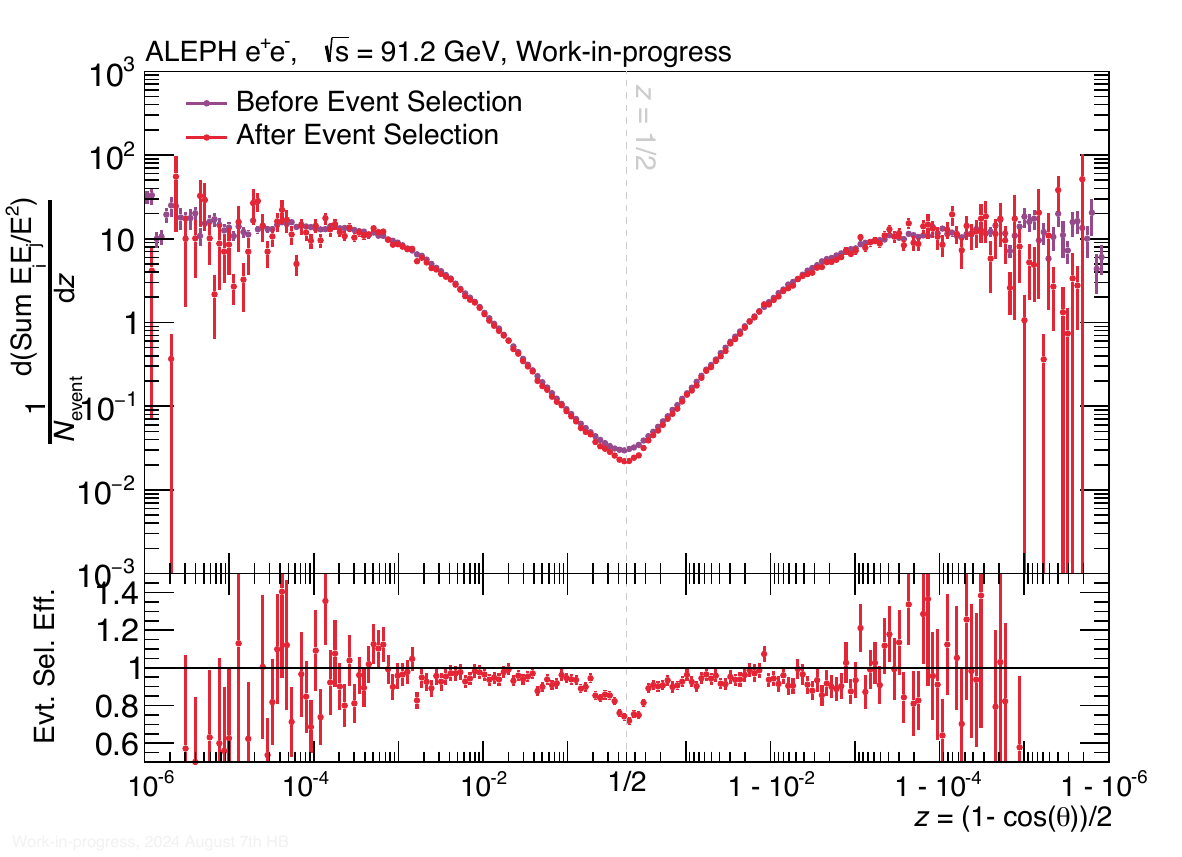}
    \caption{EEC distributions before and after event and track selection for different sphericity cuts. The sphericity cuts for the various figures are as follows: (Top Left): $7\pi/36 < \theta_{\rm sphericity} < 29\pi/36$, (Top Middle): $9\pi/36 < \theta_{\rm sphericity} < 27\pi/36$, (Top Right): $11\pi/36 < \theta_{\rm sphericity} < 25\pi/36$, , (Bottom Left): $13\pi/36 < \theta_{\rm sphericity} < 23\pi/36$, (Bottom Middle): $15\pi/36 < \theta_{\rm sphericity} < 21\pi/36$, (Bottom Right): $17\pi/36 < \theta_{\rm sphericity} < 19\pi/36$. The ratios of the EEC distributions following the event selection to the distributions before any event selection (defined as the event selection efficiency) are shown in the bottom panels. }
    \label{fig:sphericitySweep}
\end{figure}

As mentioned in Section  \ref{sec:evtselcorr}, the dip seen in the track and event selection efficiencies in Figure \ref{fig:evtSelCorr} is related to the sphericity cut chosen for this analysis. This is illustrated in Figure \ref{fig:sphericitySweep}, where as the sphericity cut is narrowed, the dip in the event selection efficiency around $z = \frac{1}{2}$ becomes larger. Conceptually, this can be understood by the fact that the sphericity cut would remove mostly lower energy particles that occur at wide angles from the sphericity axis. Note that for the studies shown in this appendix, a smaller MC sample was used in order to save processing time. The conclusions of the study are not dependent on the statistics used for the study, and therefore are also applicable to the main analysis. 

\begin{figure}[ht!]
    \centering
    \includegraphics[width=0.49\linewidth]{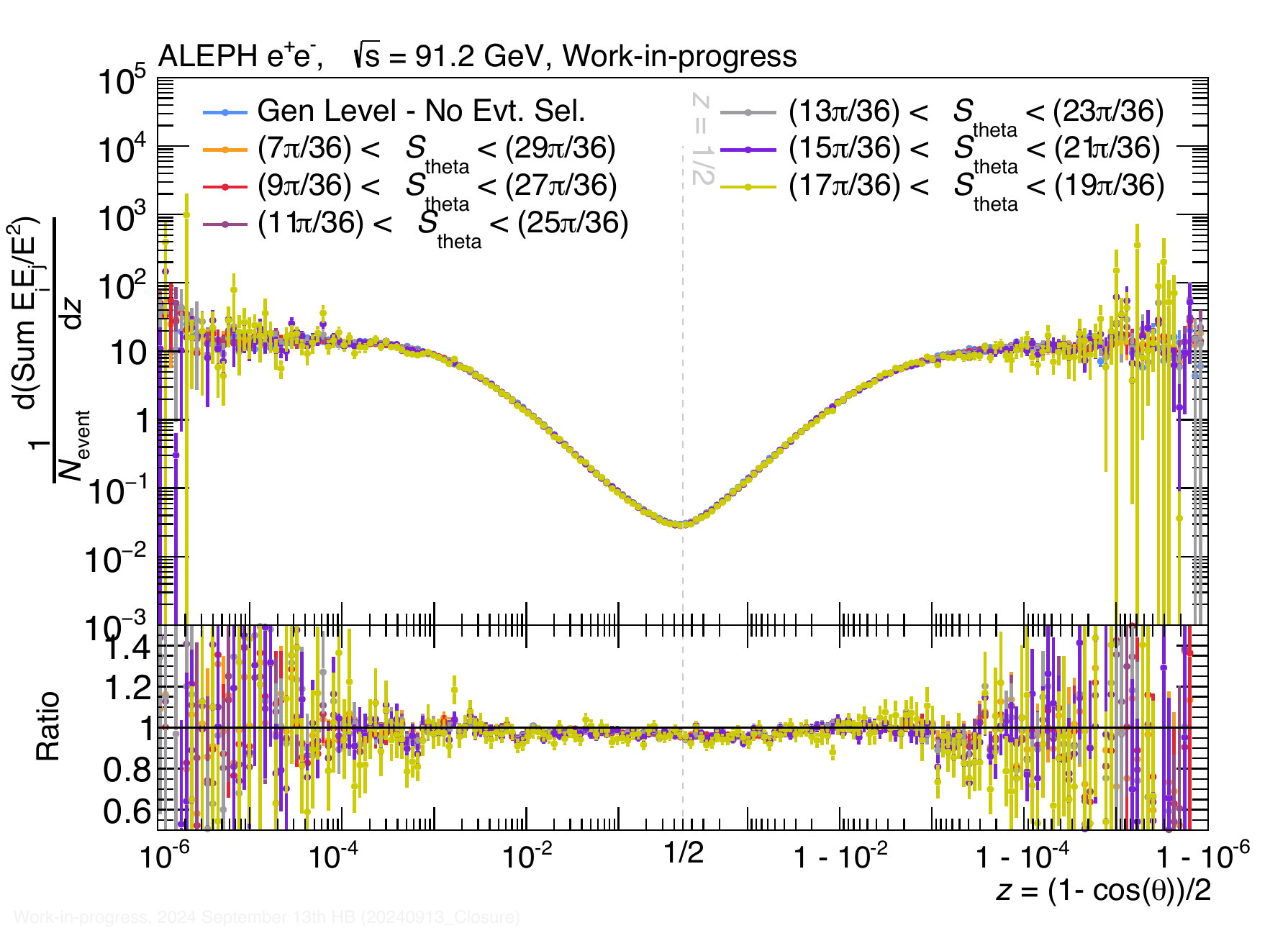}
    \includegraphics[width=0.49\textwidth]{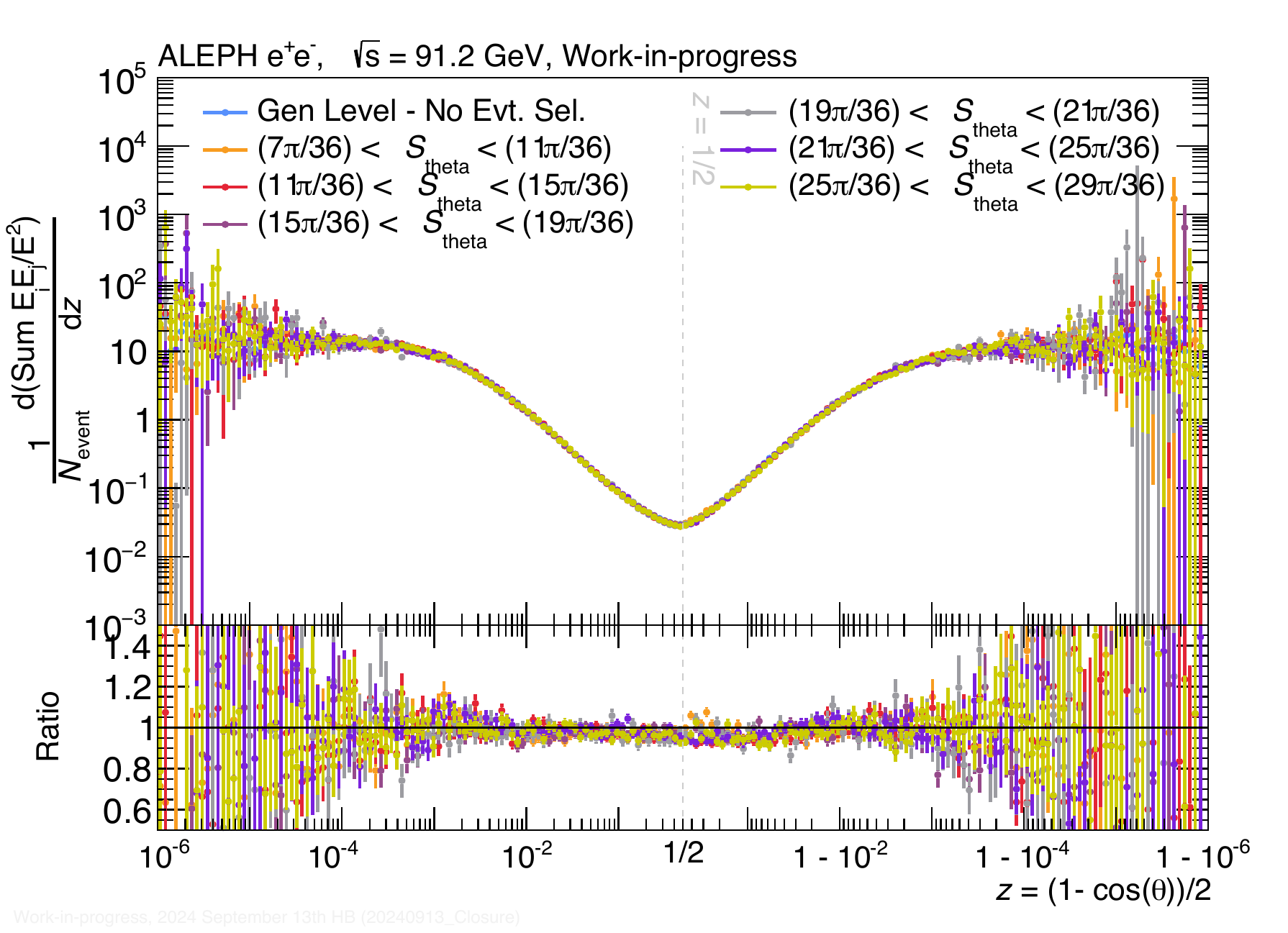}
    \caption{EEC distributions following the event selection efficiency correction using two different sweeps of sphericity cuts.}
    \label{fig:evtSelClosure}
\end{figure}

In principle, this dependence in the event selection efficiency on the sphericity cut is not an issue, provided that the same result is achieved following the correction for this efficiency regardless of the sphericity cut. In other words, the event selection efficiency exhibits good closure. Figure \ref{fig:evtSelClosure} visualizes this closure, where two different sweeping methods for different values of $\theta_{\rm sphericity}$ ($S_{\rm \theta}$). The first, shown on the left-hand side of Figure \ref{fig:evtSelClosure}, is a symmetric sweep, where the middle of the $S_{\rm \theta}$ distribution remains constant but the size of this distribution varies. The second, shown in the right panel of Figure \ref{fig:evtSelClosure} is a moving sweep where the central value of the $S_{\rm \theta}$ values used in the analysis moves for each distribution. Regardless of the sweeping method chosen, one can see in Figure \ref{fig:evtSelClosure} that the resulting distribution following the event selection efficiency distribution remains constant. Meaning that the EEC distribution corrected for the event selection efficiency does not have a strong dependence on the chosen $S_{\rm \theta}$ values for the analysis.

\section{Closure Checks of the Correction Procedure}\label{app:CorrectionClosure}
\begin{figure}[ht!]
    \centering
    \includegraphics[width=0.49\linewidth]{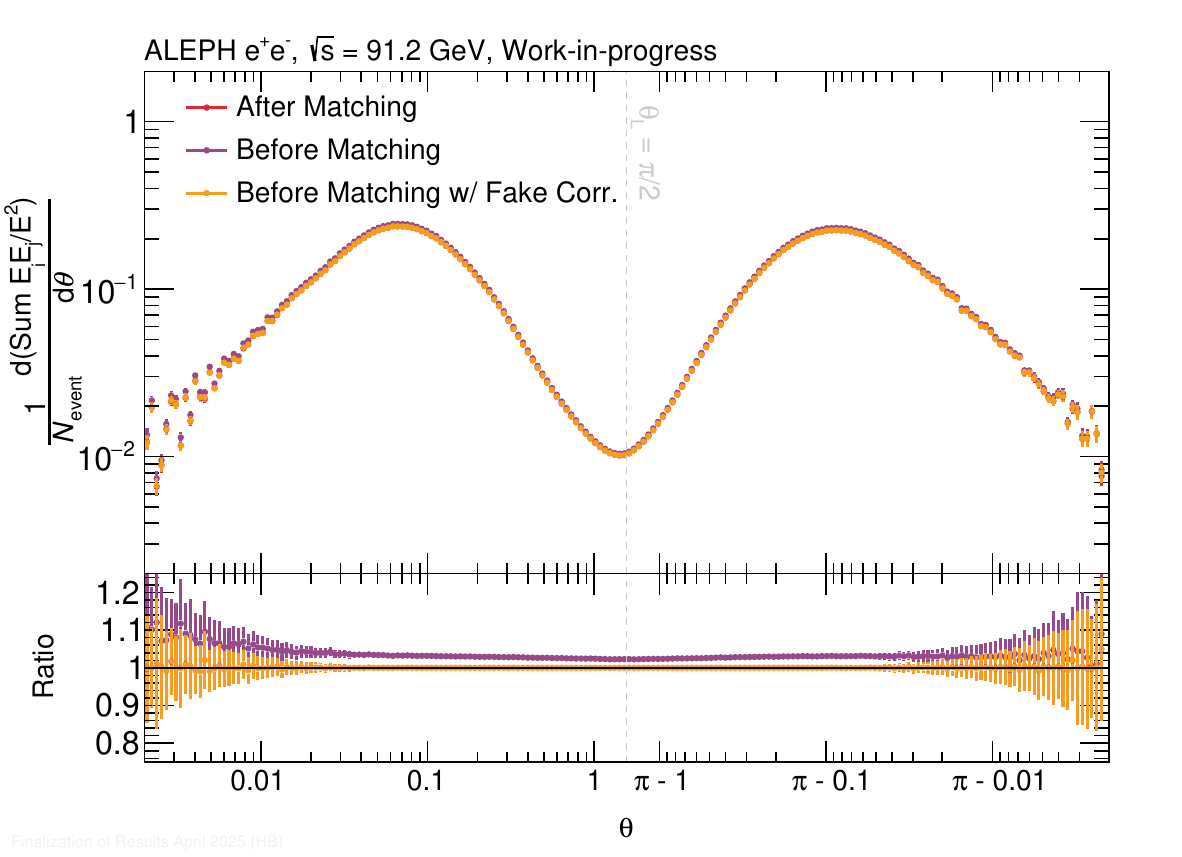}
    \includegraphics[width=0.49\linewidth]{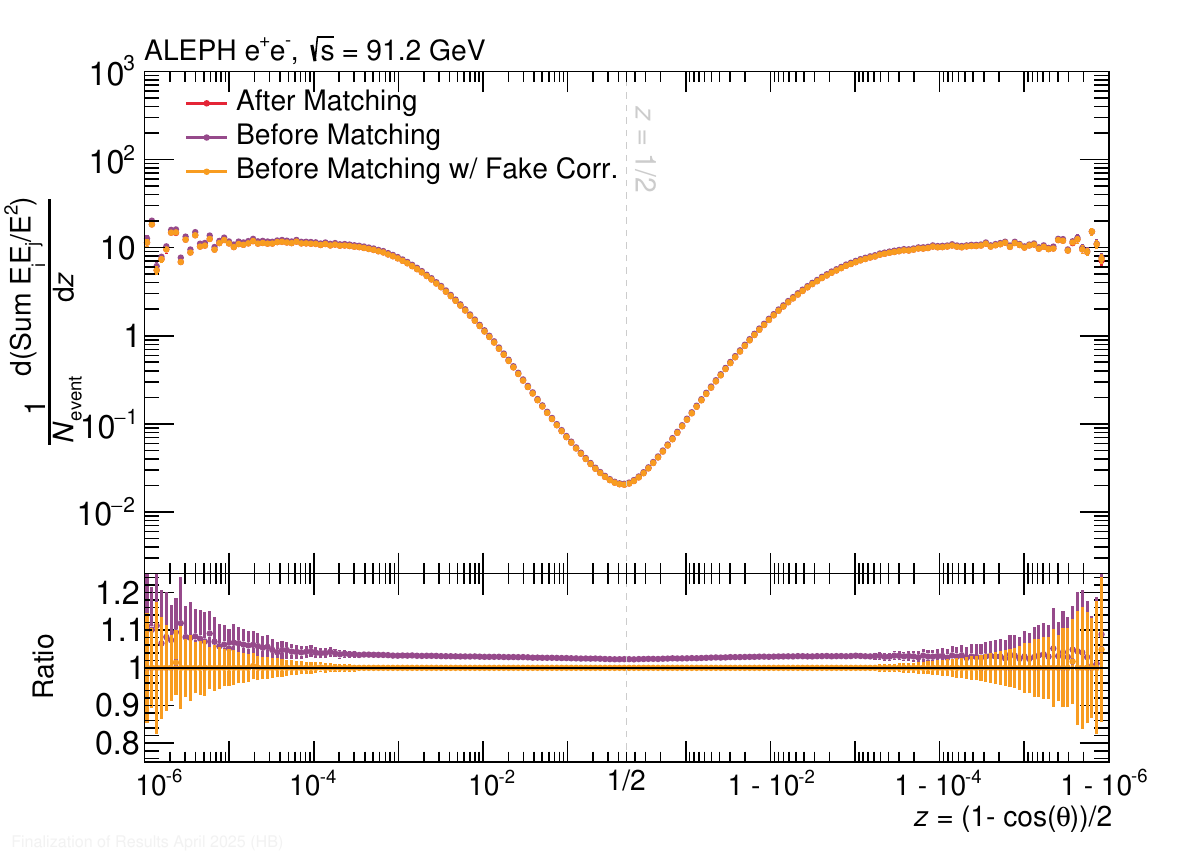}
    \caption{Closure plots for the fake correction as a function of $\theta$ (left) and $z$ (right). The red and purple curves represents the distributions after and before matching, respectively, from which the fake correction is derived. The before matching distribution with the fake correction applied is shown in yellow, demonstrating good closure.}
    \label{fig:fakeCorrectionClosure}
\end{figure}

\begin{figure}[ht!]
    \centering
    \includegraphics[width=0.49\linewidth]{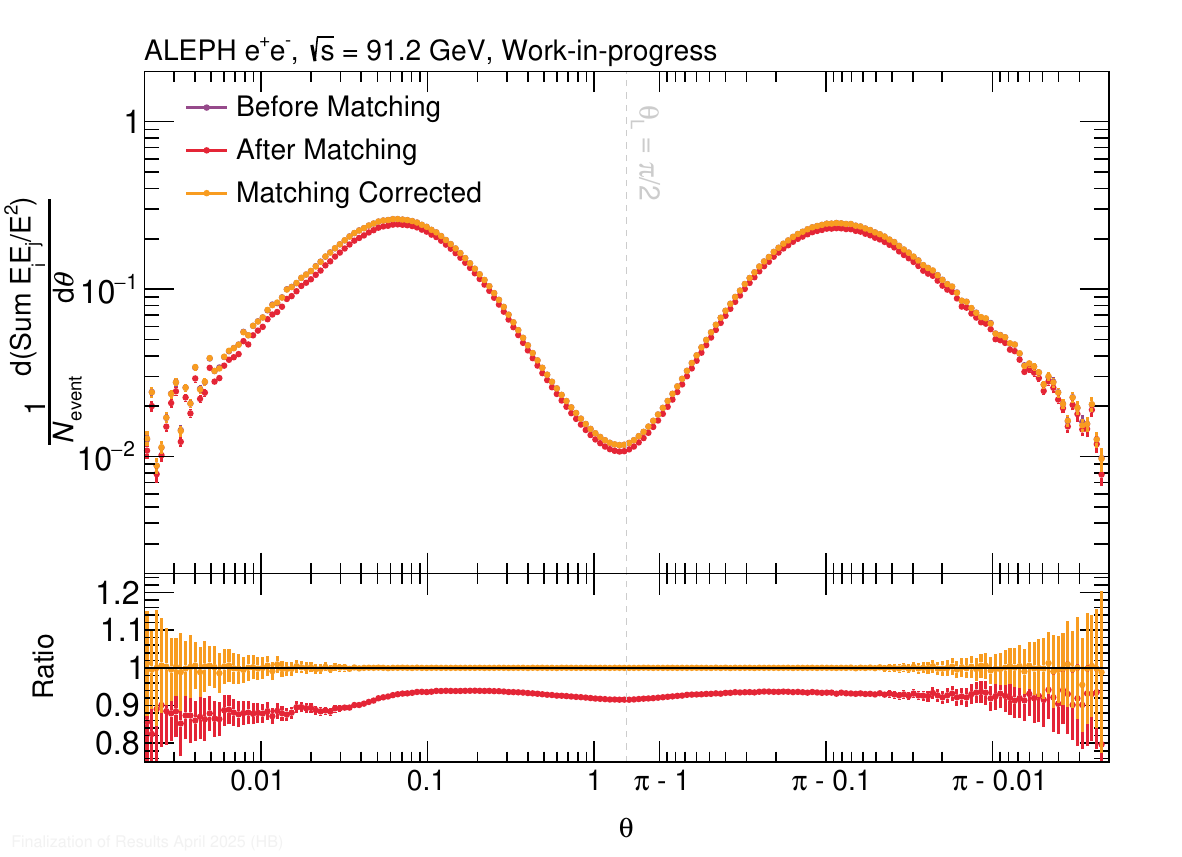}
    \includegraphics[width=0.49\linewidth]{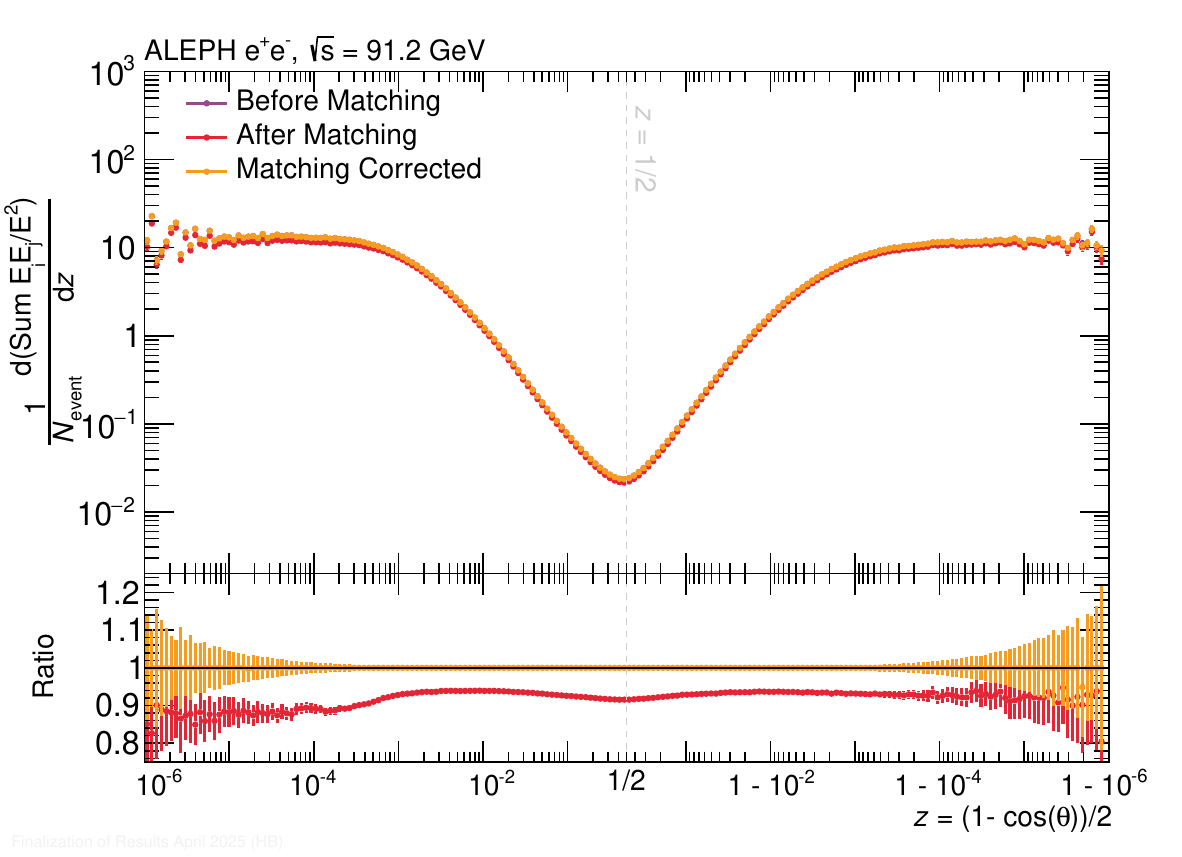}
    \caption{Closure plots for the matching efficiency correction as a function of $\theta$ (left) and $z$ (right). The red and purple curves represents the distributions after and before matching, respectively, from which the matching efficiency correction is derived. The after matching distribution with the matching efficiency correction applied is shown in yellow, demonstrating good closure.}
    \label{fig:matchingEffCorrectionClosure}
\end{figure}

\begin{figure}[ht!]
    \centering
    \includegraphics[width=0.49\linewidth]{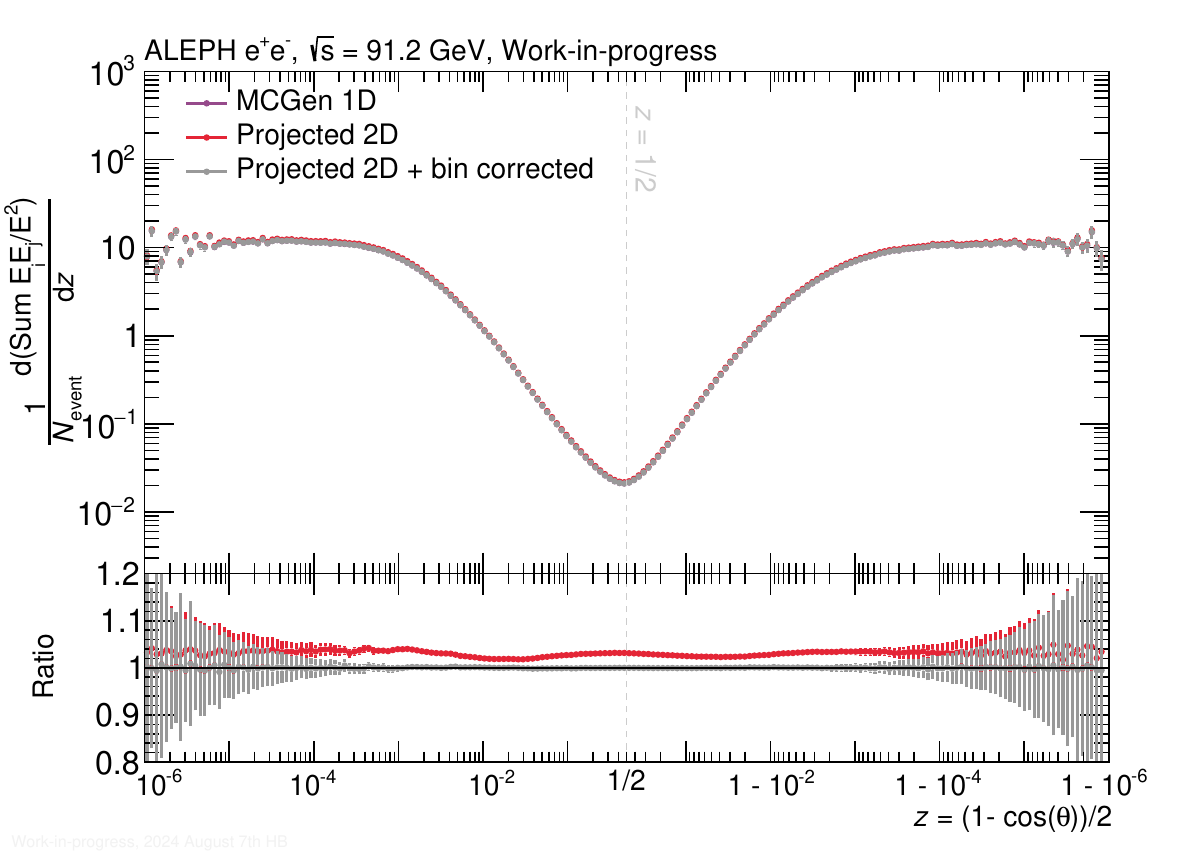}
    \includegraphics[width=0.49\linewidth]{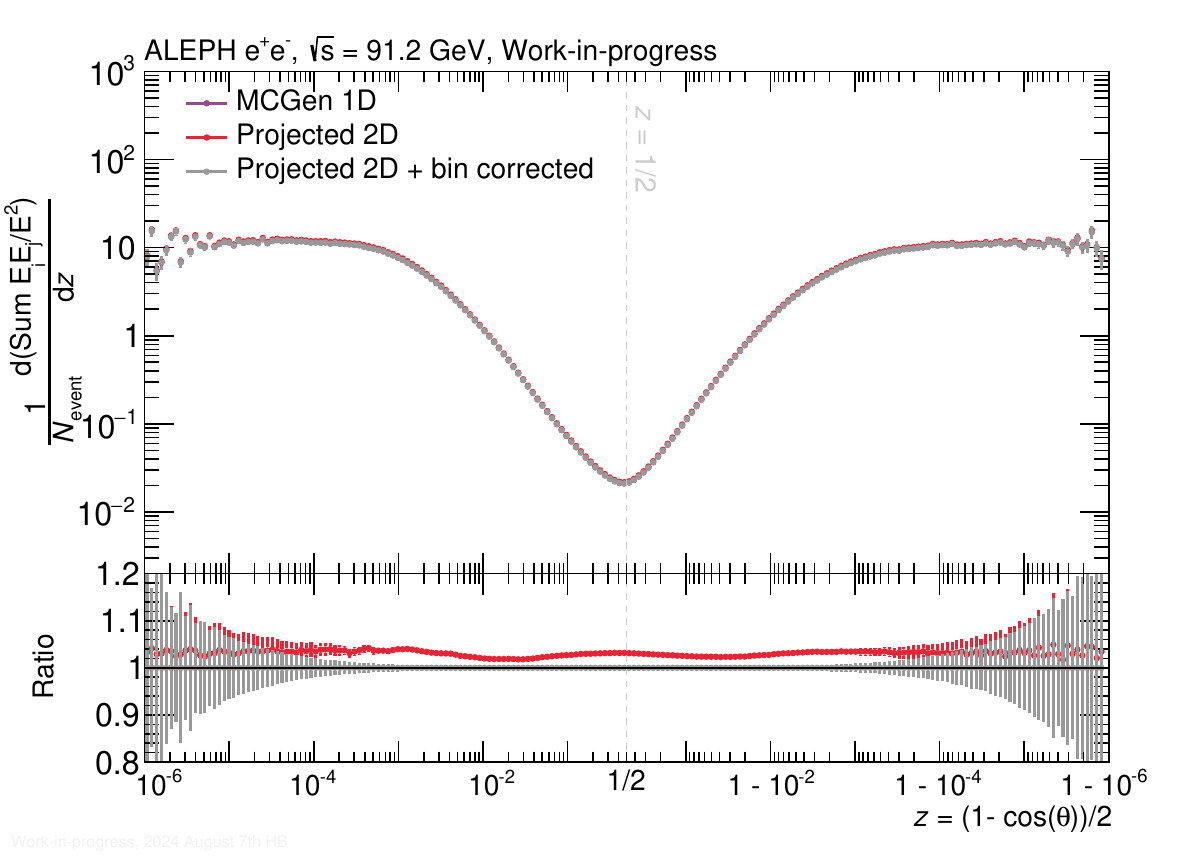}
    \caption{Closure plots for the binning correction as a function of $\theta$ (left) and $z$ (right). The purple and red curves represents the MC generator-level 1D distribution and projected 2D distributions, respectively, from which the binning correction is derived. The projected 2D version with the correction applied is shown in gray, demonstrating good closure.}
    \label{fig:binningCorrectionClosure}
\end{figure}

\begin{figure}[ht!]
    \centering
    \includegraphics[width=0.49\linewidth]{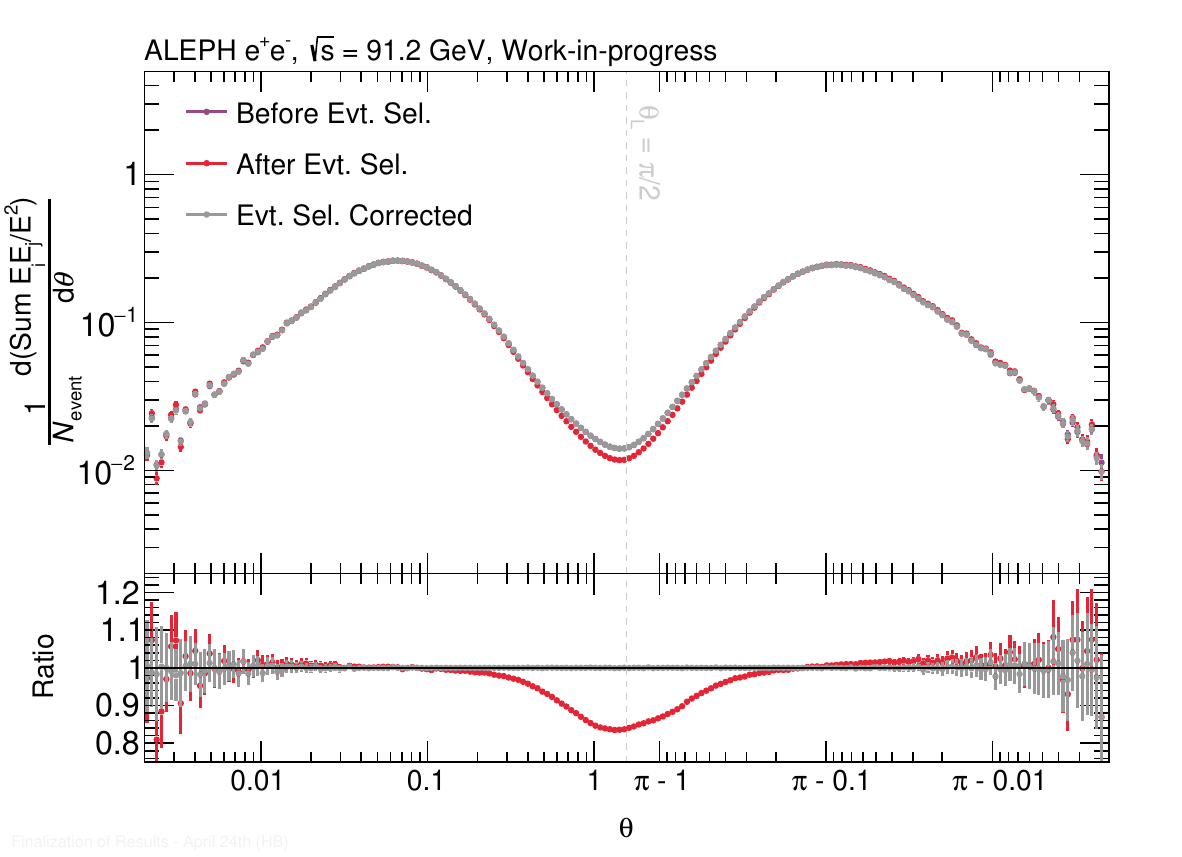}
    \includegraphics[width=0.49\linewidth]{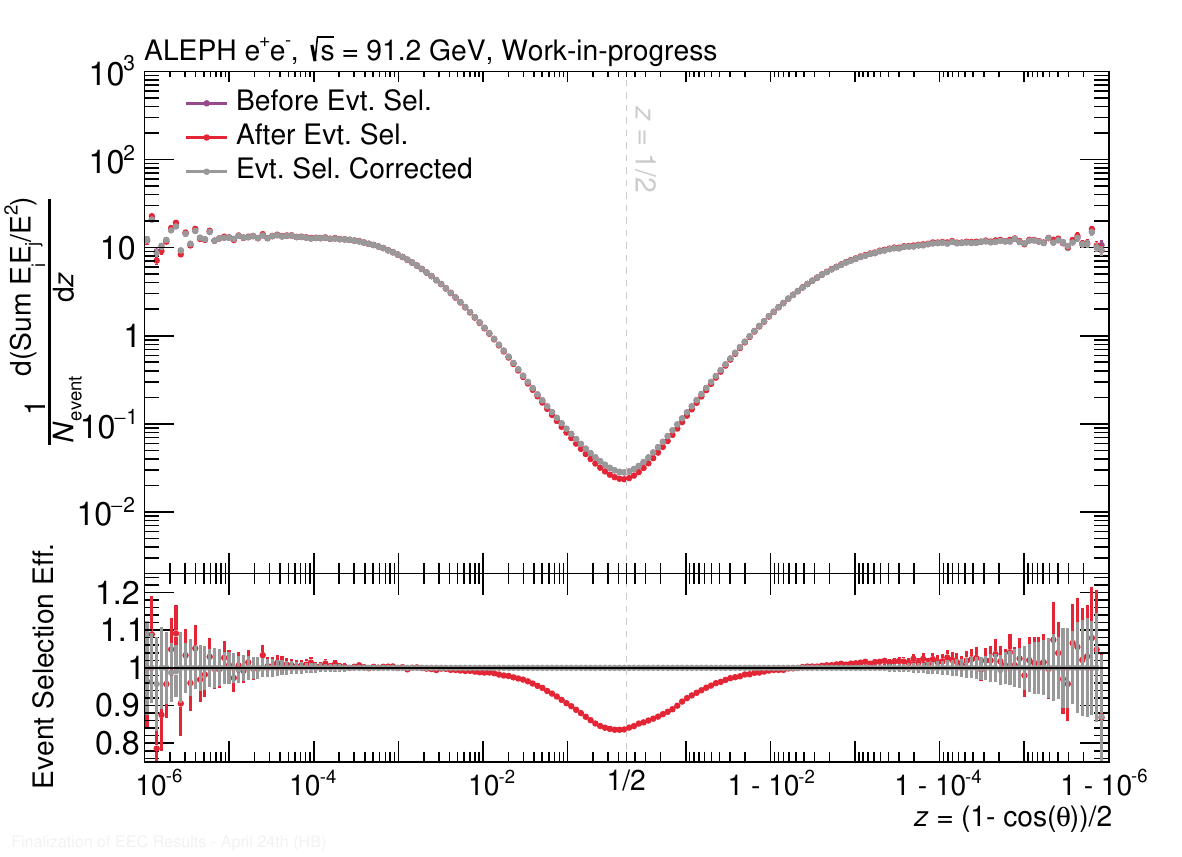}
    \caption{Closure plots for the event selection efficiency correction as a function of $\theta$ (left) and $z$ (right). The purple and red curves represents the distributions before and after event selection, respectively, from which the event selection efficiency is derived. The distribution after event selection with the event selection effiicency correction applied is shown in gray, demonstrating good closure.}
    \label{fig:eventSelectionEffCorrectionClosure}
\end{figure}

For each correction described in Section \ref{sec:effCorr}, a closure check was performed in order to ensure that the correction is being applied correctly. All efficiency corrections demonstrate good closure, indicating that the correction procedure is working as intended. The closure check for the fake correction is shown in Figure \ref{fig:fakeCorrectionClosure}. The matching efficiency correction closure is shown in Figure \ref{fig:matchingEffCorrectionClosure}. The closure check for the binning correction is shown in Figure \ref{fig:binningCorrectionClosure}. The closure check for the event selection efficiency correction is shown in Figure \ref{fig:eventSelectionEffCorrectionClosure}.

\clearpage
\section{Crosscheck of the Results with Independent Analysis Codebase}\label{app:crosscheck}
To cross-validate the primary results, we developed a fully independent codebase, which is publicly available at \url{https://github.com/jingyucms/AnalysisLEP/tree/main}. 
This cross-check differs from the main analysis in a number of aspects in the unfolding methods and the correction for the fakes and matching efficiency. 
These differences will be described in more detail in the remainder of this section. 
The results derived from this independent cross-check, performed solely for the E2C as a function of $\theta$ for simplicity, show a good agreement with the nominal results, validating the approach provided in the main analysis. 

\begin{figure}[ht!]
    \centering
    \includegraphics[width=0.45\linewidth]{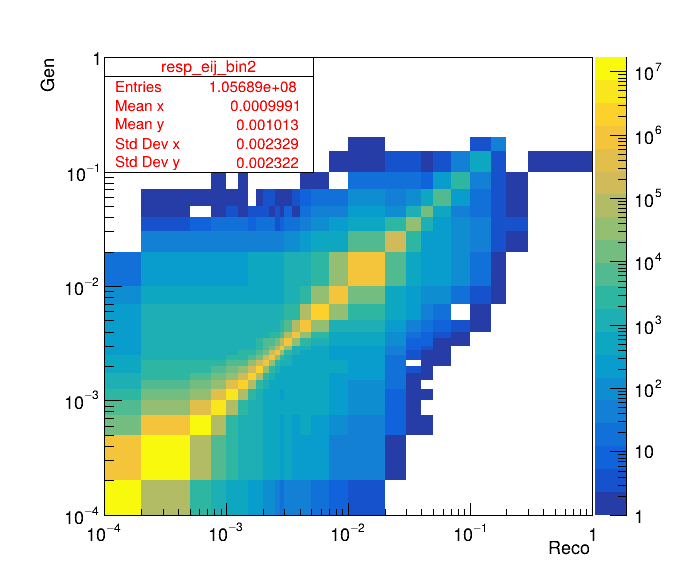}
    \includegraphics[width=0.45\linewidth]{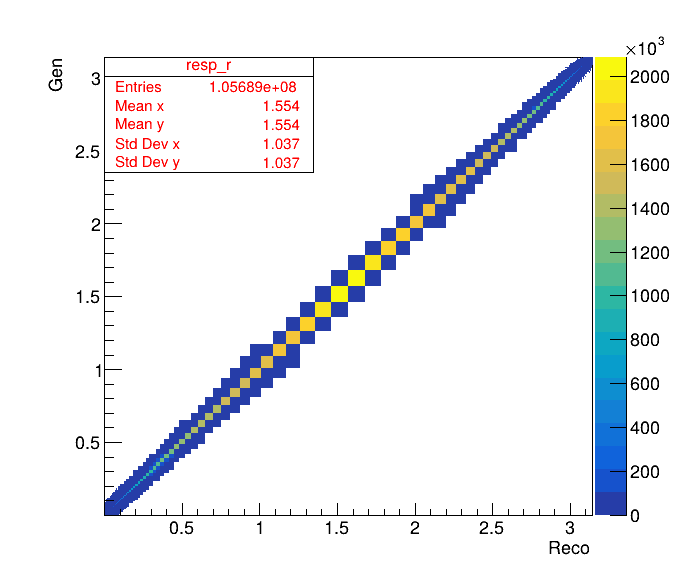}
    \caption{The response matrix, integrated over $\theta_{\rm L}$ (left) and $E_{i}E_{j}/E^{2}$ (right).}
    \label{fig:response}
\end{figure}

In the cross-check, track reconstruction inefficiencies, fake tracks, and track pairs migrating between bins of $E_{i}E_{j}/E^{2}$ and $\theta_{\rm L}$ due to finite track momentum and position resolution are corrected using the unfolding correction. 
The migration effects are characterized using a 2D response matrix that maps the generator-level distributions to detector-level ones in the archived $\textsc{pythia}$ 6 MC. 
To construct the 2D response matrix, events are selected based on the event selection criteria listed in Table~\ref{tab:SelectionSummary}. 
Detector-level charged particles are required to pass the selection requirements listed in the table. 
Generator-level charged particles are restricted to the same transverse momentum and polar angle ranges as the detector-level ones. 
The binning used in the $E_{i}E_{j}/E^{2}$ axis is the same as given in Section \ref{sec:response}:\texttt{(0.0, 0.0001, 0.0002, 0.0005, 0.00075, 0.001, 0.00125, 0.0015, 0.00175, 0.002, 0.00225, 0.0025, 0.00275, \newline 0.003, 0.0035, 0.004, 0.005, 0.007,  0.01, 0.02, 0.03, 0.04, 0.05, 0.07, \newline 0.10, 0.15, 0.20, 0.3, 1)}. 
The binning used in the $\theta_{\rm L}$ axis is also ``double log" style, and it has 200 bins ranging from $0.002$ to $\theta_{\rm L} = \pi - 0.001$. 
The selected generator- and detector-level charged tracks in the same events are matched if their difference in $\theta_{\rm L}$ is smaller than 0.05.
If one generator-level track is matched to multiple detector-level tracks or vice versa, the pair with the smallest distance in $\theta_{\rm L}$ is kept. 
The response matrices constructed with matched pairs and integrated over $E_{i}E_{j}/E^{2}$ and $\theta_{\rm L}$ are shown in Figure~\ref{fig:response}. 

\begin{figure}[ht!]
    \centering
    \includegraphics[width=0.45\linewidth]{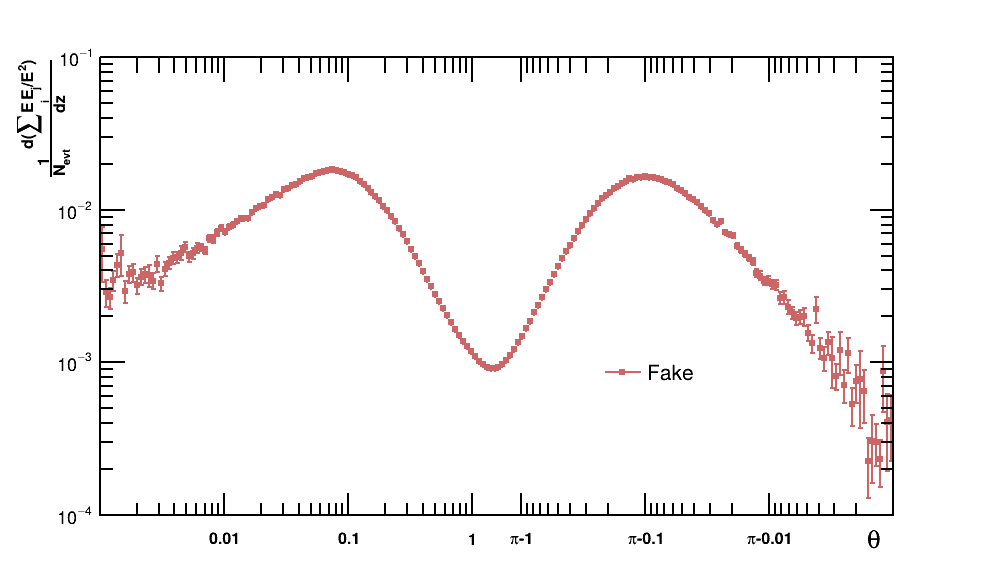}
    \includegraphics[width=0.45\linewidth]{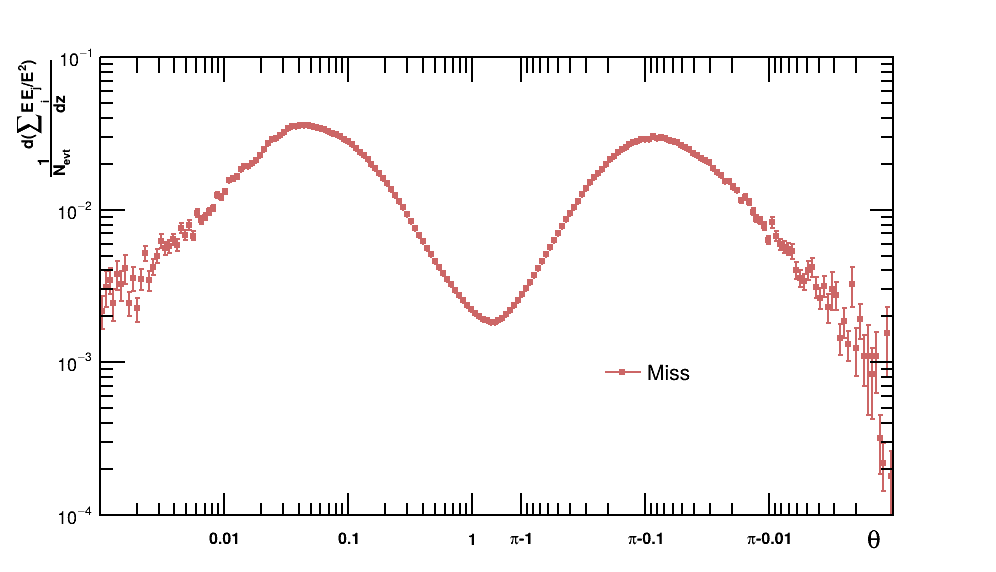}
    \caption{Fakes (left) and misses (right) from unmatched detector- and generator-level particles, respectively.}
    \label{fig:fakesAndMisses}
\end{figure}

The unmatched detector-level particles are the fake tracks. 
The unmatched generator-level particles are treated as inefficiencies (misses).
In the unfolding implementation below, fakes are treated as an additive background, and misses are treated as a multiplicative efficiency correction. 
The fakes and misses are shown in Figure~\ref{fig:fakesAndMisses}. 
The matching approach used in this crosscheck is more conservative than the one used in Section~\ref{sec:matching}. 
Therefore, it yields lower efficiency and provides a more conservative estimation of the migration effects and the final covariance matrix. 

\begin{figure}[ht!]
    \centering
    \includegraphics[width=0.75\linewidth]{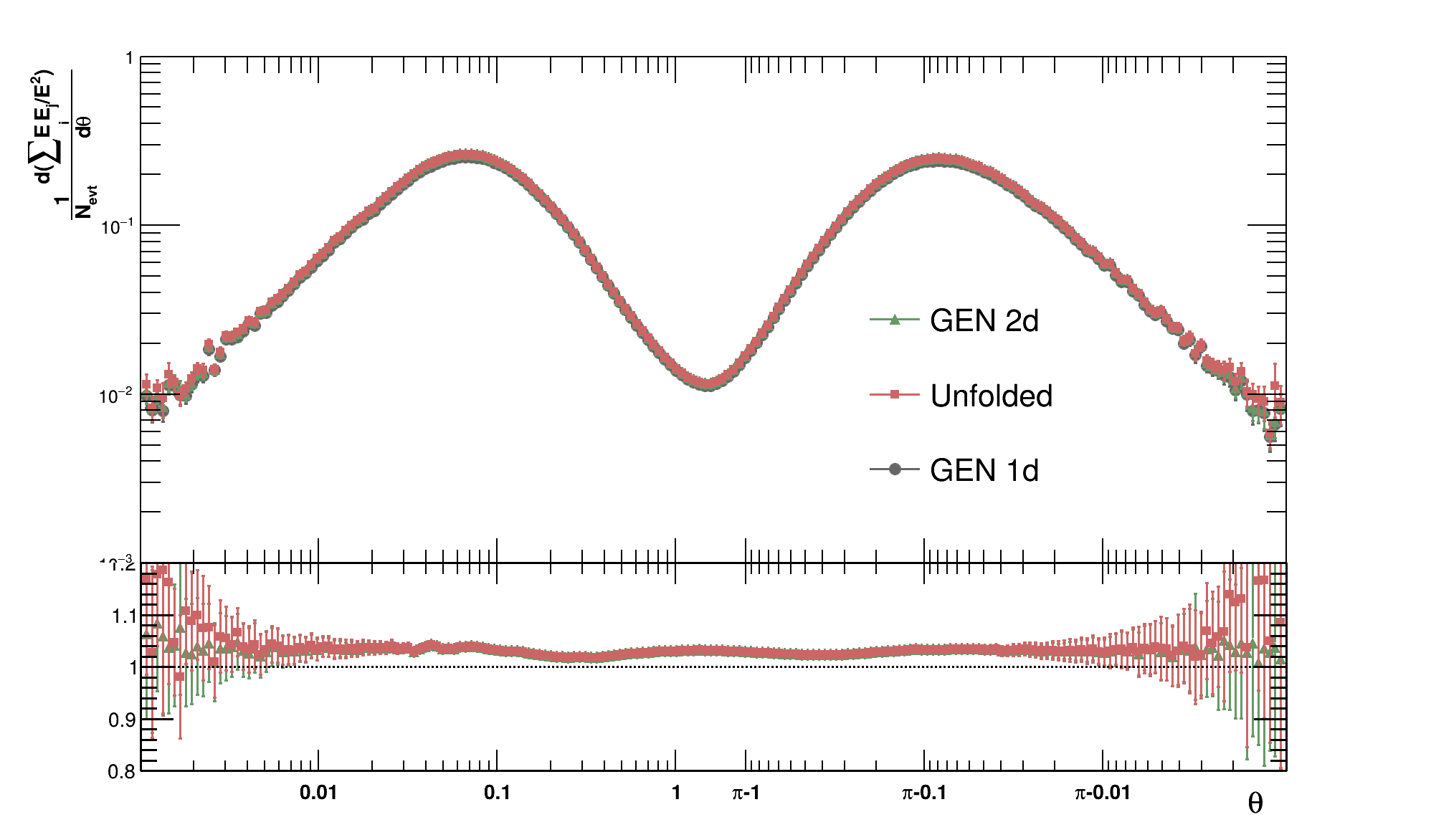}
    \caption{Comparison of the 2D unfolded (red), 2D generated (green), and 1D generated (black) distributions. The 2D distributions are projected to 1D using the procedure described in Section~\ref{sec:projections}.}
    \label{fig:trivialClosure}
\end{figure}

The unfolding is done using the D'Agostini iterative method~\cite{D'Agostini:265717} in the \texttt{RooUnfold}~\cite{Brenner:2019lmf} package. 
The D'Agostini method uses early stopping to regulate the smoothness of the unfolded distributions. 
The regularization parameter, defined by the number of iterations $N_{iter}$, is determined as follows: after each iteration, the $\chi^2/ndf$ between the distributions from the current and previous iterations is computed.
If the $\chi^2$ falls below 0.05, the current iteration is chosen as the optimal number of iterations.
Based on this criterion, $N_{iter}$ is set to 4 in this analysis. 
A trivial closure test comparing the unfolded and generated distributions using the same archived $\textsc{pythia}$ 6 MC is performed. 
Nearly perfect closure is obtained, as shown in Figure~\ref{fig:trivialClosure}. 
In the figure, the 2-dimensional $E_{i}E_{j}/E^{2}$ and $\theta_{\rm L}$ histogram is projected to one dimension taking the bin centers in each $E_{i}E_{j}/E^{2}$ bin as the approximation for the bin averages. 
This is the same procedure as the one described in section~\ref{sec:projections}. 
The generator-level distribution created in one dimension is also shown in the figure to illustrate the possible deviation of the projection.  

\begin{figure}[ht!]
    \centering
    \includegraphics[width=0.75\linewidth]{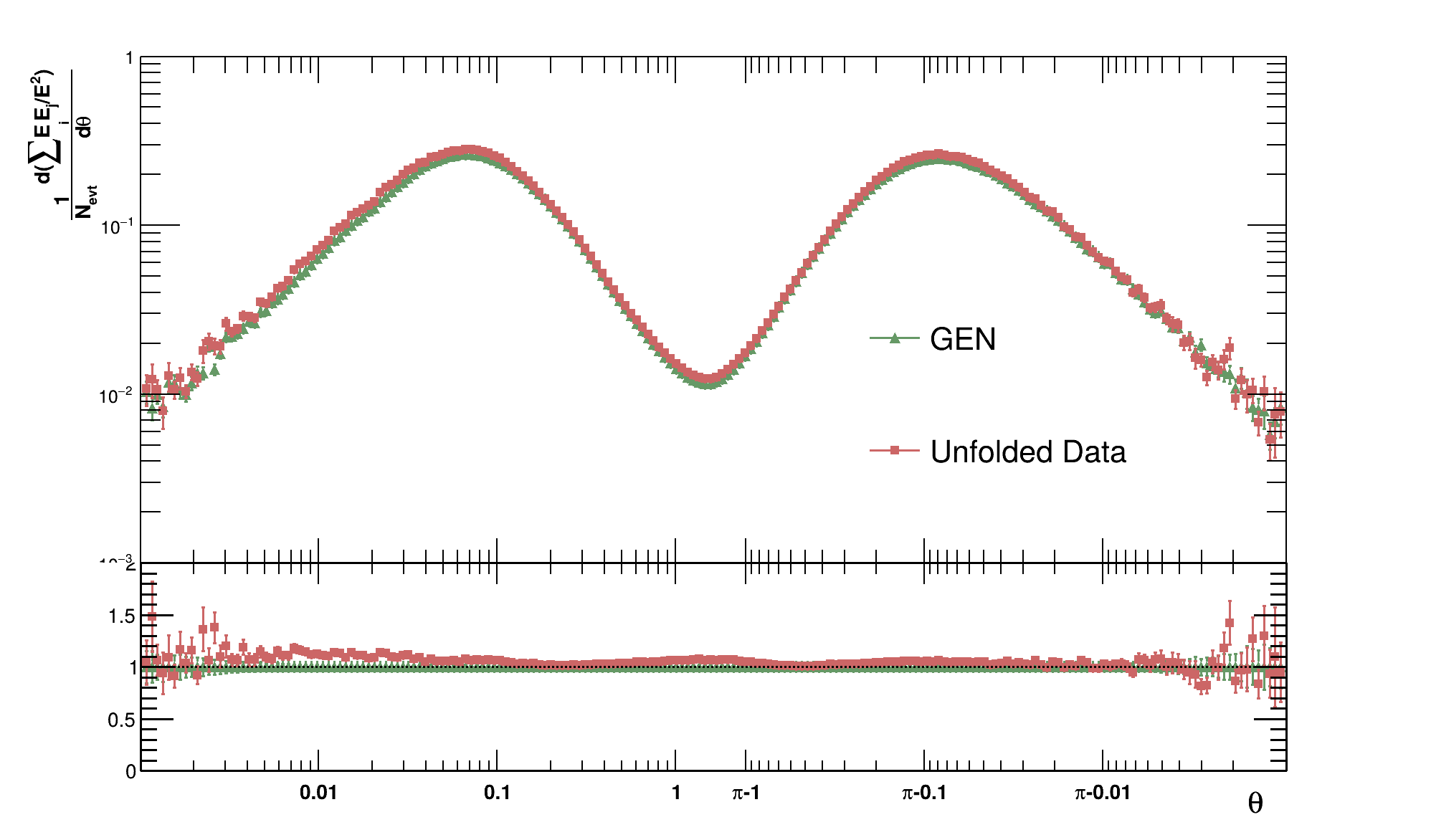}
    \caption{Unfolded ALEPH data distribution compared to generated one from the archived $\textsc{pythia}$ 6 MC. }
    \label{fig:unfoldData}
\end{figure}

In Figure~\ref{fig:unfoldData}, the resulting unfolded data distribution is compared to the generated one from the archived $\textsc{pythia}$ 6 MC. 
Both distributions are projected 1D. 
The main contribution of the remaining differences comes from the differences between data and archive MC at the detector level. 
Such effects are taken into account as a systematic uncertainty discussed in Section~\ref{sec:UnfSyst}. 

\begin{figure}[ht!]
    \centering
    \includegraphics[width=0.45\linewidth]{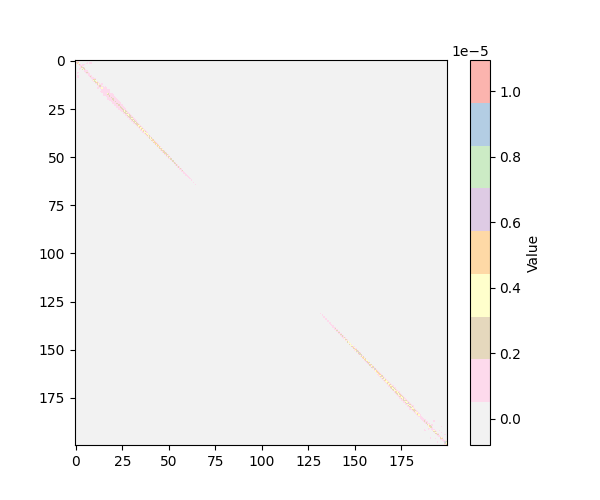}
    \includegraphics[width=0.45\linewidth]{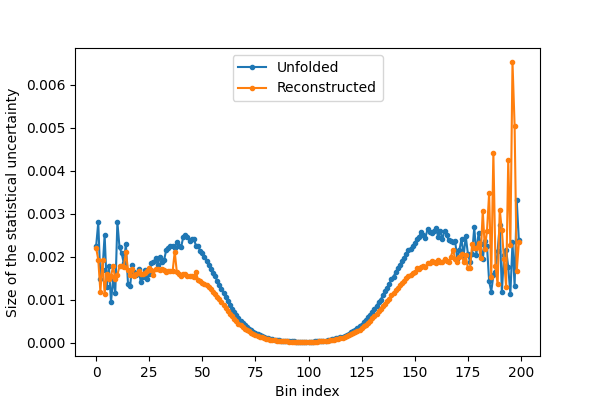}
    \caption{Left: covariance matrix of the unfolded distribution. Right: statistical uncertainty of the unfolded distribution compared to the reconstructed one. }
    \label{fig:covMatrix}
\end{figure}

The resulting covariance matrix calculated by \texttt{RooUnfold} is also shown in Figure~\ref{fig:covMatrix} left. 
Like the unfolded distribution, the 4D covariance matrix is also projected to 2D, approximating the bin averages using the bin centers in each $E_{i}E_{j}/E^{2}$ bin. 
The presence of the non-zero off-diagonal elements indicates that the unfolding procedure introduces statistical correlations to the unfolded distribution. 
As a sanity check, the statistical uncertainty, obtained from the diagonal elements of the covariance matrix, is compared to the statistical uncertainty from the reconstructed data distribution.
As shown in Figure~\ref{fig:covMatrix} right, the unfolding correction increases the statistical uncertainty by up to 50\%.

\begin{figure}[ht!]
    \centering
    \includegraphics[width=0.75\linewidth]{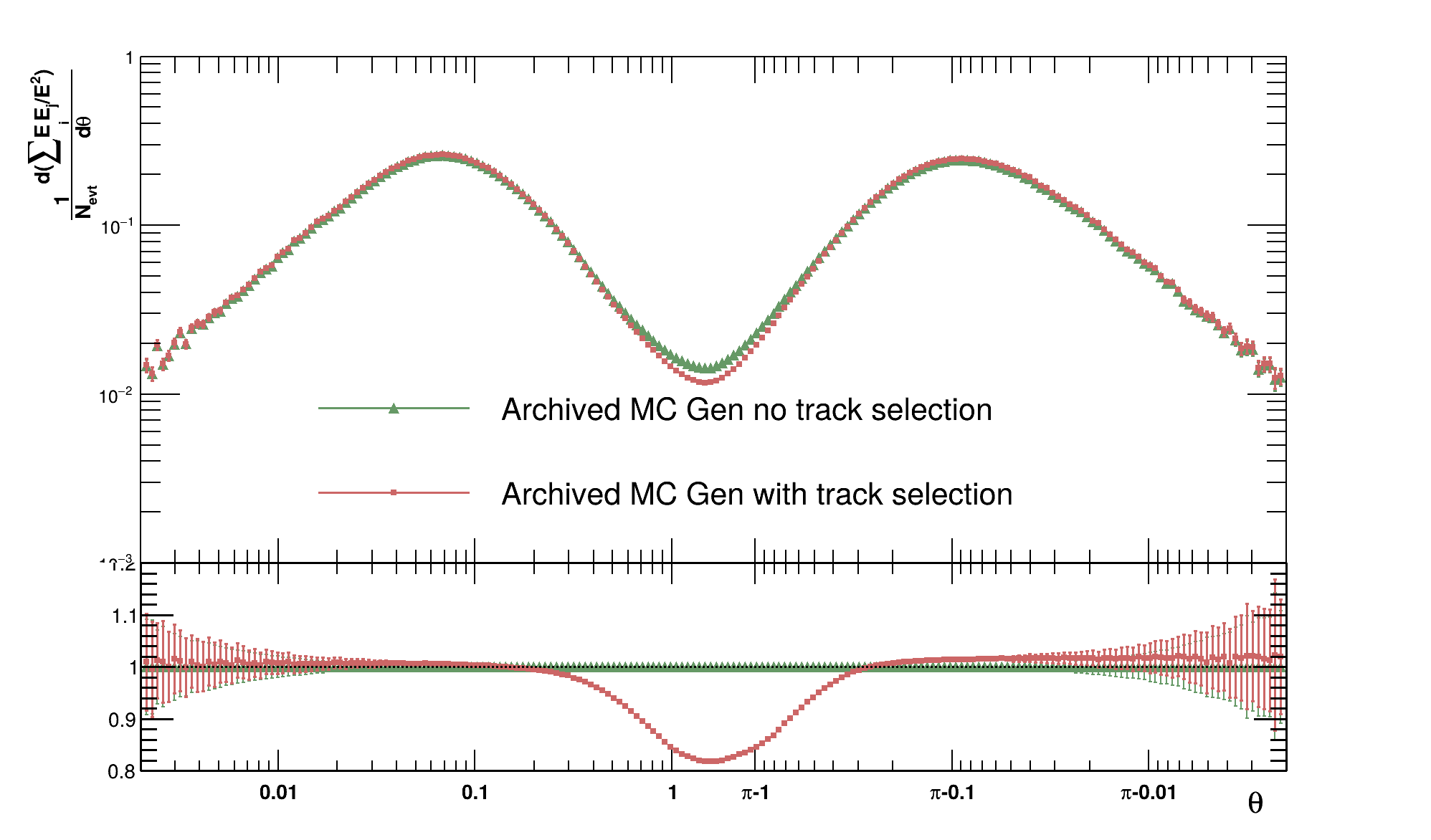}
    \caption{Comparison of the EEC distributions in the archived $\textsc{pythia}$ 6 MC with and without the track selections. Here, both distributions include the event selections.}
    \label{fig:evtcorr}
\end{figure}

\begin{figure}[ht!]
    \centering
    \includegraphics[width=0.75\linewidth]{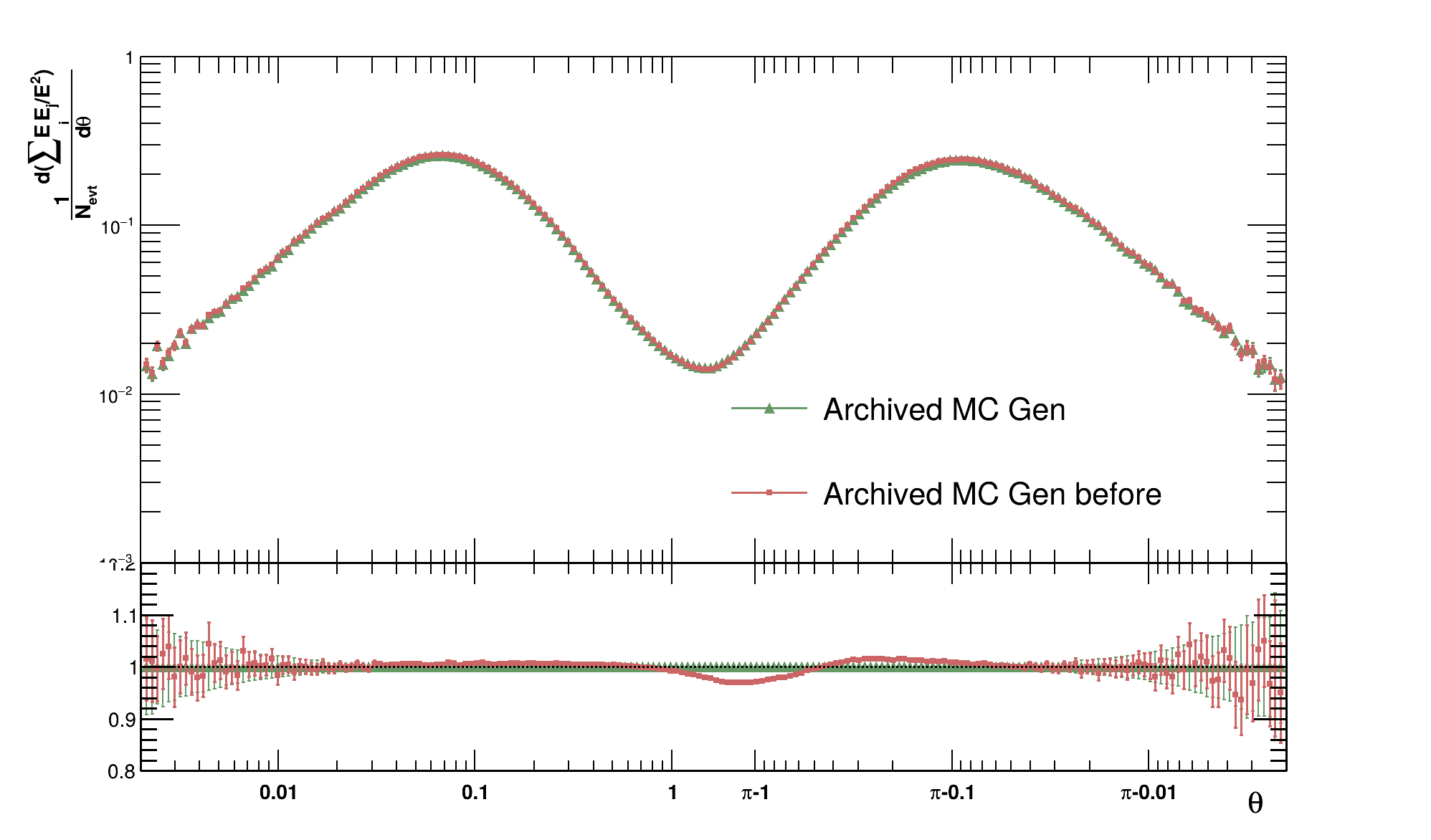}
    \caption{Comparison of the EEC distributions in the archived $\textsc{pythia}$ 6 MC with (Gen) and without (Gen before) the event selections.}
    \label{fig:trkcorr}
\end{figure}

In addition to presenting the unfolded distribution within the fiducial phase space, we correct the results to the full phase space for comparison with analytical theory predictions.
These corrections account for the track and event selections summarized in Table~\ref{tab:SelectionSummary}.
Correction factors are derived by comparing the EEC distributions in generator-level archived $\textsc{pythia}$ 6 MC, before and after the selections, and then applied bin‑by‑bin to the unfolded data. 
The comparisons are shown in Figure~\ref{fig:trkcorr} and Figure~\ref{fig:evtcorr} for track‑selection corrections and event‑selection corrections, respectively. 

\begin{figure}[ht]
    \centering
    \includegraphics[width=0.75\linewidth]{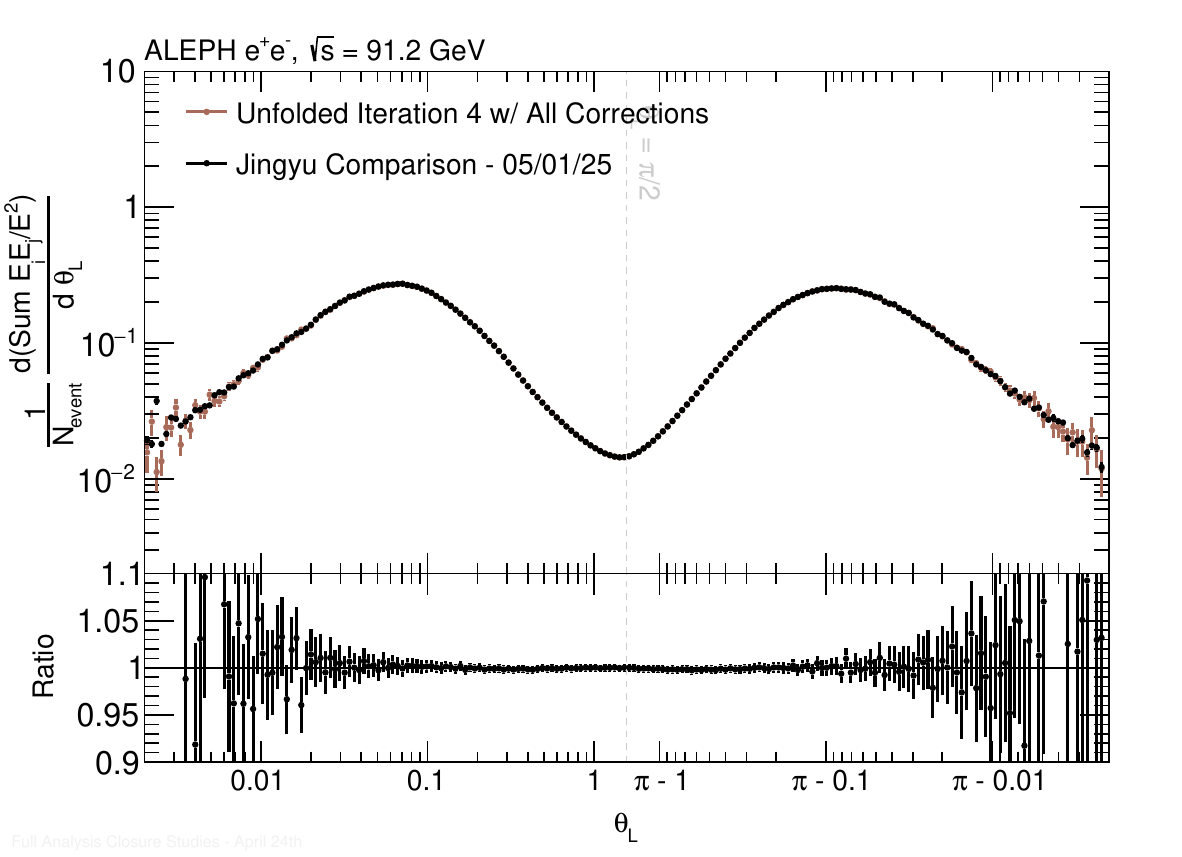}
    \caption{E2C distribution prior to binning and event selection corrections for the crosscheck analysis compared to the main analysis. The evaluated interpolation is shown as the solid black line. Note that no error bars (neither statistical nor systematic) are shown for the cross-check analysis. The ratio is shown in the bottom panel, where the error on the ratio points comes solely from the statistical error on the main analysis.}
    \label{fig:crosscheckFinalCompEvalRatio}
\end{figure}

The final results from this independent cross-check are shown and compared to the results from the nominal framework presented in the body of this note in Figure \ref{fig:crosscheckFinalCompEvalRatio}  
Note that in the final result only points from 0.006 to $\pi - 0.006$ are reported for $\theta$, but a broader range than this is reported here for the purposes of completeness. The deviation in the central values between the two procedures is shown in the bottom panel of Figure \ref{fig:crosscheckFinalCompEvalRatio}, where deviations on the order of $< 1\%$ can be seen over the kinematic range reported in the final measurement, validating the procedure. 

\end{document}